\documentclass[fleqn,usenatbib]{mnras}

\usepackage{newtxtext,newtxmath}

\usepackage[T1]{fontenc}

\DeclareRobustCommand{\VAN}[3]{#2}
\let\VANthebibliography\thebibliography
\def\thebibliography{\DeclareRobustCommand{\VAN}[3]{##3}\VANthebibliography}

\usepackage{graphicx}	
\usepackage{subcaption}
\usepackage{amsmath}	
\usepackage{anyfontsize}
\usepackage{adjustbox}
\usepackage{xcolor} 
\usepackage{enumitem}
\usepackage{algorithm}
\usepackage{algpseudocode}
\usepackage{soul}

\title[Filter Refraction]{Modeling the impact of filter-substrate refraction in the \textit{Roman} point spread function}

\author[F. Berlfein et al.]{Federico Berlfein$^{1}$\thanks{E-mail: fberlfei@andrew.cmu.edu},
Rachel Mandelbaum$^{1}$,
Tianqing Zhang$^{2}$,
Nihar Dalal$^{3,4}$,
Christopher M. Hirata$^{3,4,5}$,
\newauthor
Charuhas Shiveshwarkar$^{3,4}$
and Anthony Harbo Torres$^{3,4}$
\\
$^{1}$ McWilliams Center for Cosmology and Astrophysics, Department of Physics, Carnegie Mellon University, Pittsburgh, PA 15213, USA \\
$^{2}$ Department of Physics and Astronomy and PITT PACC, University of Pittsburgh, Pittsburgh, PA 15260, USA
\\
$^{3}$ Center for Cosmology and Astroparticle Physics, 191 West Woodruff Avenue, Columbus, OH 43210, USA
\\
$^{4}$ Department of Physics, The Ohio State University, 191 West Woodruff Avenue, Columbus, OH 43210, USA
\\
$^{5}$ Department of Astronomy, The Ohio State University, 140 West 18th Avenue, Columbus, OH 43210, USA
}

\date{Accepted XXX. Received YYY; in original form ZZZ}

\pubyear{2026}

\begin{document}
\label{firstpage}
\pagerange{\pageref{firstpage}--\pageref{lastpage}}
\maketitle


\begin{abstract}
For broadband imaging surveys, filter-substrate refraction causes light at different wavelengths to follow slightly different paths through the filter substrate before reaching the detector. This effect produces two chromatic perturbations to the point spread function (PSF): a shift in the effective focal position along the optical axis (longitudinal shift), which manifests as a defocus-like perturbation, and a wavelength-dependent displacement of the image position in the focal plane (lateral shift), which manifests as image decentering. Using image simulations, we provide the first study of these two effects independently across all eight \textit{Roman} imaging bands and over the full focal plane. We compute the resulting PSF and photometric errors from images with and without the effect included, and compare the magnitude of the effect to the \textit{Roman} science requirements. 
We find that the lateral shift is the dominant contribution, producing PSF size and ellipticity residuals in most bands of order $\sim\!0.3$--$0.4\%$. 
These exceed the \textit{Roman} science requirements for weak lensing by roughly an order of magnitude. The effect is also strongly field dependent, increasing toward the edges of the focal plane. By contrast, flux residuals remain below one third of the $1\%$ requirement for most bands, except in R062 and W146. We find the longitudinal shift to be subdominant and negligible in most bands, including the weak lensing bands. Finally, we implement the dominant lateral-shift effect in a framework suitable for large-scale image simulations and validate that the resulting PSF size and shape changes are accurately reproduced. Overall, we find that filter-substrate refraction is a relevant chromatic effect for \textit{Roman} PSF modeling, and we provide tools to model and incorporate it in large-scale image simulations. 
\end{abstract}






\section{Introduction}\label{sec:introduction}

Weak gravitational lensing (WL) is one of the primary cosmological probes \citep{Bernstein_2002, Hoekstra_2008, Mandelbaum_2018} that will be pursued by the \textit{Nancy Grace Roman Space Telescope} \citep{Spergel_2015, Akeson_2019}. 
\textit{Roman}'s combination of wide area, near-infrared imaging, and stable space-based observing conditions is designed to support, among other things, high-precision measurements of weak lensing and large-scale structure \citep{Troxel_2021, Liaudat_2023}. Making use of the full statistical power of these observations, however, requires exceptional control of instrumental systematics, among which point spread function (PSF) modeling is particularly important \citep{Liaudat_2023}. 
Since an observed galaxy image is the convolution of the true surface-brightness distribution of the galaxy with the PSF, inaccuracies in the modeled PSF can propagate directly into biases in galaxy shape measurements and hence into the inferred weak lensing signal \citep{Mandelbaum_2018}. For the \textit{Roman} High-Latitude Wide Area Survey (HLWAS) in particular, stringent requirements on knowledge of PSF size and ellipticity \citep{SRD_2023} reflect the need to keep such errors below the level at which they would bias cosmological constraints.

An important problem in PSF modeling is that broadband PSFs are intrinsically chromatic \citep{Meyers_2015b, Carlsten_2018, Eriksen_2018, Schutt_2025}. Understanding the instrumental or astrophysical effects generating chromatic effects on the PSF is therefore essential for constructing an accurate PSF model for weak lensing analyses. The origin and importance of these effects, however, differ substantially between space-based and ground-based telescopes, since not only the optical design of the telescope changes, but also the role of the atmosphere. In ground-based surveys, the PSF is strongly influenced by atmospheric effects \citep{McKechnie_1992, Meyers_2015b}, whereas for space-based telescopes, such as \textit{Roman}, the PSF is dominated by diffraction \citep{Cypriano_2010, Carlsten_2018}. As a result, the PSF in space-based telescopes can be more strongly wavelength dependent than in ground-based surveys \citep{Cypriano_2010}, making chromatic PSF effects a particularly important concern.

Although the dominant chromaticity of the \textit{Roman} PSF arises from diffraction, the telescope is not completely free of refractive elements. Within the \textit{Roman} Wide Field Instrument (WFI), the filters are among the few transmissive elements in the optical system, and are made of fused-silica substrates \citep{Cromey_2023} whose refractive index varies with wavelength. As a result, photons at different wavelengths can follow slightly different paths as they pass through the filter substrate, introducing an additional chromatic perturbation on top of the already wavelength-dependent PSF. It is therefore important to understand how filter-substrate refraction enters the \textit{Roman} PSF, and assess whether the resulting perturbations are large enough to require modeling or correction for precision cosmological analysis with \textit{Roman}.

Previous work has already shown that chromatic PSF effects can be important for precision weak lensing surveys, although looking at entirely different chromatic effects from the one considered in this work. 
For ground-based surveys, wavelength-dependent refraction has been studied primarily in the context of atmospheric differential chromatic refraction and chromatic seeing \citep{Plazas_2012, Meyers_2015b}, which if left uncorrected can produce centroid shifts and PSF distortions large enough to exceed the systematic-error budgets of surveys such as DES \citep{DES_2005} and LSST \citep{LSST_2009,LSST_2019}. For space-based missions, related work has instead focused on chromatic PSF biases arising from the wavelength dependence of the optical response and from color differences between stars and galaxies used for PSF calibration in the context of wavelength-dependent diffraction effects  \citep{Eriksen_2018, Berlfein_2025}. In both cases, chromatic PSF effects were shown to be a significant source of PSF modeling error for Stage~IV\footnote{Stage IV surveys should collectively achieve or exceed a factor of 10 gain over Stage II experiments in the dark energy figure of merit \citep{Albrecht_2006}.} weak lensing measurements. 

Recent work has also studied chromatic effects associated with the \textit{Roman} WFI filters themselves. \cite{Switzer_2025} characterized the field dependence of the WFI filter transmission, showing that angle-of-incidence effects and variations in filter coating thickness across the field of view produce spatially varying shifts in the effective bandpass. This characterization models the chromatic amplitude response of the filter by integrating transmission over the pupil, 
and is therefore distinct from the chromatic phase 
effects arising from wavelength-dependent refraction through the filter substrate. For Euclid, \cite{EuclidSchirmer_2022} characterized the spatial variability of the Near-Infrared Spectrometer and Photometer (NISP) filter transmission and angle-of-incidence effects, showing that these produce bandpass blueshifts with implications for photometric redshift accuracy. Separately, chromatic effects arising from coating thickness variations in the dichroic mirror have also been characterized and modeled for Euclid \citep{Baron_2023}. These studies make clear that chromaticity can be an important source of systematic error in imaging surveys, but the specific effect considered in this work is physically distinct: it arises from wavelength-dependent refraction from the \textit{Roman} WFI filter substrate.

Given \textit{Roman}'s diffraction-limited nature, filter-substrate refraction has received comparatively less attention in the context of PSF modeling. 
Existing large-scale image simulations of the \textit{Roman} WFI have used highly realistic PSF models\footnote{Realistic \textit{Roman} PSF models can be found in \texttt{STPSF} \citep{Perrin_2014} and in the \texttt{GalSim} \citep{Galsim} \textit{Roman} module \citep{Kannawadi_2016}.} that include a wide range of relevant effects, including telescope aberrations, detector effects, and observing conditions \citep{Troxel_2022, OU_2025}. However, they do not currently include a dedicated treatment of wavelength-dependent refraction by the filter substrate. This effect is potentially important for two reasons. First, because \textit{Roman} is diffraction limited already, the contributions from refraction become coupled with the pre-existing wavelength dependence of the PSF. Second, the magnitude of the effect is expected to vary with both field position and wavelength, making it potentially relevant for PSF calibration and image simulation.

In this paper, we isolate and quantify the impact of filter-substrate refraction on the \textit{Roman} PSF. We show that the effect can be understood in terms of two wavelength-dependent geometric quantities: a lateral chromatic shift, which induces a wavelength-dependent displacement of the image position in the focal plane, and a longitudinal chromatic shift, which moves the location of best focus position for a given photon along the optical axis by an amount depending on its wavelength. We derive an analytic formalism for these shifts, relate them to standard wavefront-aberration terms, and evaluate them using \textit{Roman}-specific optical parameters. We then validate the analytic model against a full \textit{Roman} ray-tracing calculation, propagate the resulting perturbations into PSF size, shape, and point-source flux errors using image simulations, and finally develop a practical implementation suitable for large-scale image simulations.

The remainder of this paper is organized as follows. In Sec.~\ref{sec:refraction_formalism}, we develop the formalism for chromatic refraction in the \textit{Roman} filter substrate and describe the \textit{Roman}-specific optical inputs required to evaluate it. In Sec.~\ref{sec:image_sims}, we describe how these chromatic shifts are incorporated into \textit{Roman} PSF image simulations. In Sec.~\ref{sec:results}, we validate the analytic model against ray tracing and quantify the resulting PSF and photometric errors across bands and field position. In Sec.~\ref{sec:practical_implementation}, we present a practical implementation of the dominant effect for large-scale image simulations. We summarize our conclusions in Sec.~\ref{sec:conclusion}.
\section{Refraction Formalism}\label{sec:refraction_formalism}

In this section, we establish the formalism used throughout the paper to model the effect of filter-substrate refraction on the \textit{Roman} PSF. The main objective is to describe how the wavelength dependence of the filter refractive index produces chromatic shifts in both the image position and the location of best focus, and to express those shifts in a form that can later be incorporated into image simulations. We first outline the physical origin of the effect and derive an analytic prescription to calculate the chromatic shifts. We then relate these shifts to the corresponding wavefront-aberrations, and finally describe the \textit{Roman}-specific optical and design specifications needed to calculate the relevant quantities.

\subsection{Description of problem}\label{subsec:desc_problem}

The point spread function (PSF) of a broadband imaging system is intrinsically chromatic \citep{Meyers_2015b,Carlsten_2018, Eriksen_2018}: photons of different wavelengths do not, in general, follow exactly the same path through the optical system. For a diffraction-limited, space-based telescope such as \textit{Roman}, this wavelength dependence is especially important \citep{Cypriano_2010,Berlfein_2025}. Although \textit{Roman} does not suffer from atmospheric refraction \citep[e.g.][]{Meyers_2015b}, it is not free of refractive elements. In particular, the WFI filters are among the main transmissive elements in the optical system, so wavelength-dependent refraction in the filter substrate can perturb an already chromatic PSF. In this work, we focus on how refraction in the \textit{Roman} WFI filter substrate modifies the PSF, which becomes an error in the PSF model if not properly accounted for.

\begin{figure}
    \centering
    \includegraphics[width=0.99\linewidth]{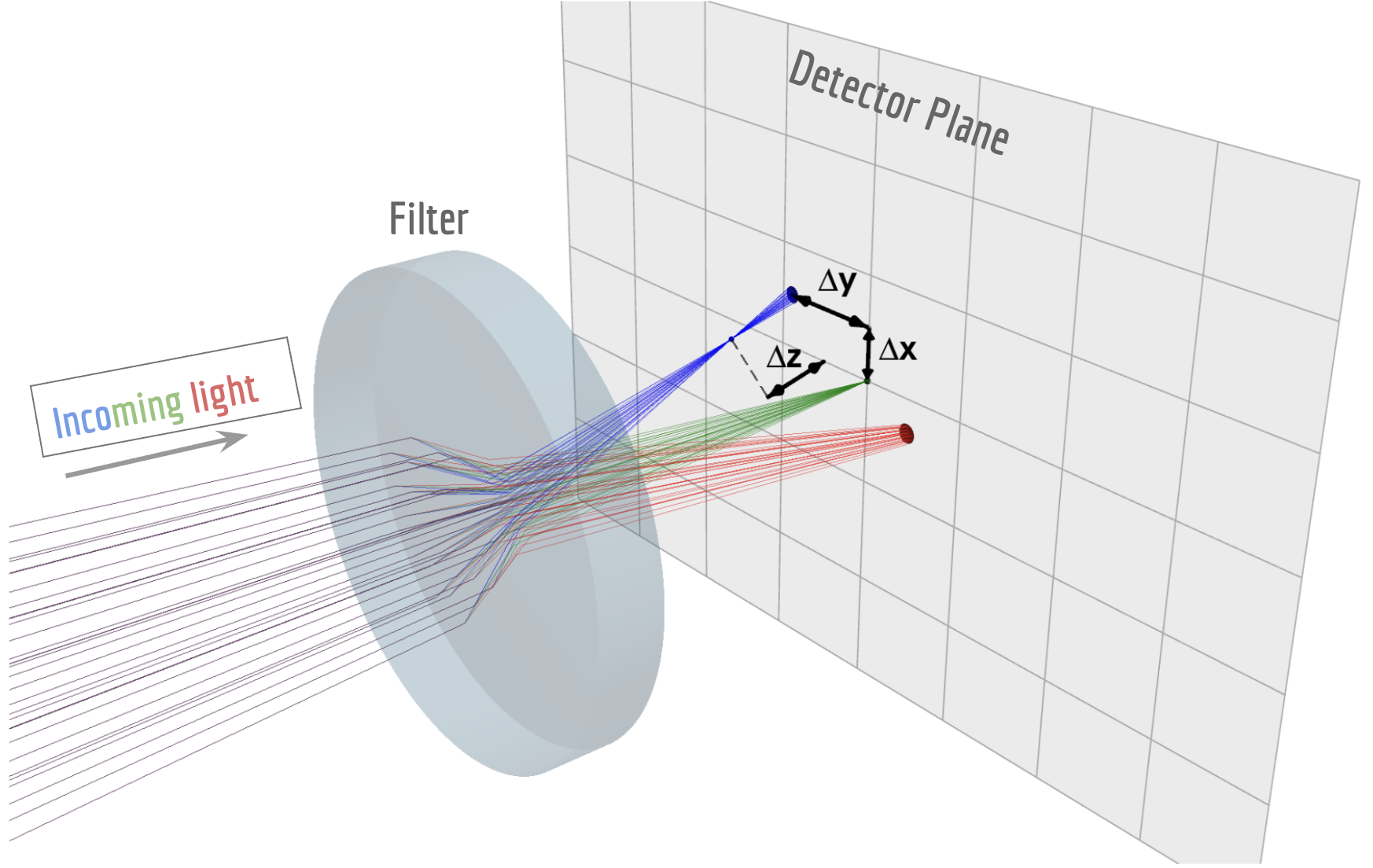}
    \caption{Schematic illustration of chromatic refraction in the \textit{Roman} WFI filter substrate, strongly exaggerated for visualization purposes. As shown, light rays of different wavelengths refract by slightly different amounts as they pass through the curved transmissive filter, causing them to propagate in different directions and reach their focus position at different (wavelength-dependent) locations relative to the nominal detector plane. This produces both wavelength-dependent lateral shifts in the focal plane, denoted $\Delta x$ and $\Delta y$, and a longitudinal shift in the best-focus position along the optical axis, denoted $\Delta z$. In the broadband PSF, these chromatic displacements manifest as wavelength-dependent image translations and defocus-like perturbations.}
    \label{fig:effect_visualization}

\end{figure}

\begin{figure*}
    \centering
    \includegraphics[width=1.0\linewidth]{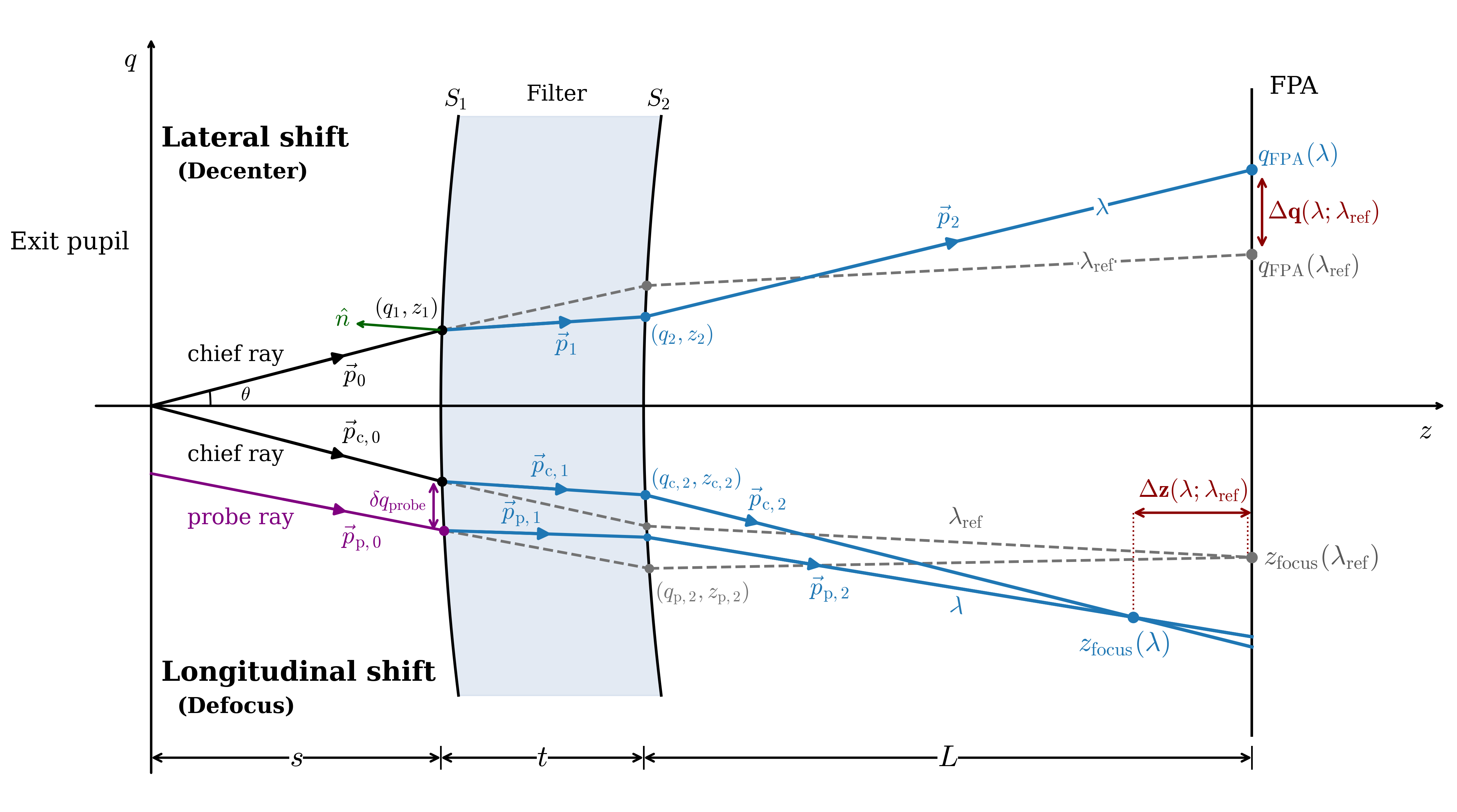}
    \caption{
    Schematic 2-D diagram of the notation and ray geometry described in Sec.~\ref{subsec:an_prescription} to model filter-substrate refraction. The horizontal and vertical axis represent the optical-axis coordinate $z$ and transverse coordinate $q$, respectively. The incident beam is represented by its chief ray, which leaves the exit pupil at an angle $\theta$ relative to the optical axis. The filter is represented by two curved refracting surfaces, $S_1$ and $S_2$, separated by the central thickness $t$. The exit pupil is located a distance $s$ from $S_1$, and the focal plane array (FPA) is located a distance $L$ from $S_2$.  The upper half shows the single-chief-ray construction used to compute the lateral displacement, $\Delta q(\lambda;\lambda_{\rm ref})$, after tracing the chief ray at wavelengths $\lambda$ and $\lambda_{\rm ref}$. The lower half shows the two-ray construction used to compute the longitudinal shift. The chief ray and a nearby probe ray are traced at wavelengths $\lambda$ and $\lambda_{\rm ref}$, with the longitudinal shift, $\Delta z(\lambda;\lambda_{\rm ref})$, defined as the separation between their resulting intersection positions.
    }
\label{fig:refraction_diagram}

\end{figure*}


The physical origin of the effect is straightforward. The WFI filters are transmissive fused-silica elements \citep{Cromey_2023} whose refractive index varies with wavelength. As a result, light rays at different wavelengths follow slightly different paths through the filter substrate. This produces two related chromatic shifts, illustrated schematically in Fig.~\ref{fig:effect_visualization}: a \textit{longitudinal shift}, $\Delta z$, in the effective focal position along the optical axis, and \textit{lateral shifts}, $\Delta x$ and $\Delta y$, in the image position within the focal plane \citep{Born_1980,Wyant_1992}. Both effects modify the effective broadband PSF, with the former acting primarily as a defocus-like perturbation, while the latter manifests as wavelength-dependent image decentering. The goal of this work is to quantify these shifts and determine how they translate into PSF modeling errors across the \textit{Roman} bands.

These effects are potentially important for \textit{Roman} PSF modeling. Accurate PSF models are required both for realistic image simulations and for downstream calibration analyses, particularly for weak lensing measurements \citep{Anderson_2000, Piotrowski_2013, Jarvis_2021, Liaudat_2023, Schutt_2025}, where stringent requirements are placed on PSF size and shape errors \citep[e.g.][]{LSST_2018, SRD_2023}. Existing PSF models used in large-scale \textit{Roman} image simulations \citep{Troxel_2022, OU_2025} typically include the telescope optics, detector effects, and other instrumental contributions, but do not include a dedicated treatment of chromatic refraction by the filter substrate. It is therefore necessary to determine the magnitude of this effect, how it varies with wavelength and field position, and whether it is large enough to require modeling and correction for \textit{Roman} science, particularly weak lensing.


\subsection{Analytic prescription}\label{subsec:an_prescription}

We now derive the geometric formalism used to predict the chromatic displacements induced by refraction in a curved filter substrate. Physically, these shifts depend on wavelength and on the local incidence angle at the filter. In practice, we infer this incidence angle $\boldsymbol{\theta}$ from the detector position $\mathbf{x}_{\rm det}$ using the field geometry. We denote this dependence  as $\boldsymbol{\theta}(\mathbf{x}_{\rm det})$. The goal is to compute, at wavelength $\lambda$ relative to a reference wavelength $\lambda_{\rm ref}$, the transverse shifts in the focal plane, $\Delta x(\lambda;\lambda_{\rm ref}, \boldsymbol{\theta}(\mathbf{x}_{\rm det})), \Delta y(\lambda;\lambda_{\rm ref}, \boldsymbol{\theta}(\mathbf{x}_{\rm det}))$, and the longitudinal shift in the best-focus position, $\Delta z(\lambda;\lambda_{\rm ref}, \boldsymbol{\theta}(\mathbf{x}_{\rm det}))$.
For readability, we use this full notation only when the dependence on reference wavelength or field position is important; otherwise, we write these quantities more compactly as $\Delta x(\lambda)$, $\Delta y(\lambda)$, and $\Delta z(\lambda)$, or as $\Delta x(\lambda;\lambda_{\rm ref})$, $\Delta y(\lambda;\lambda_{\rm ref})$, and $\Delta z(\lambda;\lambda_{\rm ref})$ when emphasizing the reference wavelength.


These shift quantities are the fundamental outputs of the formalism and are directly related to the perturbations to the PSF. The diagram in Fig.~\ref{fig:refraction_diagram} is provided as a visual guide to the notation, geometry, and derivations used throughout this section. The diagram is split in two, with the upper half showing the relevant quantities needed for the lateral shifts, while the lower half represents the construction used for the longitudinal shift. 

\subsubsection{Ray geometry and refraction through a meniscus filter}\label{subsubsec:ray_geometry}

We begin by describing how a single ray propagates through the filter substrate at a given wavelength. The filter is a transmissive optical element, so its effect on a ray is governed by refraction at its two surfaces. The strength of that refraction mainly depends on the refractive index of the substrate material, the thickness of the filter, the incidence angle of the ray, and the filter geometry. Starting from the refractive index, we model it using the Sellmeier relation \citep{Born_1980}: 
\begin{equation}
n(\lambda)=
\left[
1+
\frac{B_1 \lambda^2}{\lambda^2-C_1}+
\frac{B_2 \lambda^2}{\lambda^2-C_2}+
\frac{B_3 \lambda^2}{\lambda^2-C_3}
\right]^{1/2},
\label{eq:n_sellmeier}
\end{equation}
where the coefficients \((B_i,C_i)\) are specific to each substrate. We work in the plane containing the optical axis and chief ray (also known as the tangential or meridional plane), 
with transverse coordinate \(q\) and optical-axis coordinate \(z\). In this plane, the filter is represented by two spherical surfaces. The first (inner) surface, denoted \(S_1\), has vertex at \(z=0\), and the second (outer) surface, \(S_2\), has vertex at \(z=t\), where \(t\) is the central thickness of the filter. Their radii of curvature (ROC) are \(R_1\) and \(R_2\), respectively. The shape of each spherical surface is described by its sag,
\begin{equation}
z_{\rm sag}(q;R)=R \pm\sqrt{R^2-q^2},
\label{eq:sag}
\end{equation}
which gives the surface height \(z\) as a function of transverse position \(q\). The sign of the second term is positive if the radius of curvature is negative, and vice versa. 

To calculate refraction, we also need the surface normal to the point where the ray hits the surface. For a spherical surface, the corresponding unit normal is
\begin{equation}
\hat{\mathbf n}(q,z;R)=
\frac{\left(q/R,\; z/R-1\right)}
{\sqrt{(q/R)^2+(z/R-1)^2}}.
\label{eq:surface_normal}
\end{equation}
This vector points perpendicular to the local surface and is the quantity that enters Snell's law. Next, we specify the incoming ray. In a given meridional plane, we describe the incident ray direction by an angle \(\theta\) measured with respect to the optical axis. The corresponding unit direction vector is
\begin{equation}
\mathbf p_0 = (\sin\theta,\cos\theta).
\label{eq:chief_dir}
\end{equation}
We denote by \(s\) the distance from the pupil to the first filter surface. In the geometric construction adopted here, the incoming ray is taken to intersect the first surface at the transverse coordinate
\begin{equation}
q_1 = s\tan\theta,
\qquad
z_1 = z_{\rm sag}(q_1;R_1).
\label{eq:ray_start}
\end{equation}
This specifies the point on \(S_1\) where the ray first encounters the filter.

Once the ray reaches the surface, its direction changes according to Snell's law. In its more familiar scalar form, Snell's law states that
\begin{equation}
n_{\rm i}\sin\theta_{\rm i}
=
n_{\rm r}\sin\theta_{\rm r},
\label{eq:snell_scalar}
\end{equation}
where \(n_{\rm i}\) and \(n_{\rm r}\) are the refractive indices on the two sides of the interface, and \(\theta_{\rm i}\) and \(\theta_{\rm r}\) are the angles measured relative to the local surface normal. In our calculation, the filter surfaces are curved, so the surface normal changes from point to point across the filter. For that reason, it is more convenient to use a vector form of Snell's law, which applies the same physical law but directly updates the ray direction using the local normal at the point of intersection.

Let \(\mathbf p\) be the incident unit direction and \(\hat{\mathbf n}\) the unit normal at the surface. We define
\begin{equation}
\mu = \hat{\mathbf n}\cdot \mathbf p,
\qquad
\eta = \frac{n_{\rm r}}{n_{\rm i}}.
\end{equation}
Since \(\mathbf p\) is a unit vector, the quantity $1-\mu^2$ is the squared sine of the incidence angle relative to the local normal. The refracted direction is then written as
\begin{equation}
\mathbf p_1
=
\mathbf p_0
+
\left[
-\mu \pm\sqrt{\eta^2-\left(1-\mu^2\right)}
\right]
\,\hat{\mathbf n},
\label{eq:snell_vector}
\end{equation}
followed by normalization to unit length. The square root term is positive for $\mu > 0$ and negative for $\mu < 0$. Equation~(\ref{eq:snell_vector}) is therefore just Snell's law written in vector form for a curved surface and allows us to compute the refracted direction directly from the local geometry of the filter surface.

After refraction at the first surface, the ray propagates through the substrate along a straight line until it reaches the second surface. If a ray passes through the point \((q_1, z_1)\) with direction \(\mathbf{p}_1 = (p_{1,q}, p_{1,z})\), its path can be written as
\begin{equation}
q(\ell) = q_1 + \ell\, p_{1,q},
\qquad
z(\ell) = z_1 + \ell\, p_{1,z},
\label{eq:ray_parametric}
\end{equation}
where \(\ell\) is the distance traveled along the ray. To determine where the ray intersects the second surface, we substitute Equation~(\ref{eq:ray_parametric}) into the spherical-surface relation defined by Equation~(\ref{eq:sag}) for \(S_2\), and solve for the value of \(\ell\) at which the ray reaches that surface. This gives the next intersection point \((q_2, z_2)\). The same vector form of Snell's law is then applied again at the second surface, now for refraction from glass back into vacuum, yielding the outgoing direction $\mathbf{p}_2$.

In summary, the propagation of a ray through the filter at a given wavelength proceeds as follows: we determine the refractive index \(n(\lambda)\), locate the point where the incoming ray intersects the first surface, compute the local surface normal, refract the ray into the substrate using Snell's law, propagate it to the second surface, and refract it back out into vacuum. 

\subsubsection{Lateral chromatic shifts}\label{subsubsec:lateral_shifts}

The lateral chromatic shift follows from comparing where a given input ray lands in the focal plane depending on its wavelength. In reality, an image is composed of a bundle of converging rays associated with the ray cone. However, for small incidence angles, it is fair to assume that the overall image displacement can be well represented by the shifts in the chief ray. Therefore, to avoid having to fully ray trace a beam bundle, we approximate the lateral shifts to the image by only tracing the chief ray as portrayed by the upper half of Fig.~\ref{fig:refraction_diagram}.

For a given wavelength \(\lambda\), we begin with the incident chief ray defined in Equation~(\ref{eq:chief_dir}) and the corresponding starting point on the first filter surface given by Equation~(\ref{eq:ray_start}). We then propagate that ray through the two surfaces of the filter using the prescription described in the previous subsection: the ray is refracted at \(S_1\), propagated through the substrate to \(S_2\), refracted again as it exits the glass, and finally propagated to a plane representing the focal plane.

If the ray exits the second surface from the point \((q_2,z_2)\) with direction \(\mathbf p_2=(p_{q,2},p_{z,2})\), we propagate it to the focal plane array (FPA). Denoting by \(L\) the axial distance from the vertex of the second filter surface \(S_2\) to the focal plane, the focal plane is located at
$z_{\rm FPA}=t+L$, where $t$ is the central filter thickness as before. 
The outgoing ray intersects this plane when \(z(\ell)=z_{\rm FPA}\), which gives
\begin{equation}
\ell_{\rm FPA}=\frac{z_{\rm FPA}-z_2}{p_{z,2}}.
\label{eq:l_fpa}
\end{equation}
Then, the corresponding transverse coordinate in that plane is
\begin{equation}
q_{\rm FPA}(\lambda)=q_2+\ell_{\rm FPA}p_{q,2}.
\label{eq:q_fpa}
\end{equation}
This quantity is the predicted image position of the chief ray at wavelength \(\lambda\). The chromatic lateral shift is defined by comparing this focal-plane position to that obtained at a reference wavelength \(\lambda_{\rm ref}\). We therefore define
\begin{equation}
\Delta q(\lambda;\lambda_{\rm ref})
=
q_{\rm FPA}(\lambda)-q_{\rm FPA}(\lambda_{\rm ref}).
\label{eq:delta_q}
\end{equation}
This is the basic lateral chromatic shift in a single meridional plane. To obtain the two-dimensional shift in the focal plane, the same construction is applied independently in the two orthogonal planes. In the plane associated with the focal-plane \(x\) direction, this gives
\begin{equation}
\Delta x(\lambda;\lambda_{\rm ref})
=
x_{\rm FPA}(\lambda)-x_{\rm FPA}(\lambda_{\rm ref}),
\label{eq:delta_x}
\end{equation}
while in the orthogonal plane it gives
\begin{equation}
\Delta y(\lambda;\lambda_{\rm ref})
=
y_{\rm FPA}(\lambda)-y_{\rm FPA}(\lambda_{\rm ref}).
\label{eq:delta_y}
\end{equation}
These shifts describe how the centroid of the image moves across the focal plane as a function of wavelength. 

\subsubsection{Longitudinal chromatic shift}\label{subsubsec:long_shifts}

\begin{figure*}
    \centering
    \includegraphics[width=0.8\linewidth]{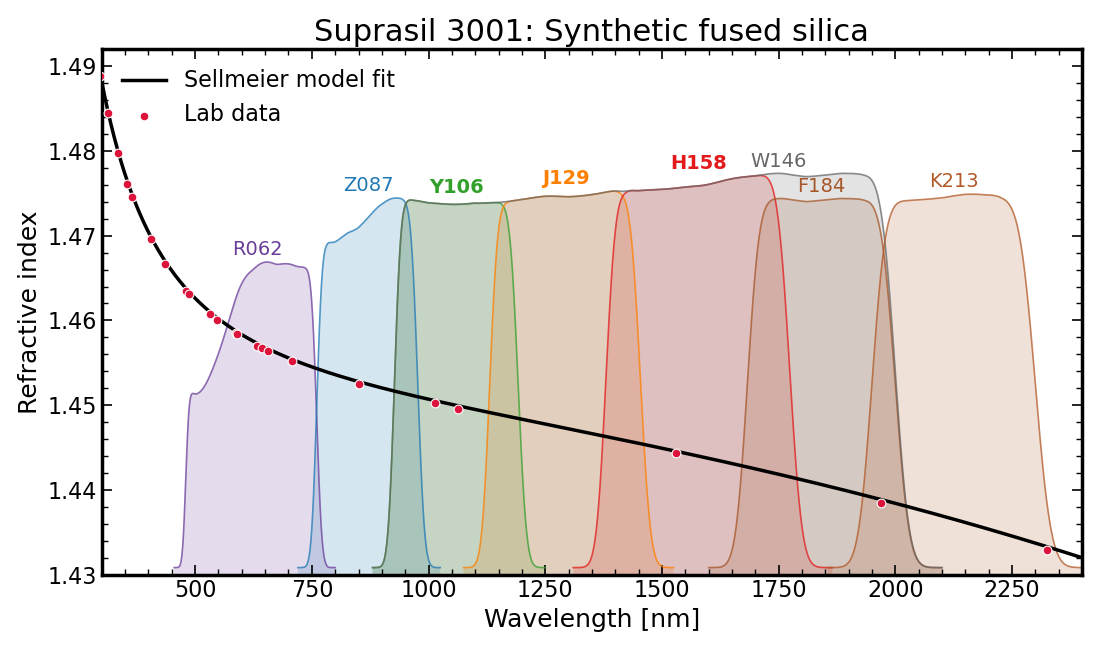}
    \caption{Refractive index of the \textit{Roman} WFI fused silica (Suprasil 3001) filter substrate  as a function of wavelength. The substrate manufacturer provides the measured refractive-index at discrete wavelengths (red points), as well as the best fit parameters to the Sellmeier relation, showed in Equation~(\ref{eq:n_sellmeier}), to construct the model fit (black solid line). The shaded regions indicate the wavelength coverage of the eight \textit{Roman} imaging bands. The figure illustrates how the refractive index varies across wavelength and within each filter. The differences within each filter will cause the wavelength-dependent distortion in the broadband PSF shown in Fig.~\ref{fig:effect_visualization}. Of particular interest are the Y106, J129, and H158 bands that will be used for weak lensing, where PSF error requirements are the most stringent. 
    }
    \label{fig:refractive_index}

\end{figure*}

The lateral shifts \(\Delta x\) and \(\Delta y\) follow from tracing a single chief ray to the focal plane. The longitudinal shift requires a different approach, as a single ray does not determine a focus position. Instead, we find where two neighboring rays intersect after passing through the filter, as portrayed by the lower half of Fig.~\ref{fig:refraction_diagram}.

We estimate this using a local two-ray construction. The first ray is the chief ray with field angle \(\theta\), as defined in Sec.~\ref{subsubsec:lateral_shifts}. The second is a probe ray, launched from a slightly displaced point on the first surface at $q_{\rm p} = q_{\rm c} + \delta q_{\rm probe}$, whose incident direction is chosen so that, in the absence of the filter, it would converge to the same nominal image point as the chief ray. Both rays are then traced through the two refracting surfaces using the prescription in Sec.~\ref{subsubsec:ray_geometry}.

After exiting the filter, the outgoing chief and probe rays travel along straight lines with slopes $m_{\rm c} = p_{{\rm c},2,q}/p_{{\rm c},2,z}$ and $m_{\rm p} = p_{{\rm p},2,q}/p_{{\rm p},2,z}$, respectively. The wavelength-dependent focus position is their intersection point,
\begin{equation}
z_{\rm focus}(\lambda)=\frac{q_{{\rm p},2}-q_{{\rm c},2}+m_{\rm c}z_{{\rm c},2}-m_{\rm p}z_{{\rm p},2}}{m_{\rm c}-m_{\rm p}},
\label{eq:z_focus}
\end{equation}
and the longitudinal chromatic shift is defined by comparing this to the focus position at a reference wavelength,
\begin{equation}
\Delta z(\lambda;\lambda_{\rm ref})=z_{\rm focus}(\lambda)-z_{\rm focus}(\lambda_{\rm ref}).
\label{eq:delta_z}
\end{equation}
Thus, \(\Delta z\) measures how the best-focus position moves along the optical axis as a function of wavelength due to refraction in the filter substrate.

\subsubsection{Connection to wavefront aberrations}\label{subsubsec:wave_aberrations}

The quantities derived above, \(\Delta x\), \(\Delta y\), and \(\Delta z\), are geometric quantities: they describe how refraction in the filter substrate shifts the image position and the location of best focus as a function of wavelength. To connect these shifts to PSF modeling, it is useful to express them in the language of wavefront aberrations.

A standard way to describe optical aberrations is through the optical path difference (OPD), which measures the departure of the wavefront from an ideal reference wavefront \citep{Wyant_1992}. If \(L(\rho,\phi)\) denotes the optical path length from the pupil to the image plane for a ray passing through pupil coordinates \((\rho,\phi)\), and \(L_0\) denotes the path length of the ideal reference spherical wave, then the OPD is defined as 
\begin{equation}
W(\rho,\phi) \equiv L(\rho,\phi)-L_0.
\label{eq:opd_def}
\end{equation}
Because the pupil is circular, we expand the OPD in Zernike polynomials,
\begin{equation}
W(\rho,\phi)=\sum_{j=1}^{\infty} a_j Z_j(\rho,\phi),
\label{eq:zernike_expand}
\end{equation}
where \(Z_j(\rho,\phi)\) denotes the \(j\)-th Zernike polynomial\footnote{See \cite{Noll_1976} for an extensive description of the Zernike polynomials and their properties.} on the unit disk and \(a_j\) is the corresponding coefficient. Throughout this work we use the Noll convention \citep{Noll_1976}, under which the polynomials satisfy \(\left\langle Z_i Z_j \right\rangle = \delta_{ij}\). 

The longitudinal shift \(\Delta z\) is naturally interpreted as a defocus term. Physically, shifting the best-focus plane along the optical axis changes the quadratic variation of the wavefront across the pupil. For a beam with focal ratio \(N=f/D\), a small axial displacement \(\Delta z\), \cite{Wyant_1992} shows that the OPD is given by:
\begin{equation}
W_{\rm def}(\rho)
=
-\frac{\Delta z}{8N^2}
\left(
\rho^2-\frac{1}{2}
\right),
\label{eq:defocus_opd}
\end{equation}
which matches the fourth Zernike mode in the Noll convention:
\begin{equation}
Z_4(\rho,\phi)=\sqrt{3}\left(2\rho^2-1\right)
=
2\sqrt{3}\left(\rho^2-\frac{1}{2}\right).
\label{eq:z4_defocus}
\end{equation}
Comparing Equations~(\ref{eq:defocus_opd}) and~(\ref{eq:z4_defocus}) shows that \(\Delta z\) maps directly onto the Noll defocus mode. Expressed in units of waves, the corresponding defocus coefficient is then:
\begin{equation}
a_4(\lambda)
=
-\frac{\Delta z(\lambda)}{16\sqrt{3}\,N^2\,\lambda}.
\label{eq:a4_defocus}
\end{equation}

The lateral shifts $\Delta x$ and $\Delta y$ have a different interpretation. A wavelength-dependent displacement of the PSF centroid can be described either as a tip/tilt of the wavefront across the pupil or, equivalently, as a translation of the image in the focal plane. In the Noll convention, the corresponding tip and tilt modes are:
\begin{equation}
Z_2(\rho,\phi)=2\rho\cos\phi,
\qquad
Z_3(\rho,\phi)=2\rho\sin\phi.
\label{eq:z23_tiptilt}
\end{equation}
These modes represent wavefront slopes in the two orthogonal directions and therefore shift the PSF centroid. In practice, however, the same effect can be implemented more simply by directly shifting the monochromatic PSF in the image plane by the predicted amounts $\Delta x$ and $\Delta y$. We adopt that approach in the image simulations discussed later. The important point is that the two descriptions are equivalent and both represent a wavelength-dependent displacement of the PSF centroid.

\subsection{\textit{Roman} optical specifications}\label{subsec:roman_specs}

To evaluate the formalism above for the \textit{Roman} WFI, we use the instrument-specific quantities summarized in Table~\ref{tab:roman_optical_specs}. These include the focal ratio, the meniscus-filter geometry, the distances between the pupil, filter, and focal plane, the focal-plane orientation, and the quantities used to convert detector position into angle of incidence at the filter.\footnote{Some optical design parameters can be found directly in \url{https://roman.gsfc.nasa.gov/science/WFI_technical.html} or in \cite{Pasquale_2018}. Parameters not found explicitly stated 
can be derived from what is in \cite{Pasquale_2018}. 
} Together with the wavelength-dependent refractive index of the filter substrate, these are the \textit{Roman}-specific inputs needed to calculate the lateral and longitudinal chromatic shifts.

For the \textit{Roman} WFI filters, we are primarily interested in shifts measured relative to the effective wavelength of each bandpass, \(\lambda_{\rm eff}\), since broadband PSF differences depend on how the monochromatic PSFs vary across the filter relative to that wavelength. Unless otherwise stated, we therefore use the shorthand \(\Delta x(\lambda)\), \(\Delta y(\lambda)\), and \(\Delta z(\lambda)\) to denote shifts defined relative to the effective wavelength of the corresponding filter.

The WFI filters are weak meniscus elements made of Suprasil 3001 fused silica \citep{Cromey_2023}, with the center bulging toward the incoming beam. The wavelength-dependent refractive index \(n(\lambda)\) is calculated using the Sellmeier coefficients provided by the manufacturer.\footnote{See \url{https://www.heraeus-covantics.com/media/Media/Documents/Products_and_Solutions/OPT/EN/Data_and_Properties_Optics_fused_silica_EN.pdf}} Fig.~\ref{fig:refractive_index} shows the refractive-index data together with the corresponding Sellmeier fit over the wavelength range spanned by the \textit{Roman} imaging bandpasses. The figure also shows that different filters sample different parts of the refractive-index curve. In particular, the slope of \(n(\lambda)\) is steeper over the wavelength range of R062 than over most of the redder imaging bands, while W146 spans a much broader wavelength range, both of which suggest a potentially enhanced chromatic refraction effect.

An additional \textit{Roman}-specific ingredient is the position of the focal plane relative to the filter. The focal plane is not exactly parallel to the local filter surface across the field, so the effective filter-to-focal-plane distance varies slightly with position. We account for this by modifying the nominal distance as \(L_{\rm eff}=L-y_{\rm FPA}\tan\theta_{\rm FPA}\), where \(y_{\rm FPA}\) is the focal-plane position in the direction of the tilt and \(\theta_{\rm FPA}\) is the inclination angle of the focal plane relative to the filter.

The angle of incidence used in the formalism is determined from the \textit{Roman} focal-plane geometry. For a given sensor chip assembly (SCA) 
and detector position, we first convert the science-frame pixel coordinates to local focal-plane coordinates \((x_{\rm FPA},y_{\rm FPA})\) using \texttt{pysiaf}\footnote{\url{https://github.com/spacetelescope/pysiaf}} \citep{Sahlmann_2019}. These coordinates are then mapped to an incidence angle at the filter using the nominal WFI plate scale, pixel size, and the demagnification between the focal plane and the exit pupil. Denoting by \(C\) the conversion from focal-plane distance in mm to incidence angle in degrees, and by \(M_{\rm pup}\) the pupil demagnification factor (the ratio of the entrance pupil diameter to the exit pupil diameter), the total angle of incidence is
\begin{equation}
\theta = \sqrt{x_{\rm FPA}^2+y_{\rm FPA}^2}\, C\, M_{\rm pup},
\label{eq:aoi_total}
\end{equation}
with components
\begin{equation}
\theta_x = x_{\rm FPA}\, C\, M_{\rm pup},
\qquad
\theta_y = y_{\rm FPA}\, C\, M_{\rm pup}.
\label{eq:aoi_components}
\end{equation}

For the \textit{Roman} WFI, the values entering \(C\) are the pixel size (\(10\,\mu{\rm m}\)) and plate scale (\(0.11~\rm arcsec/pixel\)). We adopt \(M_{\rm pup}\approx26.8\), obtained from the average pupil demagnification (along the FPA x-axis) across the field of view in the full ray-tracing model described in Sec.~\ref{subsec:ray_tracing}. Strictly speaking, the pupil is not perfectly circular and the demagnification varies across the field of view, so treating it as a single constant is an approximation adopted for simplicity. Across the field, this prescription gives angles of incidence reaching up to \(\sim 12^\circ\) near the edges of the focal plane.

Finally, we neglect the temperature dependence of the refractive index in the present analysis. Using the manufacturer-provided temperature-dependent information, we verified that the resulting change in \(n(\lambda)\) over the relevant wavelength range is at the sub-percent level, so the room-temperature Sellmeier coefficients are sufficient for measuring the chromatic effects.

\begin{table}
\centering
\caption{\textit{Roman} WFI quantities used in the filter-substrate refraction model. The table lists the \textit{Roman}-specific inputs substituted into the general formalism of Sec.~\ref{subsec:an_prescription}, including the meniscus-filter geometry, the distances relevant for ray propagation, the focal-plane orientation, and the quantities used to convert detector position into angle of incidence at the filter. 
}
\label{tab:roman_optical_specs}
\begin{tabular}{lc}
\hline
Quantity & Value \\
\hline
Focal ratio ($N$) & \(8\) \\
Central filter thickness ($t$) & \(10~\mathrm{mm}\) \\
Inner surface ROC ($R_1$) & \(-1.5~\mathrm{m}\) \\
Outer surface ROC ($R_2$) & \(-1.4993~\mathrm{m}\) \\
Exit pupil to filter distance ($s$) & \(10~\mathrm{mm}\) \\
Filter to focal plane distance ($L$) & \(0.673~\mathrm{m}\) \\
Relative focal-plane inclination ($\theta_{\text{FPA}}$) & \(13.2^\circ\) \\
Pixel size & \(10~\mu\mathrm{m}\) \\
Plate scale & \(0.11~\mathrm{arcsec/pixel}\) \\
Pupil demagnification ($M_{\rm pup}$) & \(26.8\) \\
\hline
\end{tabular}
\end{table}
\section{Simulations}\label{sec:image_sims}

In this section, we describe how the refraction formalism developed above is incorporated into \textit{Roman} PSF image simulations. The goal is to use the wavelength-dependent lateral and longitudinal shifts introduced in Sec.~\ref{sec:refraction_formalism} to build a practical implementation that can be used to generate perturbed broadband PSFs and to quantify their impact on observable image properties. Since the effect is intrinsically chromatic, the implementation must account for how the monochromatic PSF varies across the bandpass and how the refraction-induced aberrations modify that wavelength dependence.

We begin by describing how the baseline \textit{Roman} PSF is drawn and how the effective broadband PSF is constructed from monochromatic realizations across the filter bandpass. We then show how the predicted chromatic shifts are introduced into the PSF. This provides the simulation framework used in the following sections to validate the formalism and measure the resulting PSF and photometry errors.

\subsection{Drawing the \textit{Roman} PSF}\label{subsec:roman_psf}

We generate the PSF image simulations using \texttt{GalSim v2.6} \citep{Galsim}. In particular, we use the \textit{Roman} module\footnote{\url{https://galsim-developers.github.io/GalSim/_build/html/roman.html}} 
\citep{Kannawadi_2016}, which provides PSF models and instrument characteristics specific to the \textit{Roman} telescope. Throughout this work, this built-in \textit{Roman} PSF model serves as the baseline onto which we later add the chromatic aberrations induced by filter-substrate refraction.

Within \texttt{GalSim}, the \textit{Roman} PSF is constructed from tabulated Zernike coefficients for each SCA, derived from available Cycle~9 design data\footnote{see \url{https://science.nasa.gov/mission/roman-space-telescope/wfi-technical/}} 
and provided at the center and corners of each detector. For an arbitrary position within an SCA, \texttt{GalSim} interpolates these coefficients to obtain the local aberration model used to draw the PSF. The underlying Zernike coefficients are defined at a fixed reference wavelength of \(1293\,\mathrm{nm}\), which should not be confused with the bandpass effective wavelength used in the previous section when defining the chromatic shifts \(\Delta x\), \(\Delta y\), and \(\Delta z\). When a PSF is requested at wavelength \(\lambda\), \texttt{GalSim} internally rescales the aberration amplitudes from this \textit{Roman}-model reference wavelength to \(\lambda\), so any additional aberration coefficients introduced later must be specified consistently with this normalization.

We are interested in the effective PSF through a bandpass. For a source with spectral energy distribution \(S(\lambda)\) observed through a filter with transmission \(T(\lambda)\), the normalized effective broadband PSF is
\begin{equation}
{\rm PSF}_{\rm eff}(\mathbf{x})
=
\frac{\int \mathrm{d}\lambda \, S(\lambda)\,T(\lambda)\,{\rm PSF}(\mathbf{x};\lambda)}
{\int \mathrm{d}\lambda \, S(\lambda)\,T(\lambda)},
\label{eq:psf_eff_integral}
\end{equation}
where \({\rm PSF}(\mathbf{x};\lambda)\) is the PSF at focal-plane position \(\mathbf{x}\) and wavelength \(\lambda\). Because the filter-substrate refraction derived in the previous section is explicitly wavelength dependent, the perturbation must be applied before integrating over the bandpass. We therefore construct the broadband PSF manually, wavelength by wavelength.

In practice, we approximate Equation~(\ref{eq:psf_eff_integral}) on a discrete wavelength grid \(\{\lambda_i\}\) spanning the filter bandpass. At each wavelength \(\lambda_i\), we generate a monochromatic \textit{Roman} PSF at the chosen SCA and detector position, convolve it with the native \textit{Roman} pixel response, and draw it at the desired pixel scale. The resulting broadband PSF is
\begin{equation}
{\rm PSF}(\mathbf{x})=\frac{\sum_i w_i\,{\rm PSF}(\mathbf{x};\lambda_i)}{\sum_i w_i},
\label{eq:manual_psf_sum}
\end{equation}
with weights
\begin{equation}
w_i = S(\lambda_i)\,T(\lambda_i)\,\Delta\lambda_i,
\label{eq:manual_psf_weights}
\end{equation}
where \(\Delta\lambda_i\) is the wavelength-bin width. After summation, the PSF is normalized to unit flux.

Constructing the broadband PSF therefore requires a source SED. Since the PSF model will ultimately be used for galaxy shape measurements, we adopt an average galaxy SED built from the \textsc{Diffsky} \citep{Hearin2020} mock extragalactic catalogs used in the OpenUniverse 2024 image simulations \citep{OU_2025}. We apply a true magnitude cut of $H < 25$ to the catalog to remove galaxies that would generally be too faint for \textit{Roman} weak lensing cosmology\citep[see][]{Berlfein_2026}. 
For each mock galaxy, we sum the bulge, disk, and knot SED components, redshift the total SED according to the catalog redshift, and normalize it by its integrated flux through the \textit{Roman} Y106 bandpass. We then compute an average observed galaxy SED weighted by the \textit{Roman} source redshift distribution, $n(z)$, adopted in \textit{Roman}'s High-Latitude Imaging Survey (HLIS) Data Challenge 1\footnote{\url{https://github.com/CosmoLike/roman_cpip_data_challenge}} (J. Xu et al.\ in preparation), across all redshift bins. This $n(z)$-weighted mean template is used throughout this work as the source SED $S(\lambda)$. 

We verify that the discrete approximation in Equation~(\ref{eq:manual_psf_sum}) is consistent with \texttt{GalSim}'s native chromatic FFT-based drawing. To do so, we generate two broadband PSFs for each filter: one using the manual wavelength-by-wavelength summation in Equation~(\ref{eq:manual_psf_sum}), and one using the built-in \texttt{GalSim} chromatic object. In order to quantify the agreement between the two PSF drawing methods, we compare the PSF sizes using the \texttt{HSM} adaptive moments \citep{Hirata_2003,Mandelbaum_2005} module implemented in \texttt{GalSim}. The agreement will depend on the particular wavelength sampling chosen for the manual summation. Hence, we chose the wavelength sampling empirically such that the fractional size difference between the manual and native \texttt{GalSim} PSFs is below $10^{-4}$, which is much lower than the $7.2\times10^{-4}$ HLIS requirement. We find that this criterion is satisfied with a uniformly spaced wavelength grid of 20 samples spanning the full bandpass of each filter, except for W146, which needed 40 samples to account for its broader wavelength coverage. This confirms that the manual wavelength-by-wavelength approximation reproduces the \texttt{GalSim} chromatic PSF to sufficient accuracy while providing the flexibility needed to introduce wavelength-dependent aberrations. 

Because the \textit{Roman} PSF is undersampled at the native pixel scale, accurate measurements of PSF size and shape require a finer pixel grid. Throughout this work, we therefore draw the PSF at a pixel length eight times smaller than the native \textit{Roman} pixel length \citep[similar to that in][]{Cao_2024} when making quantitative measurements. For qualitative checks and visualization, however, we also render PSFs at the native pixel scale, since that is the scale at which the effect will ultimately appear in the individual \textit{Roman} images. It is important to note that in practice, galaxy shape measurements for \textit{Roman} will eventually be made on coadded oversampled images \citep{Hirata_2024}, while PSF modeling will be performed on individual exposures.

\subsection{Introducing new aberrations}\label{subsec:new_aberrations}

\begin{figure*}
    \centering
    \includegraphics[width=0.99\linewidth]{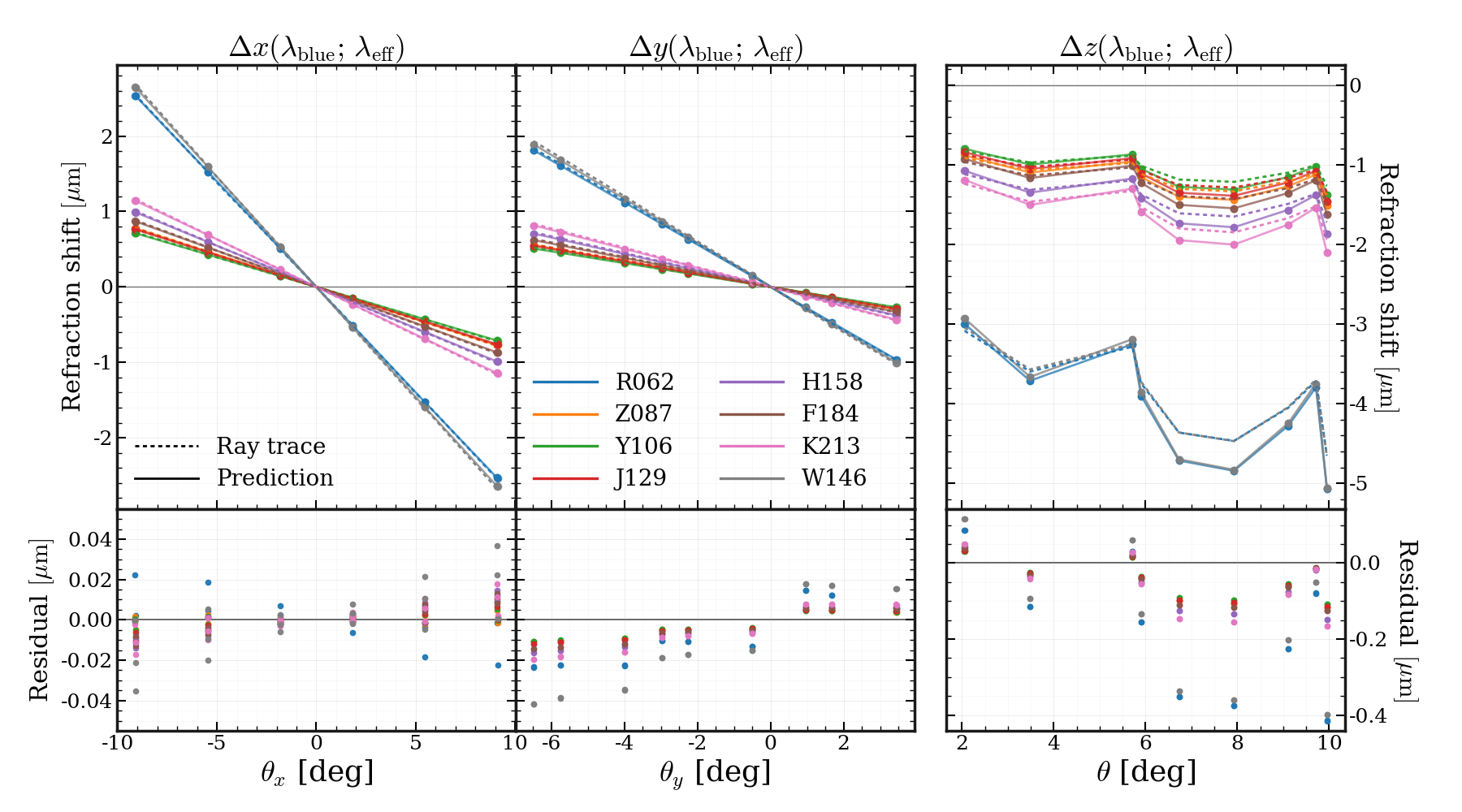}
    \caption{Comparison of the analytic predictions for the chromatic lateral and longitudinal shifts with measurements from the full \textit{Roman} ray-tracing model, evaluated at the blue edge of each filter relative to the filter effective wavelength, i.e. $\Delta x(\lambda_{\rm blue};\lambda_{\rm eff})$, $\Delta y(\lambda_{\rm blue};\lambda_{\rm eff})$, and $\Delta z(\lambda_{\rm blue};\lambda_{\rm eff})$. The top panels show the predicted shifts (solid lines) and ray-traced values (dashed lines) as a function of incidence angle; the bottom panels show the corresponding residuals. The values are computed at the centers of the 18 \textit{Roman} SCAs for all eight imaging bands. Because of the focal-plane symmetry, distinct SCA centers can have the same projected incidence angle in a given direction, so multiple points for the same filter may overlap. For the lateral shifts (left and middle panels), the agreement is very good across the focal plane, with residuals generally below $\sim 0.03\,\mu{\rm m}$ for most bands and reaching at most $\sim 0.05\,\mu{\rm m}$ for W146, far smaller than the \textit{Roman} pixel size of $10\,\mu{\rm m}$. The longitudinal shifts (right panel) are accurate to the $\sim$10\% level in the worst cases, with a slight tendency for the analytic model to overestimate the ray-traced value. Although the accuracy of the longitudinal shifts is harder to interpret directly, subsequent figures show that the PSF effects they induce are subdominant. Overall, the analytic prescription captures the main trends and amplitudes of the chromatic shifts across the \textit{Roman} focal plane very well.}
    \label{fig:ray_trace_comp}

\end{figure*}

We now describe how the wavelength-dependent refraction terms are incorporated into the manual PSF construction. For each wavelength \(\lambda_i\) in the grid, we first evaluate the predicted chromatic shifts \(\Delta x(\lambda_i)\), \(\Delta y(\lambda_i)\), and \(\Delta z(\lambda_i)\) relative to the bandpass effective wavelength, using the formalism described in Sec.~\ref{subsec:an_prescription}. These shifts are then used to modify the monochromatic \textit{Roman} PSF before it is inserted into the weighted broadband sum. In this implementation, the longitudinal and lateral terms can be enabled independently, allowing us to study the effects of defocus and decentering separately.

The longitudinal shift \(\Delta z(\lambda_i)\) is implemented as an additional defocus aberration to the existing fourth Zernike aberration in the \textit{Roman} PSF model. For each wavelength, we convert the predicted axial shift into a Noll \(Z_4\) coefficient using the defocus relation derived in Sec.\ref{subsubsec:wave_aberrations}. In practice, this coefficient is supplied through the \texttt{extra\_aberrations} argument of \texttt{roman.getPSF}, with only the defocus component modified. The normalization is expressed in units consistent with the fixed \textit{Roman} PSF reference wavelength of \(1293\,\mathrm{nm}\), which is the wavelength at which the underlying \texttt{GalSim} \textit{Roman} aberration model is defined. Thus, when the longitudinal term is enabled, each monochromatic PSF is redrawn with a wavelength-dependent additional defocus coefficient.

The lateral shifts \(\Delta x(\lambda_i)\) and \(\Delta y(\lambda_i)\) are implemented as wavelength-dependent decentering terms. Although these shifts can be interpreted formally as tip and tilt Zernike modes, the \texttt{roman} module within \texttt{GalSim} does not currently allow these Zernike coefficients to be modified. We therefore do not introduce the effect through additional Zernike terms. Instead, after generating the monochromatic PSF object, we apply the predicted displacement directly through the \texttt{GalSim} \texttt{shift} operation. In other words, before drawing, the monochromatic PSF at wavelength \(\lambda_i\) is translated by the amounts \(\Delta x(\lambda_i)\) and \(\Delta y(\lambda_i)\). Note that the \texttt{galsim.shift} operation expects its arguments in units of arcsec, so these shifts must be converted appropriately before being passed in. The final broadband PSF is obtained by inserting these perturbed monochromatic PSFs into the weighted sum of Equation~(\ref{eq:manual_psf_sum}). The main steps of this procedure are summarized in Algorithm~\ref{alg:roman_refraction}.

\begin{algorithm}
\caption{Adding filter-substrate refraction to the \textit{Roman} PSF  
}
\label{alg:roman_refraction}
\begin{algorithmic}[1]
\Require wavelength grid $\{\lambda_i\}$, filter transmission $T(\lambda)$, SED $S(\lambda)$, focal plane position $\mathbf{x}$
\State Initialize ${\rm PSF}(\mathbf{x})
\gets 0$
\For{each wavelength sample $\lambda_i$}
    \State Set and store  $w_i(\lambda) \gets S(\lambda_i)\;T(\lambda_i ) \; \Delta \lambda_i$
    \State Compute $\Delta x(\lambda_i)$, $\Delta y(\lambda_i)$, and $\Delta z(\lambda_i)$
    \If{defocus is enabled}
        \State Convert $\Delta z(\lambda_i)$ into defocus coefficient $a_4(\lambda_i)$
        \State Store monochromatic PSF$(\mathbf{x};\lambda_i)$ from \texttt{galsim.roman}%
        \Statex \hspace{\algorithmicindent} \hspace{\algorithmicindent}with $a_4(\lambda_i)$ as additional aberration in the fourth Zernike
    \EndIf
    \If{decentering is enabled}
        \State Store monochromatic PSF$(\mathbf{x};\lambda_i)$ from \texttt{galsim.roman}
        \State Shift the PSF center by $\Delta x(\lambda_i)$ and $\Delta y(\lambda_i)$
    \EndIf
    \State Update ${\rm PSF}(\mathbf{x}) \gets {\rm PSF}(\mathbf{x}) + w_i\,{\rm PSF}(\mathbf{x};\lambda_i)$
\EndFor
\State Normalize ${\rm PSF}(\mathbf{x})$ by $\sum_i w_i$
\State \Return ${\rm PSF}(\mathbf{x})$
\end{algorithmic}
\end{algorithm}


\subsection{Ray tracing}

To validate the analytic predictions from Sec.~\ref{sec:refraction_formalism}, we compare them against measurements from a geometric ray-tracing model of the full \textit{Roman} WFI optical system using \textsc{PSFSim}\footnote{\url{https://github.com/Roman-HLIS-Cosmology-PIT/PSFSim/tree/main}}. \textsc{PSFSim} uses instrumental specifications as provided by the \textit{Roman} project\footnote{RST-SYS-SPEC-0055, Revision E} to follow light rays as they reflect, refract, and intersect with the various optical components within the telescope, including the mirrors, baffles, and the filter wheel. The resultant PSF is therefore drawn with a fully spatially dependent pupil mask at a desired wavelength. The optical path differences are also interpolated at the focal plane from the same Cycle 9 design data as those in \textsc{GalSim}. At a given field point, the raytrace module returns the ray coordinates, directions, and values of the electric field at the focal plane, which are used to measure the chromatic lateral and longitudinal shifts: $\Delta x$, $\Delta y$, $\Delta z$. PSFSim has been validated by comparing the PSFs drawn at different focal plane positions and in different bands against those generated by the Space Telescope Point Spread Function (STPSF)\footnote{\url{https://github.com/spacetelescope/stpsf}}, a Python package capable of computing the simulated \textit{Roman} PSF and in part developed by the \textit{Roman} Science Operations Center. The excellent agreement between the polychromatic PSFs suggests that the ray tracing performed through the optical model in \textsc{PSFSim} is accurate. A detailed description of \textsc{PSFSim} is presented in Dalal et al. (in preparation).
\section{Results}\label{sec:results}

\begin{figure*}
    \centering
    \includegraphics[width=0.9\linewidth]{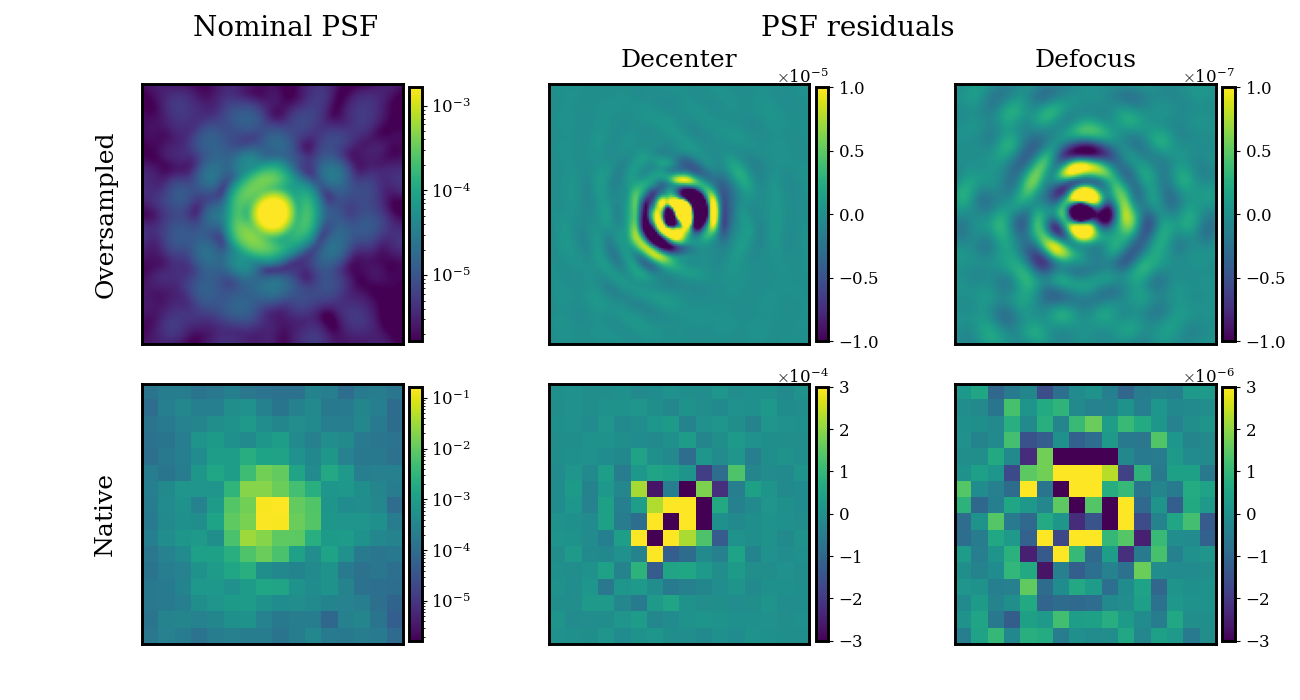}
    \caption{Example broadband \textit{Roman} PSF for an average galaxy SED (see Sec.~\ref{subsec:roman_psf}) in H158 and the corresponding image residuals induced by the two filter-substrate refraction effects: decentering and defocus. For visualization, the PSF shown is located at the edge of the focal plane, where the effect is largest. The left column shows the nominal PSF with no added refraction-induced aberrations, while the middle and right columns show the residuals obtained after adding only the wavelength-dependent decentering or defocus term, respectively, and subtracting the nominal PSF. The top row shows the oversampled PSF, drawn at a pixel scale 8 times smaller than the native \textit{Roman} scale, and the bottom row shows the same quantities at the native detector scale. The residual patterns match the expected qualitative behavior of the two perturbations: decentering produces a dipole-like residual from a wavelength-dependent shift of the PSF centroid, while defocus produces a more radially symmetric residual from a wavelength-dependent shift in the best-focus position. The color scales also show that the decentering residuals are roughly two orders of magnitude larger than the defocus residuals at the pixel level, indicating that the lateral component dominates the broadband PSF distortion. 
    }
    \label{fig:psf_img_diff}

\end{figure*}

In this section, we present the main results obtained by applying the refraction formalism to the image simulations. We first test the physical accuracy of the analytic model by comparing its predictions for the chromatic lateral and longitudinal shifts against measurements from a full \textit{Roman} ray-tracing calculation. We then use the validated model to quantify how these shifts propagate into observable PSF and photometric errors across the \textit{Roman} focal plane and imaging bands. This allows us to assess both whether the simplified geometric treatment captures the dominant behavior of filter-substrate refraction and whether the resulting perturbations are large enough to matter for \textit{Roman} science applications, particularly weak lensing and photometry.

\subsection{Comparison with ray tracing}
\label{subsec:ray_tracing}

To validate the analytic predictions from Sec.~\ref{sec:refraction_formalism}, we compare them against measurements from \textsc{PSFSim}. 
For each field point the ray tracer returns the ray coordinates and direction at the focal plane, which we use to measure the chromatic lateral and longitudinal shifts, $\Delta x,\; \Delta y,\; \Delta z$.

We perform this comparison at the blue edge of each filter bandpass, where the effect is largest, for all eight \textit{Roman} imaging bands: R062, Z087, Y106, J129, H158, F184, K213, and W146. At each SCA center, we trace a pupil-sampled ray bundle with $N=384$ rays per dimension ($384\times384$ total rays) at both $\lambda_{\text{blue}}$ and $\lambda_{\text{eff}}$. The lateral shifts are measured directly from the change in the average focal-plane position of the ray bundle, providing a direct test of the chief-ray approximation used in Sec.~\ref{subsec:an_prescription}.

The longitudinal shift is measured from the change in the best-focus plane of the traced bundle. Starting from the ray positions and slopes at the focal plane, we propagate the rays by a small axial displacement $\Delta z$,
\begin{equation}
x_i' = x_i + u_i \Delta z,
\qquad
y_i' = y_i + v_i \Delta z,
\end{equation}
where \(i\) indexes rays in the bundle, and \(u_i\) and \(v_i\) denote the ray slopes in the $x$ and $y$ directions, respectively.
We then determine the value of $\Delta z$ that minimizes the weighted second moment of the propagated bundle,
\begin{equation}
\begin{aligned}
S(\Delta z)
&=
\sum_i w_i \left[
\left(x_i'-\bar{x}'\right)^2
+
\left(y_i'-\bar{y}'\right)^2
\right] \\
&=
\sum_i w_i \left[
\left(x_i-\bar{x} + (p_i-\bar{p})\Delta z\right)^2
+
\left(y_i-\bar{y} + (q_i-\bar{q})\Delta z\right)^2
\right],
\end{aligned}
\end{equation}
here $w_i$ is unity for unmasked rays and zero otherwise, and barred quantities denote weighted means over the ray bundle. Defining
\begin{equation}
\tilde x_i = x_i - \bar{x}, \qquad \tilde y_i = y_i - \bar{y}, \qquad
\tilde p_i = p_i - \bar{p}, \qquad  \tilde q_i = q_i - \bar{q},
\end{equation}
the minimizing value is
\begin{equation}
\Delta z_{\rm best}
=
-\frac{\sum_i w_i \left(\tilde x_i \tilde p_i + \tilde y_i \tilde q_i \right)}
{\sum_i w_i \left[\tilde p_i^2 + \tilde q_i^2 \right]}.
\end{equation}
The chromatic longitudinal shift is then the difference between the best-focus positions measured at the two wavelengths.

This geometric definition is equivalent to the wavefront-based description in Sec.~\ref{subsubsec:wave_aberrations}, where the lateral shifts are encoded in the linear OPD terms and the longitudinal shift in the quadratic defocus term. We verified that the shifts measured from the ray bundle agree to sub-percent level with those obtained from independently fitting the OPD.

\begin{figure*}
    \centering
    \includegraphics[width=1\linewidth]{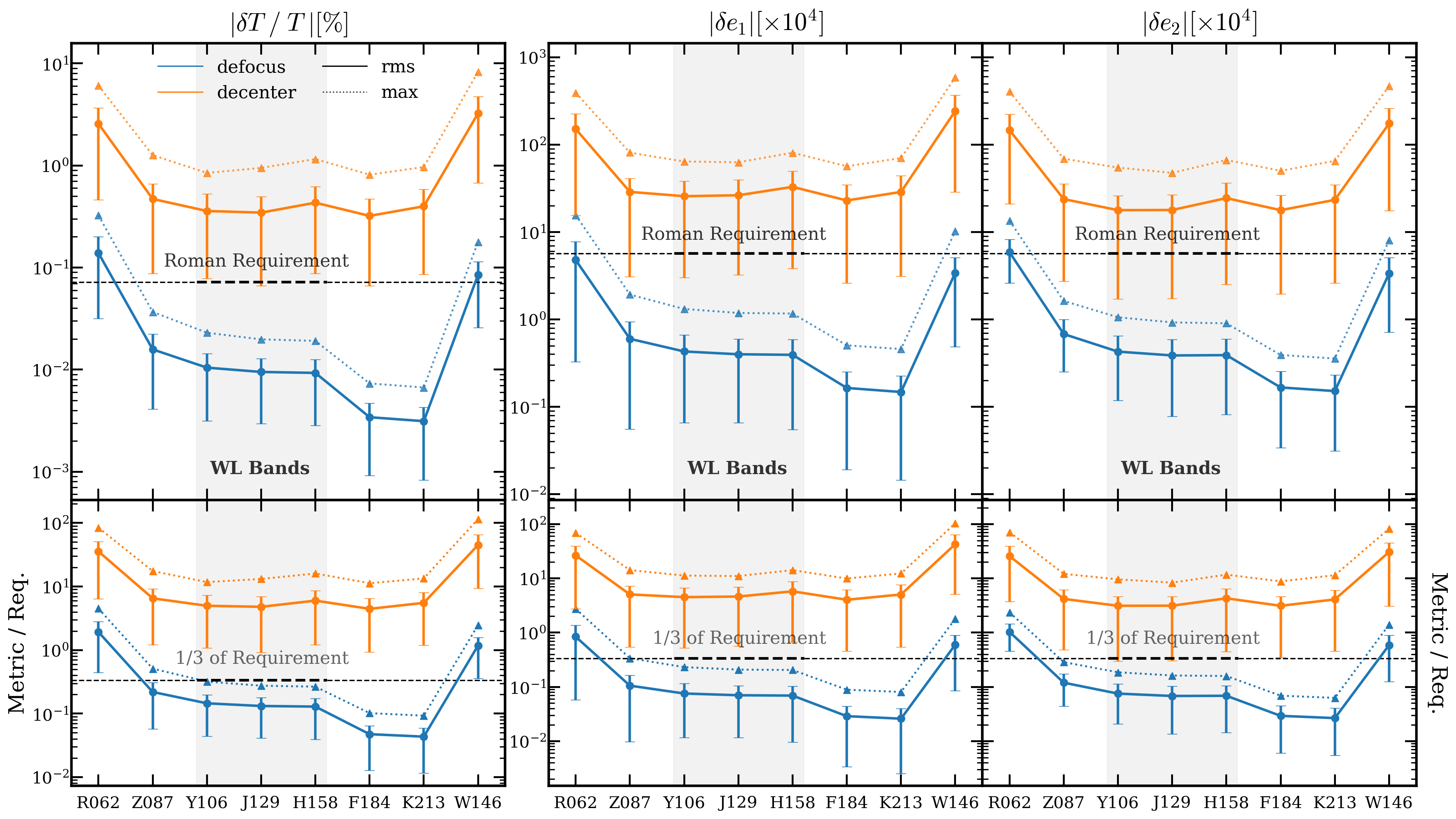}
    \caption{Filter dependence of the PSF residuals induced by filter-substrate refraction for an average galaxy SED, shown separately for the wavelength-dependent defocus term (blue) and decentering term (orange). The three columns show, from left to right, the absolute fractional error in the trace of the PSF second moments, $|\delta T / T|$, and the absolute ellipticity residuals, $|\delta e_1|$ and $|\delta e_2|$. The error bars indicate the 16th--84th percentile range of the residuals across the focal plane. In each top panel, the solid lines show the root-mean-square (RMS) of the residual over the full focal plane, while the dotted lines show the maximum residual. The bottom row shows the same quantities divided by the corresponding \textit{Roman} science requirement. The horizontal dashed line in the top row marks the requirement itself, while the dashed line in the bottom row marks one third of the requirement, which we use as an approximate per-systematic allocation threshold. The shaded vertical band highlights the weak lensing filters: Y106, J129, and H158. Across all PSF diagnostics, the decentering contribution is more than an order of magnitude larger than the defocus contribution, demonstrating that the lateral chromatic shift is the dominant source of PSF error from filter-substrate refraction. In the weak lensing bands, the defocus contribution remains below one third of the requirement for all three diagnostics, while the decentering contribution largely exceeds the \textit{Roman} weak lensing PSF size and shape requirements. 
    }
    \label{fig:psf_quant_residuals}

\end{figure*}

\begin{figure}
    \includegraphics[width=0.8\linewidth]{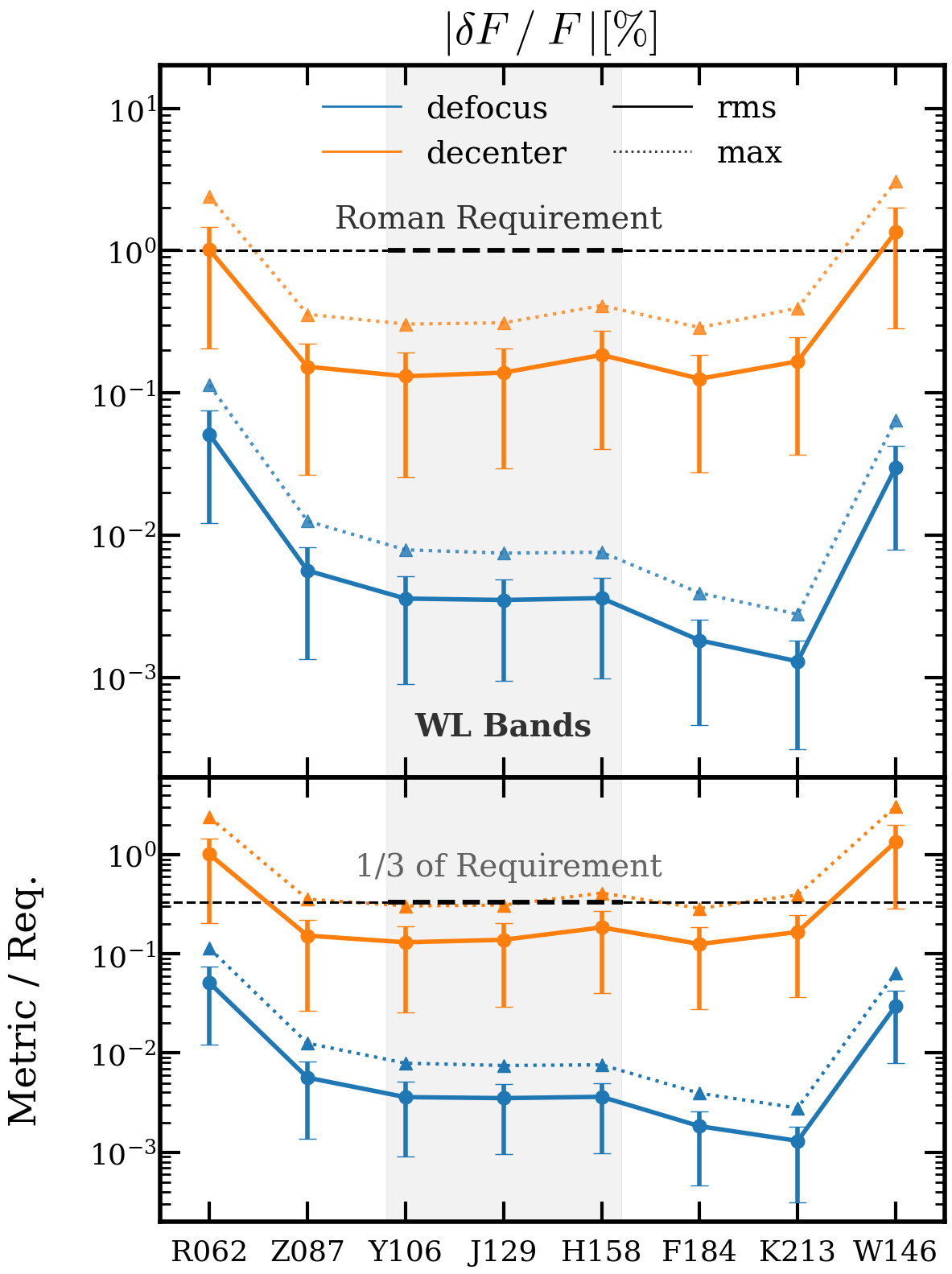}
    \caption{Similar to Fig.~\ref{fig:psf_quant_residuals} but for the point-source flux residuals. The decentering contribution is still much larger than the defocus contribution, but in this case residuals are within one third of $1\%$ flux requirement for both effects in most bands. The only exceptions are the R062 and W146 bands for decentering, where flux residuals hover around or exceed the requirement. 
    }
    \label{fig:flux_quant_residuals}

\end{figure}

Fig.~\ref{fig:ray_trace_comp} compares the ray-traced and analytic shifts across the focal plane as a function of incidence angle. As expected in the paraxial approximation, the lateral shifts scale linearly with incidence angle ($\Delta x\propto \theta_x$ and $\Delta y\propto \theta_y$). The analytic model reproduces the ray-traced lateral shifts to within \(0.03\,\mu\mathrm{m}\) for all bands except W146, where the agreement is within \(0.05\,\mu\mathrm{m}\). Given the \textit{Roman} pixel scale of \(10\,\mu\mathrm{m}\), even the largest lateral residuals are smaller than \(\sim 1/200\) of a pixel.

The longitudinal shifts are reproduced less precisely: in the largest-residual cases, the analytic model agrees with the ray tracing at roughly the \(10\%\) level and tends to slightly overestimate the shift. This is nevertheless sufficient, since the longitudinal contribution is subdominant in the PSF results shown in Sec.~\ref{subsec:psf_errors}. The field dependence of $\Delta z$ is driven mainly by the inclination of the focal plane relative to the filters, which changes the filter-to-detector distance across the field. The remaining residuals shown in the right panel of Fig.~\ref{fig:ray_trace_comp}  likely arise from the two-ray approximation of the full ray bundle used in the analytic model. This approximation does not fully capture all factors that affect the best-focus position, such as the aberrated structure of the ray bundle before it reaches the filter. 

Overall, Fig.~\ref{fig:ray_trace_comp} shows that the simplified geometric prescription captures the main trends and amplitudes of the chromatic shifts across the \textit{Roman} focal plane without requiring full ray tracing at every position and wavelength.

\subsection{PSF and flux errors}\label{subsec:psf_errors}

To quantify the impact of filter-substrate refraction on observable image properties, we evaluate the simulated \textit{Roman} PSF over the full focal plane and in all imaging bands. Specifically, we consider all 18 SCAs and all 8 \textit{Roman} filters, and on each SCA sample 100 positions arranged on a \(10\times10\) linearly spaced grid. At every sampled position, we generate three PSF realizations using the procedure described in Sec.~\ref{sec:image_sims}: (i) a reference PSF with no refraction-induced aberrations added, (ii) a PSF including only the lateral shifts, and (iii) a PSF including only the longitudinal shifts. Treating the two effects separately allows us to isolate the relative importance of decentering and defocus.

Fig.~\ref{fig:psf_img_diff} shows an example of the broadband PSF image residuals for the decentering and defocus contributions separately. In both cases, the residual pattern matches the expected qualitative behavior of the corresponding aberration: the defocus residual is approximately radially symmetric, while the decentering residual shows the expected dipole-like pattern from the wavelength-dependent displacement of the PSF centroid. The relative amplitudes are also clear from the color scale, with the decentering residual roughly two orders of magnitude larger than the defocus residual at the pixel level.

We now turn to quantitative diagnostics relevant for weak lensing and photometric measurements. Differences in PSF size and ellipticity are directly relevant for weak lensing applications \citep[see, e.g.,][]{Jarvis_2021, Schutt_2025}, where biases in the modeled PSF can propagate into shear measurements \citep{Mandelbaum_2018}. Mismatches between the true PSF and the PSF used in model fitting can also bias flux estimates, which is relevant for stellar, galaxy, and supernova photometry \citep{Astier_2013,Jimenez_2015, Lee_2023}. Here we evaluate the induced flux error only for point-source estimates; for extended sources the result would also depend on the galaxy surface brightness profiles and the chosen photometric estimator, which is beyond the scope of this work.

PSF size residuals are quantified in terms of the fractional difference in the trace of the PSF second moments \citep{Hirata_2003},
\begin{equation}
\frac{\Delta T}{T}
=
\frac{T_{\rm ab}-T_0}{T_0},
\label{eq:frac_T}
\end{equation}
where \(T \equiv I_{xx}+I_{yy}\), \(T_0\) is measured from the reference PSF, and \(T_{\rm ab}\) from the PSF with either the lateral or longitudinal aberration added. The second moments are defined as
\begin{equation}
I_{ij} = \int x_i\, x_j\, w(\mathbf{x})\,I(\mathbf{x})\,d^2x,
\label{eq:moments_def}
\end{equation}
where \(x_i \in \{x, y\}\), \(I(\mathbf{x})\) is the PSF intensity profile, and \(w(\mathbf{x})\) is a Gaussian weight function normalized such that \(\int w(\mathbf{x})\,I(\mathbf{x})\,d^2x = 1\). For the PSF shape, we use the two distortion components
\begin{equation}
e_1 = \frac{I_{xx}-I_{yy}}{I_{xx}+I_{yy}},
\qquad
e_2 = \frac{2I_{xy}}{I_{xx}+I_{yy}},
\label{eq:e12_def}
\end{equation}
and define the corresponding residuals
\begin{equation}
\Delta e_1 = e_{1,{\rm ab}} - e_{1,0},
\qquad
\Delta e_2 = e_{2,{\rm ab}} - e_{2,0},
\label{eq:delta_e}
\end{equation}
where \(e_1\) describes ellipticity along the detector \(x\)- and \(y\)-axes and \(e_2\) along axes rotated by \(45^\circ\). The PSF size and shape are measured using adaptive moments with \texttt{GalSim}'s \textsc{HSM} module.

For point-source photometry, we treat the PSF with the added aberration as the observed image and fit it using the unperturbed PSF as the model template. For a noiseless point source, the best-fit flux from least-squares matching \citep{Lee_2023} is
\begin{equation}
\hat{f}
=
\frac{\sum_{ij} I_{ij} P_{ij}}{\sum_{ij} P_{ij}^2},
\label{eq:best_fit_flux}
\end{equation}
where \(I_{ij}\) is the perturbed image and \(P_{ij}\) is the template PSF evaluated in pixel \(ij\). We then define the fractional flux residual as
\begin{equation}
\frac{\Delta F}{F}
=
\frac{\hat{f}-f_0}{f_0},
\label{eq:frac_flux}
\end{equation}
where \(f_0\) is the true input flux, equal to 1 by construction in our simulations. This test should be interpreted as a simple point-source diagnostic of PSF mismatch rather than as a direct prediction for galaxy photometry.

\subsubsection{Filter dependence}\label{subsubsec:filter_dep}

\begin{figure*}
    \centering
    \includegraphics[width=0.99\linewidth]{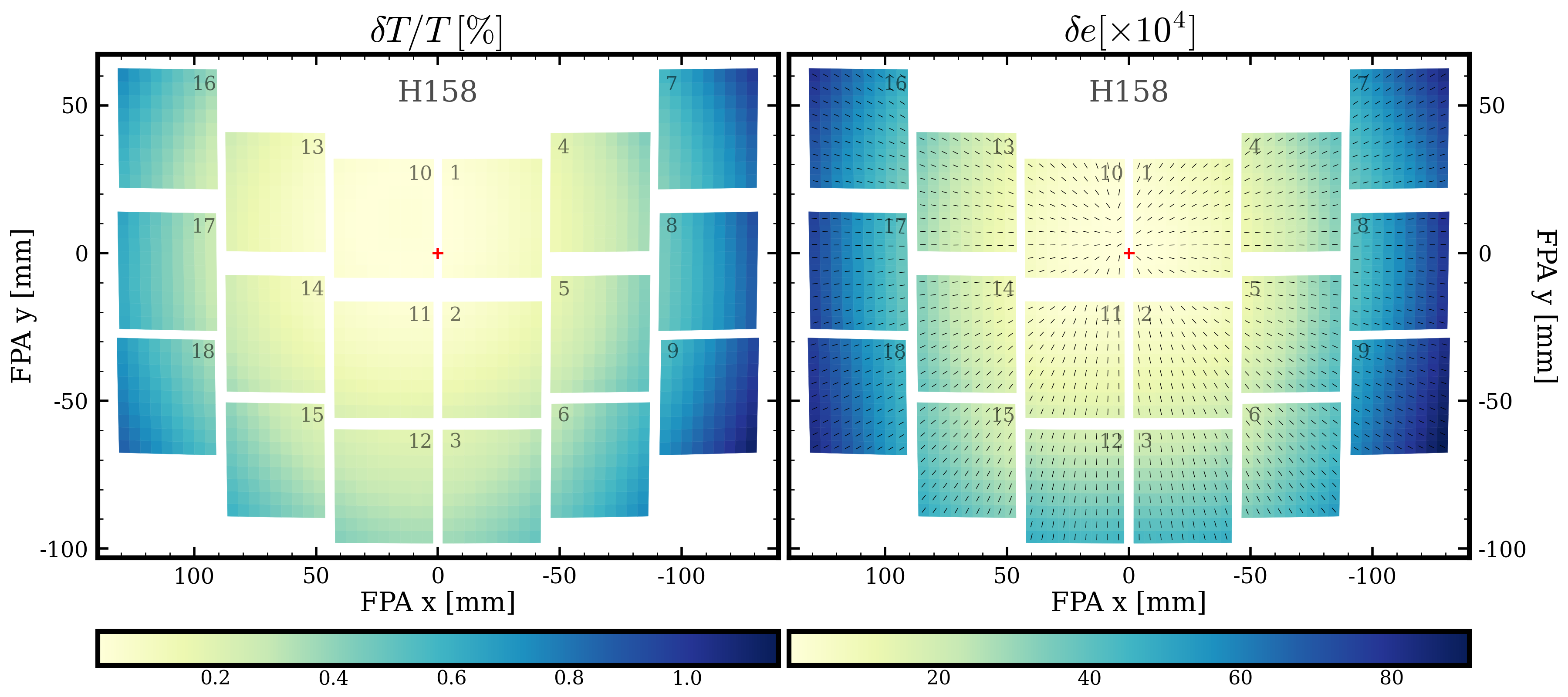}
        \caption{Spatial dependence of the dominant decentering contribution to the PSF residuals for an average galaxy SED across the full \textit{Roman} focal plane, shown here for the H158 filter. The two panels show the absolute fractional PSF size residual, $\delta T/T$, and the absolute PSF ellipticity residual, $\delta e =  \sqrt{\delta e_1^2 + \delta e_2^2}$, evaluated on a $10\times10$ grid of positions within each of the 18 SCAs. 
        In the ellipticity panel, the whiskers indicate the direction of the ellipticity residual vector. 
        In both cases, the residuals are smallest near the center of the focal plane and increase smoothly toward the outer SCAs, reaching their largest values near the field edges and corners. This behavior follows directly from the geometric origin of the effect: the angle of incidence at the filter increases with distance from the focal-plane center (shown as the red plus sign), so the wavelength-dependent lateral shifts, and hence the decentering-induced PSF errors, also grow toward larger field angles. The ellipticity residuals also show a highly coherent pattern, with an approximately radial alignment relative to the field center. The figure illustrates that the dominant PSF impact of filter-substrate refraction is not only large in magnitude, but also highly field dependent and spatially coherent across the \textit{Roman} focal plane. 
        }
    \label{fig:fov_dependence}

\end{figure*}

\begin{figure}
    \centering
    \includegraphics[width=0.9\linewidth]{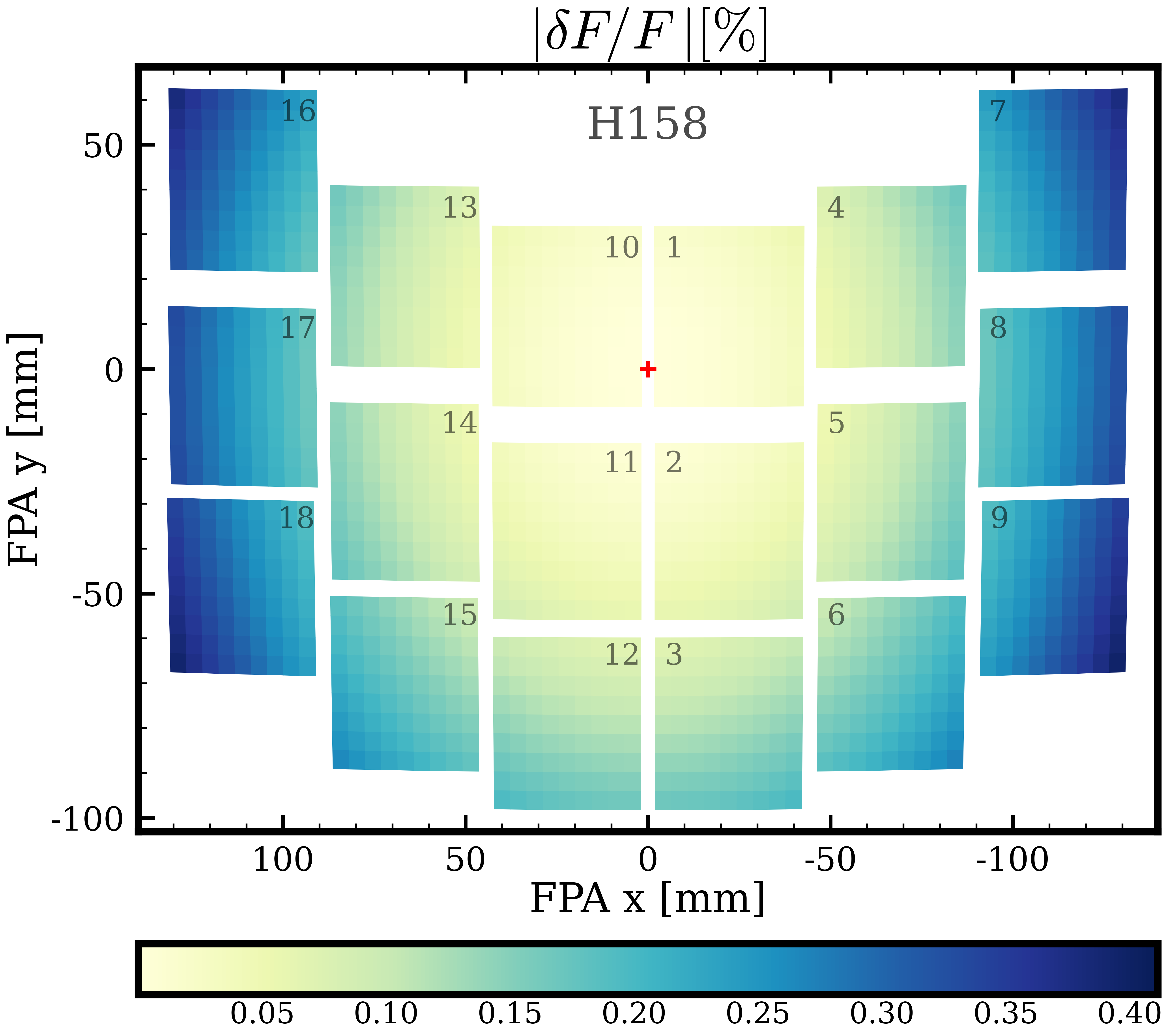}
        \caption{Similar to Fig.~\ref{fig:fov_dependence} but for the point-source flux residuals. As we can see, the flux residual closely tracking the spatial pattern of the PSF size residual, as expected for point-source photometry based on PSF fitting. For the H158 band, flux residuals can reach $\sim 0.4\%$ towards the edges of the focal plane.}
    \label{fig:flux_fov_dependence}

\end{figure}

We begin by examining how the PSF size, shape, and point-source flux residuals vary across the \textit{Roman} imaging bands. Since the relevant requirements are formulated in terms of root-mean-square (RMS) residuals, we compute the RMS of \(\Delta T/T\), \(\Delta e_1\), \(\Delta e_2\), and \(\Delta F/F\) across the full focal plane, and also track the maximum residual to quantify the worst-case behavior.

Figure~\ref{fig:psf_quant_residuals} shows the PSF size and shape residuals for all eight bands. The clearest result is that decentering is the dominant contribution: across all three PSF diagnostics, its residuals are more than an order of magnitude larger than those from defocus. 
This is consistent with the pixel-level differences shown in Fig.~\ref{fig:psf_img_diff}, where the decentering residuals are also approximately two orders of magnitude larger than those from defocus. To place these results in context, we compare them to the \textit{Roman} HLIS requirements from \cite{SRD_2023}, namely \(7.2\times10^{-4}\) for the fractional error in the trace of the second moments and \(5.7\times10^{-4}\) for each ellipticity component. The lower panels show each metric normalized by its requirement, with a horizontal line at one third of the requirement, a commonly used approximate per-systematic allocation when independent systematics are added in quadrature \citep{LSST_2018}. This more stringent reference on individual systematics is motivated by the fact that multiple effects lead to PSF errors and therefore contribute to the total PSF error budget. 
These PSF requirements are most directly relevant for the weak lensing bands Y106, J129, and H158.

For the defocus contribution, the RMS residual remains below one third of the requirement for all three PSF diagnostics in the weak lensing bands, and even the worst-case focal-plane regions remain below this threshold. The RMS exceeds one third of the requirement only for the PSF size and shape residuals in R062 and W146, which are not part of the baseline weak lensing analysis. We therefore conclude that the longitudinal component of filter-substrate refraction is negligible for weak lensing shape measurement in the baseline \textit{Roman} bands.

The decentering contribution is very different. In the weak lensing bands, the RMS PSF size residual is typically \(0.3\)–\(0.4\%\), exceeding \(1\%\) in the outer part of the focal plane, and the corresponding ellipticity residuals are likewise well above the systematic error budget. Thus, the lateral chromatic shift exceeds the allowed PSF size and shape errors by nearly an order of magnitude and is clearly relevant for weak lensing analyses. The effect is even larger in R062 and W146, where the PSF size residual reaches the \(2\)–\(3\%\) level.

The point-source flux residuals are shown separately in Fig.~\ref{fig:flux_quant_residuals}. Using the \(1\%\) \textit{Roman} point-source photometric requirement from \cite{SRD_2023}, which is primarily motivated by the need for accurate fluxes for photometric redshift estimation,  
we again find that decentering dominates over defocus. For most bands, the decentering-induced flux bias remains below one third of the requirement, with only the worst-case regions approaching that level. In R062 and W146, however, the RMS flux residual reaches the requirement and the worst cases exceed it, indicating that filter-substrate refraction can become relevant for point-source photometry in these bands. Since a similar \(1\%\) requirement also applies to the \textit{Roman} High Latitude Time Domain Survey, these results might also be relevant for supernova cosmology.

Finally, although W146 is not part of the baseline \textit{Roman} weak lensing strategy, several papers have proposed using it to increase the depth of a wide-area survey \citep[see][]{Eifler_2021,Eifler_2023, Linden_2023, Eifler_2024}. Our results highlight one of the challenges of precision shape measurement in such a broad filter: chromatic effects that are negligible in narrower bands can become non-negligible in W146, while effects already relevant in the standard weak lensing bands can be amplified far beyond requirement levels. This adds to the concerns noted in the ROTAC recommendations \citep{ROTAC_2025} and in previous work on chromatic diffraction effects \citep[see][]{Berlfein_2025,Berlfein_2026} about using W146 for weak lensing with \textit{Roman}. 

\subsubsection{Spatial dependence}\label{subsubsec:spatial_dep}

We now turn to the spatial dependence of the effect. Given the results above, we focus on the dominant decentering contribution, using the H158 filter as a representative example since it enters all tiers of the HLIS. Fig.~\ref{fig:fov_dependence} shows the spatial dependence of the PSF size and ellipticity residuals induced by decentering across the full \textit{Roman} focal plane. In both panels, the residuals are smallest near the center of the focal plane and increase smoothly toward the outer SCAs, reaching their largest values near the field edges and corners. The pattern is approximately radially symmetric, as expected from the geometric origin of the effect: the angle of incidence increases with distance from the focal-plane center, so the wavelength-dependent lateral shifts, and hence the decentering-induced PSF errors, also grow towards the edges of the focal plane. Small asymmetries are due to the intrinsic PSF variation across the focal plane.

The ellipticity residuals in Fig.~\ref{fig:fov_dependence} are also highly coherent, with an approximately radial alignment relative to the field center. This shows that the induced anisotropy is not random from detector to detector, but instead follows the same underlying field dependence as the PSF size residuals. In other words, the lateral chromatic shift does not merely broaden the effective broadband PSF; it also induces a coherent, spatially varying anisotropy across the focal plane.

The corresponding point-source flux residuals are shown separately in Fig.~\ref{fig:flux_fov_dependence}. These closely track the PSF size residuals across the field, as expected for PSF-fitting photometry since changes in the effective PSF width alter how the light is distributed across pixels and therefore bias the recovered flux. We find that the flux residuals vary coherently as we move towards the edges and corners of the focal plane, with errors as large as $\sim 0.4\%$. This level of flux residuals towards the edges of the focal plane might be more relevant for transient science, for which accurate flux measurements across the focal plane are more critical.

Although not shown in the figure, we also find spatial dependence in the defocus contribution. In that case, the pattern is less controlled by incidence angle and more by the pre-existing detector-dependent aberration structure, since the same added defocus produces a larger change where the native defocus is already larger. Because this component is subdominant overall, we defer further discussion of its spatial dependence to Appendix~\ref{app:defocus}.


\subsubsection{SED dependence}
The PSF and flux errors shown in Sec.~\ref{subsubsec:filter_dep} and Sec.~\ref{subsubsec:spatial_dep} are computed using the average galaxy SED described in Sec.~\ref{subsec:roman_psf}. However, because the broadband PSF depends on the source SED, the resulting errors will also depend on this choice. To understand the sensitivity of our results to the assumed SED, we repeat the simulations and analysis in the weak lensing bands using two extreme cases: a high redshift ($z\sim2.5$) SED with a much steeper slope than the average galaxy SED across the \textit{Roman} wavelengths, and a lower redshift ($z\sim0.5$) SED with a flatter slope across the same wavelengths. We then compare the resulting RMS and maximum PSF and flux residuals to those obtained with the average SED and find that they differ by at most $10\%$ for the high-redshift SED and $5\%$ for the low-redshift SED. Even for these extreme cases, this variation is much smaller than the spatial variation across the focal plane shown in Figs.~\ref{fig:fov_dependence} and \ref{fig:flux_fov_dependence}. This indicates that the filter refraction residuals are much more sensitive to the focal-plane position than to the source SED, justifying our use of an average galaxy SED to quantify them.

\section{Practical implementation in large-scale image simulations}\label{sec:practical_implementation}

The previous sections used \textsc{GalSim}'s FFT-based drawing method to quantify the impact of filter-substrate refraction on the \textit{Roman} PSF. That approach is well suited for small and controlled studies, since it allows the chromatic perturbation to be inserted explicitly at each wavelength before integrating over the bandpass. However, this manual wavelength-by-wavelength construction is not a practical approach for large-scale image simulations, where one typically draws many objects over wide areas. In that regime, the computational cost of repeatedly generating and summing monochromatic PSFs across the bandpass becomes infeasible.

Large-scale image simulations commonly rely on photon-shooting rather than FFT-based image drawing \citep[e.g.][]{Kitching_2010,OU_2025}. In \texttt{GalSim}, photon-shooting generates a Monte Carlo realization of the image by sampling photons from the surface-brightness profile of the object and then accumulating those photons onto the detector grid\footnote{See \url{https://galsim-developers.github.io/GalSim/_build/html/photon.html}}. Photon operators can then modify quantities such as the photon positions, directions, or fluxes before the photons reach the detector plane and are binned into pixels. This provides a simple way to implement the chromatic shifts described in this work without the need to explicitly draw and sum many monochromatic PSFs as we did in the FFT procedure.

Motivated by this, we implement the dominant refraction-induced effect---the lateral chromatic decentering---using a photon operator that directly fits within the photon-shooting framework. We do not include the defocus-like effect in this practical implementation, since the results in Sec.~\ref{subsec:psf_errors} showed that it is subdominant for point-source photometry in all bands and also subdominant for PSF size and shape in most bands, including the weak lensing bands. The goal of this section is therefore to implement a custom photon operator to apply the wavelength-dependent lateral shifts during photon-shooting and to verify that it reproduces a PSF with the same properties (size and shape) as that of the FFT-based method.

\subsection{Photon operator}\label{subsec:photon_operator}

\begin{figure*}
    \centering
    \includegraphics[width=0.99\linewidth]{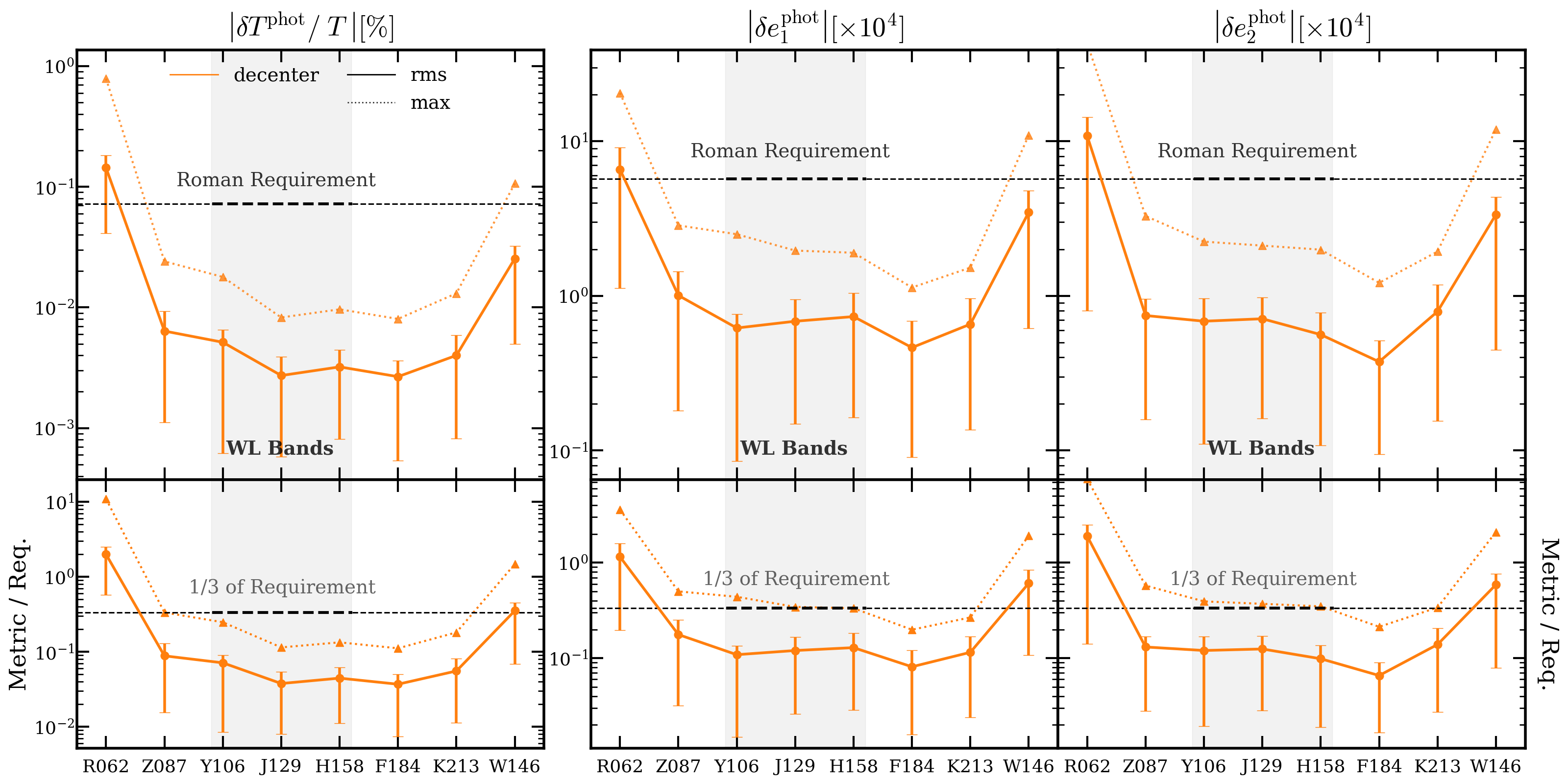}
    \caption{Comparison of the practical photon-shooting implementation of the decentering effect with the FFT-based reference calculation for an average galaxy SED. The three columns show the differences between the photon-shooting and FFT implementations for the PSF size and shape diagnostics, namely $|\delta T^{\rm phot}/T|$, $|\delta e^{\rm phot}_1|$, and $|\delta e^{\rm phot}_2|$. The lines, colors and panels are similar to those in Fig.~\ref{fig:psf_quant_residuals}. Overall, the agreement between the photon-shooting and FFT implementations is very good, with the RMS residuals in the weak lensing bands remaining well below one third of the requirement for all three diagnostics and the worst-case residuals staying at or around that level. The largest deviations occur in the R062 and W146 filters, where stronger chromatic variation across the bandpass makes the photon-operator approximation less exact. Nevertheless, the results show that the custom photon operator reproduces the dominant chromatic decentering effect with sufficient accuracy for large-scale \textit{Roman} image simulations.}
    \label{fig:phot_shooting}

\end{figure*}

In \texttt{GalSim}, a photon operator is an object that acts directly on the sampled photons during photon-shooting. Rather than modifying the final rendered image, a photon operator changes the photon properties themselves---for example, their positions or directions---before they are accumulated onto the image grid. This makes photon operators a natural way of introducing instrumental effects within the photon-shooting framework.

Our implementation is built around a custom \texttt{GalSim} photon operator that applies the predicted lateral chromatic shift according to each photon's assigned wavelength. The wavelength of each photon is assigned statistically from the object's SED and the filter throughput. For a photon of wavelength $\lambda$ and incidence angle $\boldsymbol{\theta}(\mathbf{x}_{\rm det})$, 
we first evaluate the geometric shifts, $\Delta x(\lambda), \; \Delta y(\lambda)$, relative to the effective wavelength of the bandpass, using the same refraction formalism described in Sec.~\ref{subsec:an_prescription}. 
These shifts are then converted into pixel units and applied to the photon positions $(x_i, y_i)$:
\begin{equation}
x_i \rightarrow x_i + \Delta x(\lambda_i),
\qquad
y_i \rightarrow y_i + \Delta y(\lambda_i).
\end{equation}
Conceptually, this reproduces the same physical effect as in the FFT implementation, with photons at different wavelengths being displaced by different amounts. One important caveat, however, is that this is not exactly equivalent to the FFT approach described in Sec.~\ref{sec:image_sims}. In the photon-shooting case, the photons being modified are sampled from the broadband profile after integration over the bandpass, so each photon represents a contribution to the broadband image rather than to the monochromatic image at its own wavelength. As a result, one expects larger deviations between the photon-operator and FFT implementations in filters with stronger chromatic PSF variation across the bandpass, such as the W146, or in filters where the refractive index changes more rapidly over the wavelength range of the band, such as the R062 filter. We note this statement also depends on how the wavelength grid is sampled for the FFT implementation. Hence, we use the same wavelength sampling for the FFT as described in Sec.~\ref{subsec:roman_psf}, which closely approximates \texttt{GalSim}'s internal chromatic drawing. 

\subsection{Comparison with FFT}\label{subsec:fft_comp}

To validate the photon-operator implementation, we compare it directly to the FFT-based method described in Section~\ref{sec:image_sims}, evaluating PSFs drawn with and without the added decentering effect across the \textit{Roman} focal plane. For computational efficiency, we use a \(5\times 5\) grid of positions on each SCA, corresponding to 25 sampled positions per detector across all 18 SCAs. For the photon-shooting calculation, we draw the PSF 20 times using 5 million photons each and average the resulting moments to suppress Poisson noise, which we verified is sufficient to achieve convergence in the measured moments. 

To isolate differences arising from the chromatic decentering model rather than from intrinsic differences between the two rendering methods, we subtract the baseline no-aberration difference between photon-shooting and FFT. The resulting comparison diagnostics are:
\begin{equation}
\frac{\Delta T^{\rm phot}}{T} =\frac{\left(T^{\rm phot}_{\rm dec} - T^{\rm FFT}_{\rm dec}\right) -
\left(T^{\rm phot}_{0} - T^{\rm FFT}_{0}\right)}{
T^{\rm FFT}_{0}},
\label{eq:delta_fracT_phot_fft}
\end{equation}
\begin{equation}
\Delta e_i^{\rm phot}=\left(e_{i,{\rm dec}}^{\rm phot}- e_{i,{\rm dec}}^{\rm FFT}\right)-\left(e_{i,0}^{\rm phot} - e_{i,0}^{\rm FFT}\right), \qquad i\in\{1,2\},
\label{eq:delta_e_phot_fft}
\end{equation}
\begin{equation}
\frac{\Delta F^{\rm phot}}{F} =
\frac{\left(F^{\rm phot}_{\rm dec} - F^{\rm FFT}_{\rm dec}\right) - \left(F^{\rm phot}_{0} - F^{\rm FFT}_{0}\right)}{F^{\rm FFT}_{\rm dec}},
\label{eq:delta_fracF_phot_fft}
\end{equation}
where superscripts denote the drawing method and subscripts dec or $0$ denote PSFs drawn with or without decentering. Note that in photon-shooting the pixel convolution is tied to the drawing scale, unlike the FFT method where the pixel response is applied separately. We therefore draw both versions at the same oversampled pixel scale without convolving by the native pixel response, applying the same approximation to both methods so that their relative agreement isolates the chromatic decentering model rather than pixel-convolution differences.

The results are shown in Fig.~\ref{fig:phot_shooting}. 
Overall, the agreement is very good: for most filters the RMS residuals in all three metrics remain comfortably below one third of the \textit{Roman} requirement, while even the maximum residuals typically remain near or below this threshold. This is especially true in the weak lensing bands Y106, J129, and H158, where the RMS residuals in \(\delta T/T\), \(\delta e_1\), and \(\delta e_2\) are all well below one third of the corresponding requirements and even the worst-case residuals remain at or around that level. These results confirm that the photon operator reproduces the dominant chromatic decentering effect to the accuracy needed for weak lensing image simulations in the baseline \textit{Roman} bands.

The largest deviations occur in R062 and W146, where the RMS residuals approach the requirement and the maximum residuals can exceed the one third-allocation line by a substantial margin. This is consistent with the caveat discussed in Sec.~\ref{subsec:photon_operator}: the photon-operator approach is not exactly equivalent to the FFT implementation because the photons being shifted are sampled from the broadband profile rather than from monochromatic PSFs before bandpass integration, so discrepancies are expected to be larger in filters with stronger chromatic variation. This is the case for R062, where the refractive index varies more rapidly over the filter wavelength range, and for W146, whose very broad bandpass contains substantial PSF variation. 

Although not shown in Fig.~\ref{fig:phot_shooting}, we also examined the point-source flux residuals. For all bands, including R062 and W146, the corresponding photon-shooting versus FFT differences remain well below the adopted photometric requirement. This indicates that, even where the agreement in PSF size and shape is less precise, the remaining differences are still small enough that the induced point-source photometry mismatch is subdominant.

These results, in combination with the validation described in Sec.~\ref{sec:results}, show that this implementation provides an accurate and practical way to include the dominant filter-substrate refraction effect in photon-shooting simulations. In summary, we applied a two-step verification process: the analytic chromatic shift approximation was first validated against ray tracing and then used to construct FFT-drawn PSFs that include the desired chromatic effect; the FFT-drawn PSFs were then compared to the photon-shooting PSFs, with both implementations using the same analytic approximation, to validate the latter. 
We therefore conclude that the photon operator is sufficiently accurate for use in large-scale \textit{Roman} image simulations, particularly for the baseline weak lensing bands where precise modeling of the decentering effect is most important.

The photon operator introduced here is currently being incorporated into the \textit{Roman} image simulation
package \texttt{roman\_imsim}\footnote{https://github.com/DukeCosmology/\texttt{roman\_imsim}}, which uses the \texttt{GalSim}  configuration framework for large-scale image simulations of the \textit{Roman} WFI instrument \citep[e.g.][]{OU_2025}. 


\section{Conclusion}\label{sec:conclusion}

The \textit{Roman Space Telescope} is designed to deliver high-precision measurements of weak gravitational lensing and large-scale structure. This level of precision demands exceptional control of instrumental systematics, particularly those that affect PSF modeling. One of these systematics comes from wavelength-dependent refraction in the \textit{Roman} WFI filters, which can introduce perturbations to the PSF. This effect, currently absent from large-scale \textit{Roman} image simulations, has not been independently quantified or studied for \textit{Roman}. In this work, we quantified the impact of filter-substrate refraction on the \textit{Roman} PSF and established a practical approach for including it in future image simulation efforts.

We showed that filter-substrate refraction can be decomposed into two wavelength-dependent geometric quantities: a lateral chromatic shift, which displaces the image centroid in the focal plane as a function of wavelength, and a longitudinal chromatic shift, which moves the location of best focus along the optical axis. We developed an analytic prescription for both quantities using \textit{Roman}-specific optical parameters and validated it against a full WFI ray-tracing calculation, then propagated the resulting shifts into PSF and photometric diagnostics through image simulations. Our analysis leads to the following main conclusions:

\begin{itemize}[leftmargin=*]
    \item \textbf{The analytic refraction model presented in 
    Sec.~\ref{sec:refraction_formalism} reproduces to high accuracy the chromatic shifts measured from a full \textit{Roman} ray-tracing calculation.} Despite some simplifying assumptions, the lateral chromatic shifts predicted by our model and those calculated from ray tracing agree to within $1/200$ of a pixel or better in the scenario were the effect is the largest. The longitudinal shifts agree within $10\%$, which is an acceptable level given their subdominant effect on PSF and photometric residuals.

    \item \textbf{The lateral chromatic shift (decentering) is the dominant contribution to the broadband PSF error, while the longitudinal chromatic shift (defocus) is generally subdominant and below \textit{Roman} requirements.} The decentering term produces substantially larger PSF size and shape residuals than the defocus term, indicating that the dominant impact of filter-substrate refraction is through wavelength-dependent image translation rather than a shift in the best-focus position. The defocus contribution is therefore not a limiting systematic for \textit{Roman} weak lensing and does not require dedicated treatment in image simulations or PSF modeling.

    \item \textbf{The induced PSF and flux errors depend strongly on both filter and field position, increasing toward the outer focal plane.} This spatial trend follows directly from the geometric origin of the effect, since the angle of incidence at the filter grows with distance from the focal-plane center. The spatial pattern of the decentering-induced ellipticity residuals is highly coherent, showing an approximately radial alignment across the focal plane.

    \item \textbf{In the baseline weak lensing bands Y106, J129, and H158, the decentering contribution exceeds the \textit{Roman} PSF size and shape requirements by nearly an order of magnitude, implying that this effect must be modeled for precision weak lensing analyses.} 
    Across these bands, the RMS PSF size residual is typically $\sim 0.3$--$0.4\%$, rising above $1\%$ in the outer regions of the focal plane. The corresponding RMS ellipticity residuals are typically of order $\sim 40\times10^{-4}$ and can approach $\sim 100\times10^{-4}$ toward the field edges. These residuals are considerably larger than the \textit{Roman} HLIS requirements of $0.072\%$ for the fractional error on the PSF size and $5.7\times10^{-4}$ for each ellipticity component. By contrast, the impact on point-source photometry remains below the requirements for most bands, becoming relevant only in R062 and W146.

    \item \textbf{The dominant decentering effect can be accurately incorporated into large-scale image simulations via a custom photon operator.} We implement the predicted lateral chromatic shifts, validated against ray tracing as described in the first bullet point, as a \texttt{GalSim} photon operator that directly displaces photon positions during shooting according to each photon's wavelength. Comparison with the FFT-based approach shows that this implementation reproduces the PSF size and shape residuals with accuracy well within the \textit{Roman} requirements in the weak lensing bands, making it suitable for incorporation into large-scale \textit{Roman} image simulations. 
\end{itemize}

The results also highlight a broader lesson about chromatic effects in the W146 filter. The W146 band shows substantially larger chromatic refraction-induced PSF errors than the weak lensing bands: PSF size residuals reaching the $2$--$3\%$ level, point-source flux residuals exceeding the $1\%$ requirement, and defocus-like effects becoming relevant for size and shape characterization of the PSF. This reinforces existing concerns, raised in the context of chromatic diffraction rather than refraction, about using the W146 filter for precision shape measurement \citep{Berlfein_2025,Berlfein_2026}. 

Taken together, this work establishes that filter-substrate refraction is a non-negligible source of chromatic PSF error for \textit{Roman}, and that its dominant manifestation, wavelength-dependent lateral decentering, must be accounted for in any realistic PSF model aimed at precision weak lensing science. The analytic formalism and photon-operator implementation presented here provide the tools needed to incorporate this effect into the next generation of \textit{Roman} image simulations so as to test and validate any corrections for the effect. Looking further ahead, the model predictions are also testable once the telescope is in-flight. Observations of stars with a range of spectral energy distributions across the focal plane could provide an empirical verification of the refraction model, testing its predicted wavelength and field dependence as one component of the broader end-to-end response characterization of the instrument. In addition, understanding how filter-substrate refraction affects empirical PSF modeling frameworks such as \texttt{PIFF} \footnote{https://github.com/rmjarvis/Piff}, which is expected to be used for PSF modeling for \textit{Roman} in-flight data, will also be an important next step.


\section*{Acknowledgments}

This paper has undergone internal review in the \textit{Roman} High Latitude Imaging Survey Cosmology Project Infrastructure Team (PIT). We would like to thank Ami Choi and Kaili Cao for helpful comments and feedback during the review process.
This work was supported in part by the OpenUniverse effort, which is funded by NASA under JPL Contract Task 70-711320, “Maximizing Science Exploitation of Simulated Cosmological Survey Data Across Surveys”; and in part by the “Maximizing Cosmological Science with the \textit{Roman} High Latitude Imaging Survey” \textit{Roman} Project Infrastructure Team (NASA grant 22-ROMAN11-0011). TZ is supported by Schmidt Sciences.

\section*{Data Availability}
The \texttt{GalSim} package is publicly available at \url{https://github.com/GalSim-developers/GalSim}. This research made use of \texttt{pysiaf}, an open source Python package for Handling of Science Instrument Aperture Files (SIAF) for space telescopes \citep{Sahlmann_2019}. The \texttt{Diffsky} extragalactic catalog and stellar catalog are available through the NASA/IPAC
Infrared Science Archive (IRSA) at \url{https://irsa.ipac.caltech.edu/data/theory/openuniverse2024/overview.html}.



\bibliographystyle{mnras}
\bibliography{main}

@ARTICLE{Troxel_2022,
       author = {{Troxel}, M.~A. and {Lin}, C. and {Park}, A. and {Hirata}, C. and {Mandelbaum}, R. and {Jarvis}, M. and {Choi}, A. and {Givans}, J. and {Higgins}, M. and {Sanchez}, B. and {Yamamoto}, M. and {Awan}, H. and {Chiang}, J. and {Dor{\'e}}, O. and {Walter}, C.~W. and {Zhang}, T. and {Cohen-Tanugi}, J. and {Gawiser}, E. and {Hearin}, A. and {Heitmann}, K. and {Ishak}, M. and {Kovacs}, E. and {Mao}, Y. -Y. and {Wood-Vasey}, M. and {Becker}, Matt and {Meyers}, Josh and {Melchior}, Peter and {LSST Dark Energy Science Collaboration}},
        title = "{A joint Roman Space Telescope and Rubin Observatory synthetic wide-field imaging survey}",
      journal = {\mnras},
         year = 2023,
        month = jun,
       volume = {522},
       number = {2},
        pages = {2801-2820},
          doi = {10.1093/mnras/stad664},
archivePrefix = {arXiv},
       eprint = {2209.06829},
 primaryClass = {astro-ph.IM},
       adsurl = {https://ui.adsabs.harvard.edu/abs/2023MNRAS.522.2801T}
}

@ARTICLE{Galsim,
       author = {{Rowe}, B.~T.~P. and {Jarvis}, M. and {Mandelbaum}, R. and {Bernstein}, G.~M. and {Bosch}, J. and {Simet}, M. and {Meyers}, J.~E. and {Kacprzak}, T. and {Nakajima}, R. and {Zuntz}, J. and {Miyatake}, H. and {Dietrich}, J.~P. and {Armstrong}, R. and {Melchior}, P. and {Gill}, M.~S.~S.},
        title = "{GALSIM: The modular galaxy image simulation toolkit}",
      journal = {Astronomy and Computing},
         year = 2015,
        month = apr,
       volume = {10},
        pages = {121-150},
          doi = {10.1016/j.ascom.2015.02.002},
archivePrefix = {arXiv},
       eprint = {1407.7676},
 primaryClass = {astro-ph.IM},
       adsurl = {https://ui.adsabs.harvard.edu/abs/2015A&C....10..121R}
}

@ARTICLE{Hirata_2024,
       author = {{Hirata}, Christopher M. and {Yamamoto}, Masaya and {Laliotis}, Katherine and {Macbeth}, Emily and {Troxel}, M.~A. and {Zhang}, Tianqing and {Cao}, Kaili and {Choi}, Ami and {Givans}, Jahmour and {Heitmann}, Katrin and {Ishak}, Mustapha and {Jarvis}, Mike and {Kovacs}, Eve and {Long}, Heyang and {Mandelbaum}, Rachel and {Park}, Andy and {Porredon}, Anna and {Walter}, Christopher W. and {Wood-Vasey}, W. Michael},
        title = "{Simulating image coaddition with the Nancy Grace Roman Space Telescope - I. Simulation methodology and general results}",
      journal = {\mnras},
         year = 2024,
        month = feb,
       volume = {528},
       number = {2},
        pages = {2533-2561},
          doi = {10.1093/mnras/stae182},
archivePrefix = {arXiv},
       eprint = {2303.08749},
 primaryClass = {astro-ph.IM},
       adsurl = {https://ui.adsabs.harvard.edu/abs/2024MNRAS.528.2533H}
}

@ARTICLE{Hearin2020,
       author = {{Hearin}, Andrew and {Korytov}, Danila and {Kovacs}, Eve and {Benson}, Andrew and {Aung}, Han and {Bradshaw}, Christopher and {Campbell}, Duncan and {LSST Dark Energy Science Collaboration}},
        title = "{Generating synthetic cosmological data with GalSampler}",
      journal = {\mnras},
         year = 2020,
        month = jul,
       volume = {495},
       number = {4},
        pages = {5040-5051},
          doi = {10.1093/mnras/staa1495},
archivePrefix = {arXiv},
       eprint = {1909.07340},
 primaryClass = {astro-ph.CO},
       adsurl = {https://ui.adsabs.harvard.edu/abs/2020MNRAS.495.5040H}
}

@ARTICLE{Mandelbaum_2018,
       author = {{Mandelbaum}, Rachel},
        title = "{Weak Lensing for Precision Cosmology}",
      journal = {\araa},
         year = 2018,
        month = sep,
       volume = {56},
        pages = {393-433},
          doi = {10.1146/annurev-astro-081817-051928},
archivePrefix = {arXiv},
       eprint = {1710.03235},
 primaryClass = {astro-ph.CO},
       adsurl = {https://ui.adsabs.harvard.edu/abs/2018ARA&A..56..393M}
}

@ARTICLE{Hoekstra_2008,
       author = {{Hoekstra}, Henk and {Jain}, Bhuvnesh},
        title = "{Weak Gravitational Lensing and Its Cosmological Applications}",
      journal = {Annual Review of Nuclear and Particle Science},
         year = 2008,
        month = nov,
       volume = {58},
       number = {1},
        pages = {99-123},
          doi = {10.1146/annurev.nucl.58.110707.171151},
archivePrefix = {arXiv},
       eprint = {0805.0139},
 primaryClass = {astro-ph},
       adsurl = {https://ui.adsabs.harvard.edu/abs/2008ARNPS..58...99H}
}

@article{Liaudat_2023,
    author = "Liaudat, Tobias and Starck, Jean-Luc and Kilbinger, Martin",
    title = "{Point spread function modelling for astronomical telescopes: a review focused on weak gravitational lensing studies}",
    eprint = "2306.07996",
    archivePrefix = "arXiv",
    primaryClass = "astro-ph.IM",
    doi = "10.3389/fspas.2023.1158213",
    journal = "Front. Astron. Space Sci.",
    volume = "10",
    pages = "1158213",
    year = "2023"
}

@ARTICLE{Piotrowski_2013,
       author = {{Piotrowski}, L.~W. and {Batsch}, T. and {Czyrkowski}, H. and {Cwiok}, M. and {Dabrowski}, R. and {Kasprowicz}, G. and {Majcher}, A. and {Majczyna}, A. and {Malek}, K. and {Mankiewicz}, L. and {Nawrocki}, K. and {Opiela}, R. and {Siudek}, M. and {Sokolowski}, M. and {Wawrzaszek}, R. and {Wrochna}, G. and {Zaremba}, M. and {{\.Z}arnecki}, A.~F.},
        title = "{PSF modelling for very wide-field CCD astronomy}",
      journal = {\aap},
         year = 2013,
        month = mar,
       volume = {551},
          eid = {A119},
        pages = {A119},
          doi = {10.1051/0004-6361/201219230},
archivePrefix = {arXiv},
       eprint = {1302.0145},
 primaryClass = {astro-ph.IM},
       adsurl = {https://ui.adsabs.harvard.edu/abs/2013A&A...551A.119P}
}

@ARTICLE{Jarvis_2021,
       author = {{Jarvis}, M. and {Bernstein}, G.~M. and {Amon}, A. and {Davis}, C. and {L{\'e}get}, P.~F. and {Bechtol}, K. and {Harrison}, I. and {Gatti}, M. and {Roodman}, A. and {Chang}, C. and {Chen}, R. and {Choi}, A. and {Desai}, S. and {Drlica-Wagner}, A. and {Gruen}, D. and {Gruendl}, R.~A. and {Hernandez}, A. and {MacCrann}, N. and {Meyers}, J. and {Navarro-Alsina}, A. and {Pandey}, S. and {Plazas}, A.~A. and {Secco}, L.~F. and {Sheldon}, E. and {Troxel}, M.~A. and {Vorperian}, S. and {Wei}, K. and {Zuntz}, J. and {Abbott}, T.~M.~C. and {Aguena}, M. and {Allam}, S. and {Avila}, S. and {Bhargava}, S. and {Bridle}, S.~L. and {Brooks}, D. and {Carnero Rosell}, A. and {Carrasco Kind}, M. and {Carretero}, J. and {Costanzi}, M. and {da Costa}, L.~N. and {De Vicente}, J. and {Diehl}, H.~T. and {Doel}, P. and {Everett}, S. and {Flaugher}, B. and {Fosalba}, P. and {Frieman}, J. and {Garc{\'\i}a-Bellido}, J. and {Gaztanaga}, E. and {Gerdes}, D.~W. and {Gutierrez}, G. and {Hinton}, S.~R. and {Hollowood}, D.~L. and {Honscheid}, K. and {James}, D.~J. and {Kent}, S. and {Kuehn}, K. and {Kuropatkin}, N. and {Lahav}, O. and {Maia}, M.~A.~G. and {March}, M. and {Marshall}, J.~L. and {Melchior}, P. and {Menanteau}, F. and {Miquel}, R. and {Ogando}, R.~L.~C. and {Paz-Chinch{\'o}n}, F. and {Rykoff}, E.~S. and {Sanchez}, E. and {Scarpine}, V. and {Schubnell}, M. and {Serrano}, S. and {Sevilla-Noarbe}, I. and {Smith}, M. and {Suchyta}, E. and {Swanson}, M.~E.~C. and {Tarle}, G. and {Varga}, T.~N. and {Walker}, A.~R. and {Wester}, W. and {Wilkinson}, R.~D. and {DES Collaboration}},
        title = "{Dark Energy Survey year 3 results: point spread function modelling}",
      journal = {\mnras},
         year = 2021,
        month = feb,
       volume = {501},
       number = {1},
        pages = {1282-1299},
          doi = {10.1093/mnras/staa3679},
archivePrefix = {arXiv},
       eprint = {2011.03409},
 primaryClass = {astro-ph.IM},
       adsurl = {https://ui.adsabs.harvard.edu/abs/2021MNRAS.501.1282J}
}

@ARTICLE{Anderson_2000,
       author = {{Anderson}, Jay and {King}, Ivan R.},
        title = "{Toward High-Precision Astrometry with WFPC2. I. Deriving an Accurate Point-Spread Function}",
      journal = {\pasp},
         year = 2000,
        month = oct,
       volume = {112},
       number = {776},
        pages = {1360-1382},
          doi = {10.1086/316632},
archivePrefix = {arXiv},
       eprint = {astro-ph/0006325},
 primaryClass = {astro-ph},
       adsurl = {https://ui.adsabs.harvard.edu/abs/2000PASP..112.1360A}
}

@INPROCEEDINGS{Wyant_1992,
       author = {{Wyant}, James C. and {Creath}, Katherine},
        title = "{Basic Wavefront Aberration Theory for Optical Metrology}",
    booktitle = {Applied Optics and Optical Engineering, Volume XI},
        series = 1,
         year = 1992,
       editor = {{Shannon}, Robert R. and {Wyant}, James C.},
       volume = {11},
        month = jan,
        pages = {2},
       adsurl = {https://ui.adsabs.harvard.edu/abs/1992aooe...11....2W}
}

@ARTICLE{Cypriano_2010,
       author = {{Cypriano}, E.~S. and {Amara}, A. and {Voigt}, L.~M. and {Bridle}, S.~L. and {Abdalla}, F.~B. and {R{\'e}fr{\'e}gier}, A. and {Seiffert}, M. and {Rhodes}, J.},
        title = "{Cosmic shear requirements on the wavelength dependence of telescope point spread functions}",
      journal = {\mnras},
         year = 2010,
        month = jun,
       volume = {405},
       number = {1},
        pages = {494-502},
          doi = {10.1111/j.1365-2966.2010.16461.x},
archivePrefix = {arXiv},
       eprint = {1001.0759},
 primaryClass = {astro-ph.CO},
       adsurl = {https://ui.adsabs.harvard.edu/abs/2010MNRAS.405..494C}
}

@INPROCEEDINGS{Perrin_2014,
       author = {{Perrin}, Marshall D. and {Sivaramakrishnan}, Anand and {Lajoie}, Charles-Philippe and {Elliott}, Erin and {Pueyo}, Laurent and {Ravindranath}, Swara and {Albert}, Lo{\"\i}c.},
        title = "{Updated point spread function simulations for JWST with WebbPSF}",
    booktitle = {Space Telescopes and Instrumentation 2014: Optical, Infrared, and Millimeter Wave},
         year = 2014,
       editor = {{Oschmann}, Jr., Jacobus M. and {Clampin}, Mark and {Fazio}, Giovanni G. and {MacEwen}, Howard A.},
       series = {Society of Photo-Optical Instrumentation Engineers (SPIE) Conference Series},
       volume = {9143},
        month = aug,
          eid = {91433X},
        pages = {91433X},
          doi = {10.1117/12.2056689},
       adsurl = {https://ui.adsabs.harvard.edu/abs/2014SPIE.9143E..3XP}
}

@article{Meyers_2015b,
    author = "Meyers, Joshua E. and Burchat, Patricia R.",
    title = "{Impact of Atmospheric Chromatic Effects on Weak Lensing Measurements}",
    eprint = "1409.6273",
    archivePrefix = "arXiv",
    primaryClass = "astro-ph.CO",
    doi = "10.1088/0004-637X/807/2/182",
    journal = "Astrophys. J.",
    volume = "807",
    number = "2",
    pages = "182",
    year = "2015",
    month = 07
}

@ARTICLE{Plazas_2012,
       author = {{Plazas}, Andr{\'e}s Alejandro and {Bernstein}, Gary},
        title = "{Atmospheric Dispersion Effects in Weak Lensing Measurements}",
      journal = {\pasp},
         year = 2012,
        month = oct,
       volume = {124},
       number = {920},
        pages = {1113},
          doi = {10.1086/668294},
archivePrefix = {arXiv},
       eprint = {1204.1346},
 primaryClass = {astro-ph.IM},
       adsurl = {https://ui.adsabs.harvard.edu/abs/2012PASP..124.1113P}
}

@ARTICLE{Eriksen_2018,
       author = {{Eriksen}, Martin and {Hoekstra}, Henk},
        title = "{Implications of a wavelength-dependent PSF for weak lensing measurements}",
      journal = {\mnras},
         year = 2018,
        month = jul,
       volume = {477},
       number = {3},
        pages = {3433-3448},
          doi = {10.1093/mnras/sty830},
archivePrefix = {arXiv},
       eprint = {1707.04334},
 primaryClass = {astro-ph.CO},
       adsurl = {https://ui.adsabs.harvard.edu/abs/2018MNRAS.477.3433E}
}

@ARTICLE{Carlsten_2018,
       author = {{Carlsten}, S.~G. and {Strauss}, Michael A. and {Lupton}, Robert H. and {Meyers}, Joshua E. and {Miyazaki}, Satoshi},
        title = "{Wavelength-dependent PSFs and their impact on weak lensing measurements}",
      journal = {\mnras},
         year = 2018,
        month = sep,
       volume = {479},
       number = {2},
        pages = {1491-1504},
          doi = {10.1093/mnras/sty1636},
archivePrefix = {arXiv},
       eprint = {1804.04277},
 primaryClass = {astro-ph.IM},
       adsurl = {https://ui.adsabs.harvard.edu/abs/2018MNRAS.479.1491C}
}

@ARTICLE{Bernstein_2002,
       author = {{Bernstein}, G.~M. and {Jarvis}, M.},
        title = "{Shapes and Shears, Stars and Smears: Optimal Measurements for Weak Lensing}",
      journal = {\aj},
         year = 2002,
        month = feb,
       volume = {123},
       number = {2},
        pages = {583-618},
          doi = {10.1086/338085},
archivePrefix = {arXiv},
       eprint = {astro-ph/0107431},
 primaryClass = {astro-ph},
       adsurl = {https://ui.adsabs.harvard.edu/abs/2002AJ....123..583B}
}

@ARTICLE{Hirata_2003,
       author = {{Hirata}, Christopher and {Seljak}, Uro{\v{s}}},
        title = "{Shear calibration biases in weak-lensing surveys}",
      journal = {\mnras},
         year = 2003,
        month = aug,
       volume = {343},
       number = {2},
        pages = {459-480},
          doi = {10.1046/j.1365-8711.2003.06683.x},
archivePrefix = {arXiv},
       eprint = {astro-ph/0301054},
 primaryClass = {astro-ph},
       adsurl = {https://ui.adsabs.harvard.edu/abs/2003MNRAS.343..459H}
}

@ARTICLE{Mandelbaum_2005,
       author = {{Mandelbaum}, Rachel and {Hirata}, Christopher M. and {Seljak}, Uro{\v{s}} and {Guzik}, Jacek and {Padmanabhan}, Nikhil and {Blake}, Cullen and {Blanton}, Michael R. and {Lupton}, Robert and {Brinkmann}, Jonathan},
        title = "{Systematic errors in weak lensing: application to SDSS galaxy-galaxy weak lensing}",
      journal = {\mnras},
         year = 2005,
        month = aug,
       volume = {361},
       number = {4},
        pages = {1287-1322},
          doi = {10.1111/j.1365-2966.2005.09282.x},
archivePrefix = {arXiv},
       eprint = {astro-ph/0501201},
 primaryClass = {astro-ph},
       adsurl = {https://ui.adsabs.harvard.edu/abs/2005MNRAS.361.1287M}
}

@ARTICLE{Schutt_2025,
       author = {{Schutt}, T. and {Jarvis}, M. and {Roodman}, A. and {Amon}, A. and {Becker}, M.~R. and {Gruendl}, R.~A. and {Yamamoto}, M. and {Bechtol}, K. and {Bernstein}, G.~M. and {Gatti}, M. and {Rykoff}, E.~S. and {Sheldon}, E. and {Troxel}, M.~A. and {Abbott}, T.~M.~C. and {Aguena}, M. and {Andrade-Oliveira}, F. and {Brooks}, D. and {Rosell}, A. Carnero and {Carretero}, J. and {Chang}, C. and {Choi}, A. and {Crocce}, M. and {da Costa}, L.~N. and {Davis}, T.~M. and {De Vicente}, J. and {Desai}, S. and {Diehl}, H.~T. and {Doel}, P. and {Fert{\'e}}, A. and {Frieman}, J. and {Garc{\'\i}a-Bellido}, J. and {Gaztanaga}, E. and {Gruen}, D. and {Gutierrez}, G. and {Hinton}, S.~R. and {Hollowood}, D.~L. and {Honscheid}, K. and {Kuehn}, K. and {Lahav}, O. and {Lee}, S. and {Lima}, M. and {Marshall}, J.~L. and {Mena-Fern{\'a}ndez}, J. and {Miquel}, R. and {Mohr}, J.~J. and {Myles}, J. and {Ogando}, R.~L.~C. and {Pieres}, A. and {Malag{\'o}n}, A.~A. Plazas and {Porredon}, A. and {Samuroff}, S. and {Sanchez}, E. and {Cid}, D. Sanchez and {Sevilla-Noarbe}, I. and {Smith}, M. and {Suchyta}, E. and {Tarle}, G. and {Vikram}, V. and {Walker}, A.~R. and {Weaverdyck}, N.},
        title = "{Dark Energy Survey Year 6 Results: Point-Spread Function Modeling}",
      journal = {The Open Journal of Astrophysics},
         year = 2025,
        month = mar,
       volume = {8},
          eid = {26},
        pages = {26},
          doi = {10.33232/001c.132299},
archivePrefix = {arXiv},
       eprint = {2501.05781},
 primaryClass = {astro-ph.CO},
       adsurl = {https://ui.adsabs.harvard.edu/abs/2025OJAp....8E..26S}
}

@ARTICLE{OU_2025,
       author = {{OpenUniverse} and {LSST Dark Energy Science Collaboration} and {Roman HLIS Project Infrastructure} and {Roman Rapid Project Infrastructure Team} and {Roman Supernova Cosmology Project Infrastructure Team} and {Alarcon}, A. and {Aldoroty}, L. and {Beltz-Mohrmann}, G. and {Bera}, A. and {Blazek}, J. and {Bogart}, J. and {Braeunlich}, G. and {Broughton}, A. and {Cao}, K. and {Chiang}, J. and {Chisari}, N.~E. and {Desai}, V. and {Fang}, Y. and {Galbany}, L. and {Hearin}, A. and {Heitmann}, K. and {Hirata}, C. and {Hounsell}, R. and {Jain}, B. and {Jarvis}, M. and {Jencson}, J. and {Kannawadi}, A. and {Kasliwal}, M.~K. and {Kessler}, R. and {Kiessling}, A. and {Knop}, R. and {Kovacs}, E. and {Laher}, R. and {Laliotis}, K. and {Lin}, C. and {Lopes}, I. and {MacBeth}, E. and {Mahabal}, A. and {Mandelbaum}, R. and {Masiero}, J. and {Mau}, S. and {Meehan}, C. and {Meyers}, J. and {Moraes}, B. and {Paladini}, R. and {Pearl}, A. and {Malagon}, A. Plazas and {Rose}, B. and {Rubin}, D. and {Rusholme}, B. and {Santos}, A. and {{\v{S}}ar{\v{c}}evi{\'c}}, N. and {Scolnic}, D. and {Singhal}, J. and {Troxel}, M.~A. and {van Alfen}, N. and {van Dyke}, S. and {Walter}, C.~W. and {Wu}, T. and {Yamamoto}, M. and {Yan}, L. and {Zhang}, T.},
        title = "{OpenUniverse2024: a shared, simulated view of the sky for the next generation of cosmological surveys}",
      journal = {\mnras},
         year = 2025,
        month = dec,
       volume = {544},
       number = {4},
        pages = {3799-3823},
          doi = {10.1093/mnras/staf1833},
archivePrefix = {arXiv},
       eprint = {2501.05632},
 primaryClass = {astro-ph.CO},
       adsurl = {https://ui.adsabs.harvard.edu/abs/2025MNRAS.544.3799O}
}

@ARTICLE{DES_2005,
       author = {{The Dark Energy Survey Collaboration}},
        title = "{The Dark Energy Survey}",
      journal = {arXiv e-prints},
         year = 2005,
        month = oct,
          eid = {astro-ph/0510346},
        pages = {astro-ph/0510346},
          doi = {10.48550/arXiv.astro-ph/0510346},
archivePrefix = {arXiv},
       eprint = {astro-ph/0510346},
 primaryClass = {astro-ph},
       adsurl = {https://ui.adsabs.harvard.edu/abs/2005astro.ph.10346T}
}

@ARTICLE{LSST_2019,
       author = {{Ivezi{\'c}}, {\v{Z}}eljko and {Kahn}, Steven M. and {Tyson}, J. Anthony and {Abel}, Bob and {Acosta}, Emily and {Allsman}, Robyn and {Alonso}, David and {AlSayyad}, Yusra and {Anderson}, Scott F. and {Andrew}, John and {Angel}, James Roger P. and {Angeli}, George Z. and {Ansari}, Reza and {Antilogus}, Pierre and {Araujo}, Constanza and {Armstrong}, Robert and {Arndt}, Kirk T. and {Astier}, Pierre and {Aubourg}, {\'E}ric and {Auza}, Nicole and {Axelrod}, Tim S. and {Bard}, Deborah J. and {Barr}, Jeff D. and {Barrau}, Aurelian and {Bartlett}, James G. and {Bauer}, Amanda E. and {Bauman}, Brian J. and {Baumont}, Sylvain and {Bechtol}, Ellen and {Bechtol}, Keith and {Becker}, Andrew C. and {Becla}, Jacek and {Beldica}, Cristina and {Bellavia}, Steve and {Bianco}, Federica B. and {Biswas}, Rahul and {Blanc}, Guillaume and {Blazek}, Jonathan and {Blandford}, Roger D. and {Bloom}, Josh S. and {Bogart}, Joanne and {Bond}, Tim W. and {Booth}, Michael T. and {Borgland}, Anders W. and {Borne}, Kirk and {Bosch}, James F. and {Boutigny}, Dominique and {Brackett}, Craig A. and {Bradshaw}, Andrew and {Brandt}, William Nielsen and {Brown}, Michael E. and {Bullock}, James S. and {Burchat}, Patricia and {Burke}, David L. and {Cagnoli}, Gianpietro and {Calabrese}, Daniel and {Callahan}, Shawn and {Callen}, Alice L. and {Carlin}, Jeffrey L. and {Carlson}, Erin L. and {Chandrasekharan}, Srinivasan and {Charles-Emerson}, Glenaver and {Chesley}, Steve and {Cheu}, Elliott C. and {Chiang}, Hsin-Fang and {Chiang}, James and {Chirino}, Carol and {Chow}, Derek and {Ciardi}, David R. and {Claver}, Charles F. and {Cohen-Tanugi}, Johann and {Cockrum}, Joseph J. and {Coles}, Rebecca and {Connolly}, Andrew J. and {Cook}, Kem H. and {Cooray}, Asantha and {Covey}, Kevin R. and {Cribbs}, Chris and {Cui}, Wei and {Cutri}, Roc and {Daly}, Philip N. and {Daniel}, Scott F. and {Daruich}, Felipe and {Daubard}, Guillaume and {Daues}, Greg and {Dawson}, William and {Delgado}, Francisco and {Dellapenna}, Alfred and {de Peyster}, Robert and {de Val-Borro}, Miguel and {Digel}, Seth W. and {Doherty}, Peter and {Dubois}, Richard and {Dubois-Felsmann}, Gregory P. and {Durech}, Josef and {Economou}, Frossie and {Eifler}, Tim and {Eracleous}, Michael and {Emmons}, Benjamin L. and {Fausti Neto}, Angelo and {Ferguson}, Henry and {Figueroa}, Enrique and {Fisher-Levine}, Merlin and {Focke}, Warren and {Foss}, Michael D. and {Frank}, James and {Freemon}, Michael D. and {Gangler}, Emmanuel and {Gawiser}, Eric and {Geary}, John C. and {Gee}, Perry and {Geha}, Marla and {Gessner}, Charles J.~B. and {Gibson}, Robert R. and {Gilmore}, D. Kirk and {Glanzman}, Thomas and {Glick}, William and {Goldina}, Tatiana and {Goldstein}, Daniel A. and {Goodenow}, Iain and {Graham}, Melissa L. and {Gressler}, William J. and {Gris}, Philippe and {Guy}, Leanne P. and {Guyonnet}, Augustin and {Haller}, Gunther and {Harris}, Ron and {Hascall}, Patrick A. and {Haupt}, Justine and {Hernandez}, Fabio and {Herrmann}, Sven and {Hileman}, Edward and {Hoblitt}, Joshua and {Hodgson}, John A. and {Hogan}, Craig and {Howard}, James D. and {Huang}, Dajun and {Huffer}, Michael E. and {Ingraham}, Patrick and {Innes}, Walter R. and {Jacoby}, Suzanne H. and {Jain}, Bhuvnesh and {Jammes}, Fabrice and {Jee}, M. James and {Jenness}, Tim and {Jernigan}, Garrett and {Jevremovi{\'c}}, Darko and {Johns}, Kenneth and {Johnson}, Anthony S. and {Johnson}, Margaret W.~G. and {Jones}, R. Lynne and {Juramy-Gilles}, Claire and {Juri{\'c}}, Mario and {Kalirai}, Jason S. and {Kallivayalil}, Nitya J. and {Kalmbach}, Bryce and {Kantor}, Jeffrey P. and {Karst}, Pierre and {Kasliwal}, Mansi M. and {Kelly}, Heather and {Kessler}, Richard and {Kinnison}, Veronica and {Kirkby}, David and {Knox}, Lloyd and {Kotov}, Ivan V. and {Krabbendam}, Victor L. and {Krughoff}, K. Simon and {Kub{\'a}nek}, Petr and {Kuczewski}, John and {Kulkarni}, Shri and {Ku}, John and {Kurita}, Nadine R. and {Lage}, Craig S. and {Lambert}, Ron and {Lange}, Travis and {Langton}, J. Brian and {Le Guillou}, Laurent and {Levine}, Deborah and {Liang}, Ming and {Lim}, Kian-Tat and {Lintott}, Chris J. and {Long}, Kevin E. and {Lopez}, Margaux and {Lotz}, Paul J. and {Lupton}, Robert H. and {Lust}, Nate B. and {MacArthur}, Lauren A. and {Mahabal}, Ashish and {Mandelbaum}, Rachel and {Markiewicz}, Thomas W. and {Marsh}, Darren S. and {Marshall}, Philip J. and {Marshall}, Stuart and {May}, Morgan and {McKercher}, Robert and {McQueen}, Michelle and {Meyers}, Joshua and {Migliore}, Myriam and {Miller}, Michelle and {Mills}, David J. and {Miraval}, Connor and {Moeyens}, Joachim and {Moolekamp}, Fred E. and {Monet}, David G. and {Moniez}, Marc and {Monkewitz}, Serge and {Montgomery}, Christopher and {Morrison}, Christopher B. and {Mueller}, Fritz and {Muller}, Gary P. and {Mu{\~n}oz Arancibia}, Freddy and {Neill}, Douglas R. and {Newbry}, Scott P. and {Nief}, Jean-Yves and {Nomerotski}, Andrei and {Nordby}, Martin and {O'Connor}, Paul and {Oliver}, John and {Olivier}, Scot S. and {Olsen}, Knut and {O'Mullane}, William and {Ortiz}, Sandra and {Osier}, Shawn and {Owen}, Russell E. and {Pain}, Reynald and {Palecek}, Paul E. and {Parejko}, John K. and {Parsons}, James B. and {Pease}, Nathan M. and {Peterson}, J. Matt and {Peterson}, John R. and {Petravick}, Donald L. and {Libby Petrick}, M.~E. and {Petry}, Cathy E. and {Pierfederici}, Francesco and {Pietrowicz}, Stephen and {Pike}, Rob and {Pinto}, Philip A. and {Plante}, Raymond and {Plate}, Stephen and {Plutchak}, Joel P. and {Price}, Paul A. and {Prouza}, Michael and {Radeka}, Veljko and {Rajagopal}, Jayadev and {Rasmussen}, Andrew P. and {Regnault}, Nicolas and {Reil}, Kevin A. and {Reiss}, David J. and {Reuter}, Michael A. and {Ridgway}, Stephen T. and {Riot}, Vincent J. and {Ritz}, Steve and {Robinson}, Sean and {Roby}, William and {Roodman}, Aaron and {Rosing}, Wayne and {Roucelle}, Cecille and {Rumore}, Matthew R. and {Russo}, Stefano and {Saha}, Abhijit and {Sassolas}, Benoit and {Schalk}, Terry L. and {Schellart}, Pim and {Schindler}, Rafe H. and {Schmidt}, Samuel and {Schneider}, Donald P. and {Schneider}, Michael D. and {Schoening}, William and {Schumacher}, German and {Schwamb}, Megan E. and {Sebag}, Jacques and {Selvy}, Brian and {Sembroski}, Glenn H. and {Seppala}, Lynn G. and {Serio}, Andrew and {Serrano}, Eduardo and {Shaw}, Richard A. and {Shipsey}, Ian and {Sick}, Jonathan and {Silvestri}, Nicole and {Slater}, Colin T. and {Smith}, J. Allyn and {Smith}, R. Chris and {Sobhani}, Shahram and {Soldahl}, Christine and {Storrie-Lombardi}, Lisa and {Stover}, Edward and {Strauss}, Michael A. and {Street}, Rachel A. and {Stubbs}, Christopher W. and {Sullivan}, Ian S. and {Sweeney}, Donald and {Swinbank}, John D. and {Szalay}, Alexander and {Takacs}, Peter and {Tether}, Stephen A. and {Thaler}, Jon J. and {Thayer}, John Gregg and {Thomas}, Sandrine and {Thornton}, Adam J. and {Thukral}, Vaikunth and {Tice}, Jeffrey and {Trilling}, David E. and {Turri}, Max and {Van Berg}, Richard and {Vanden Berk}, Daniel and {Vetter}, Kurt and {Virieux}, Francoise and {Vucina}, Tomislav and {Wahl}, William and {Walkowicz}, Lucianne and {Walsh}, Brian and {Walter}, Christopher W. and {Wang}, Daniel L. and {Wang}, Shin-Yawn and {Warner}, Michael and {Wiecha}, Oliver and {Willman}, Beth and {Winters}, Scott E. and {Wittman}, David and {Wolff}, Sidney C. and {Wood-Vasey}, W. Michael and {Wu}, Xiuqin and {Xin}, Bo and {Yoachim}, Peter and {Zhan}, Hu},
        title = "{LSST: From Science Drivers to Reference Design and Anticipated Data Products}",
      journal = {\apj},
         year = 2019,
        month = mar,
       volume = {873},
       number = {2},
          eid = {111},
        pages = {111},
          doi = {10.3847/1538-4357/ab042c},
archivePrefix = {arXiv},
       eprint = {0805.2366},
 primaryClass = {astro-ph},
       adsurl = {https://ui.adsabs.harvard.edu/abs/2019ApJ...873..111I}
}

@ARTICLE{Spergel_2015,
       author = {{Spergel}, D. and {Gehrels}, N. and {Baltay}, C. and {Bennett}, D. and {Breckinridge}, J. and {Donahue}, M. and {Dressler}, A. and {Gaudi}, B.~S. and {Greene}, T. and {Guyon}, O. and {Hirata}, C. and {Kalirai}, J. and {Kasdin}, N.~J. and {Macintosh}, B. and {Moos}, W. and {Perlmutter}, S. and {Postman}, M. and {Rauscher}, B. and {Rhodes}, J. and {Wang}, Y. and {Weinberg}, D. and {Benford}, D. and {Hudson}, M. and {Jeong}, W. -S. and {Mellier}, Y. and {Traub}, W. and {Yamada}, T. and {Capak}, P. and {Colbert}, J. and {Masters}, D. and {Penny}, M. and {Savransky}, D. and {Stern}, D. and {Zimmerman}, N. and {Barry}, R. and {Bartusek}, L. and {Carpenter}, K. and {Cheng}, E. and {Content}, D. and {Dekens}, F. and {Demers}, R. and {Grady}, K. and {Jackson}, C. and {Kuan}, G. and {Kruk}, J. and {Melton}, M. and {Nemati}, B. and {Parvin}, B. and {Poberezhskiy}, I. and {Peddie}, C. and {Ruffa}, J. and {Wallace}, J.~K. and {Whipple}, A. and {Wollack}, E. and {Zhao}, F.},
        title = "{Wide-Field InfrarRed Survey Telescope-Astrophysics Focused Telescope Assets WFIRST-AFTA 2015 Report}",
      journal = {arXiv e-prints},
         year = 2015,
        month = mar,
          eid = {arXiv:1503.03757},
        pages = {arXiv:1503.03757},
          doi = {10.48550/arXiv.1503.03757},
archivePrefix = {arXiv},
       eprint = {1503.03757},
 primaryClass = {astro-ph.IM},
       adsurl = {https://ui.adsabs.harvard.edu/abs/2015arXiv150303757S}
}

@ARTICLE{Eifler_2021,
       author = {{Eifler}, Tim and {Simet}, Melanie and {Krause}, Elisabeth and {Hirata}, Christopher and {Huang}, Hung-Jin and {Fang}, Xiao and {Miranda}, Vivian and {Mandelbaum}, Rachel and {Doux}, Cyrille and {Heinrich}, Chen and {Huff}, Eric and {Miyatake}, Hironao and {Hemmati}, Shoubaneh and {Xu}, Jiachuan and {Rogozenski}, Paul and {Capak}, Peter and {Choi}, Ami and {Dor{\'e}}, Olivier and {Jain}, Bhuvnesh and {Jarvis}, Mike and {Kruk}, Jeffrey and {MacCrann}, Niall and {Masters}, Dan and {Rozo}, Eduardo and {Spergel}, David N. and {Troxel}, Michael and {von der Linden}, Anja and {Wang}, Yun and {Weinberg}, David H. and {Wenzl}, Lukas and {Wu}, Hao-Yi},
        title = "{Cosmology with the Roman Space Telescope: synergies with the Rubin Observatory Legacy Survey of Space and Time}",
      journal = {\mnras},
         year = 2021,
        month = oct,
       volume = {507},
       number = {1},
        pages = {1514-1527},
          doi = {10.1093/mnras/stab533},
archivePrefix = {arXiv},
       eprint = {2004.04702},
 primaryClass = {astro-ph.CO},
       adsurl = {https://ui.adsabs.harvard.edu/abs/2021MNRAS.507.1514E}
}

@ARTICLE{Albrecht_2006,
       author = {{Albrecht}, Andreas and {Bernstein}, Gary and {Cahn}, Robert and {Freedman}, Wendy L. and {Hewitt}, Jacqueline and {Hu}, Wayne and {Huth}, John and {Kamionkowski}, Marc and {Kolb}, Edward W. and {Knox}, Lloyd and {Mather}, John C. and {Staggs}, Suzanne and {Suntzeff}, Nicholas B.},
        title = "{Report of the Dark Energy Task Force}",
      journal = {arXiv e-prints},
         year = 2006,
        month = sep,
          eid = {astro-ph/0609591},
        pages = {astro-ph/0609591},
          doi = {10.48550/arXiv.astro-ph/0609591},
archivePrefix = {arXiv},
       eprint = {astro-ph/0609591},
 primaryClass = {astro-ph},
       adsurl = {https://ui.adsabs.harvard.edu/abs/2006astro.ph..9591A}
}

@ARTICLE{Kannawadi_2016,
       author = {{Kannawadi}, Arun and {Shapiro}, Charles A. and {Mandelbaum}, Rachel and {Hirata}, Christopher M. and {Kruk}, Jeffrey W. and {Rhodes}, Jason D.},
        title = "{The Impact of Interpixel Capacitance in CMOS Detectors on PSF Shapes and Implications for WFIRST}",
      journal = {\pasp},
         year = 2016,
        month = sep,
       volume = {128},
       number = {967},
        pages = {095001},
          doi = {10.1088/1538-3873/128/967/095001},
archivePrefix = {arXiv},
       eprint = {1512.01570},
 primaryClass = {astro-ph.IM},
       adsurl = {https://ui.adsabs.harvard.edu/abs/2016PASP..128i5001K}
}

@ARTICLE{Cao_2024,
       author = {{Cao}, Kaili and {Hirata}, Christopher M. and {Laliotis}, Katherine and {Yamamoto}, Masaya and {Macbeth}, Emily and {Troxel}, M.~A.},
        title = "{Simulating Image Coaddition with the Nancy Grace Roman Space Telescope. III. Software Improvements and New Linear Algebra Strategies}",
      journal = {\apjs},
         year = 2025,
        month = apr,
       volume = {277},
       number = {2},
          eid = {55},
        pages = {55},
          doi = {10.3847/1538-4365/adb580},
archivePrefix = {arXiv},
       eprint = {2410.05442},
 primaryClass = {astro-ph.IM},
       adsurl = {https://ui.adsabs.harvard.edu/abs/2025ApJS..277...55C}
}

@ARTICLE{LSST_2018,
       author = {{The LSST Dark Energy Science Collaboration} and {Mandelbaum}, Rachel and {Eifler}, Tim and {Hlo{\v{z}}ek}, Ren{\'e}e and {Collett}, Thomas and {Gawiser}, Eric and {Scolnic}, Daniel and {Alonso}, David and {Awan}, Humna and {Biswas}, Rahul and {Blazek}, Jonathan and {Burchat}, Patricia and {Chisari}, Nora Elisa and {Dell'Antonio}, Ian and {Digel}, Seth and {Frieman}, Josh and {Goldstein}, Daniel A. and {Hook}, Isobel and {Ivezi{\'c}}, {\v{Z}}eljko and {Kahn}, Steven M. and {Kamath}, Sowmya and {Kirkby}, David and {Kitching}, Thomas and {Krause}, Elisabeth and {Leget}, Pierre-Fran{\c{c}}ois and {Marshall}, Philip J. and {Meyers}, Joshua and {Miyatake}, Hironao and {Newman}, Jeffrey A. and {Nichol}, Robert and {Rykoff}, Eli and {Sanchez}, F. Javier and {Slosar}, An{\v{z}}e and {Sullivan}, Mark and {Troxel}, M.~A.},
        title = "{The LSST Dark Energy Science Collaboration (DESC) Science Requirements Document}",
      journal = {arXiv e-prints},
         year = 2018,
        month = sep,
          eid = {arXiv:1809.01669},
        pages = {arXiv:1809.01669},
          doi = {10.48550/arXiv.1809.01669},
archivePrefix = {arXiv},
       eprint = {1809.01669},
 primaryClass = {astro-ph.CO},
       adsurl = {https://ui.adsabs.harvard.edu/abs/2018arXiv180901669T}
}

@ARTICLE{ROTAC_2025,
       author = {{Roman Observations Time Allocation Committee} and {Core Community Survey Definition Committees}},
        title = "{Roman Observations Time Allocation Committee: Final Report and Recommendations}",
      journal = {arXiv e-prints},
         year = 2025,
        month = may,
          eid = {arXiv:2505.10574},
        pages = {arXiv:2505.10574},
          doi = {10.48550/arXiv.2505.10574},
archivePrefix = {arXiv},
       eprint = {2505.10574},
 primaryClass = {astro-ph.IM},
       adsurl = {https://ui.adsabs.harvard.edu/abs/2025arXiv250510574O}
}

@manual{SRD_2023,
       author = {{Kruk}, Jeffrey and others},
        title = "{Science Requirements Document}",
         year = 2023,
        url = {https://roman.gsfc.nasa.gov/science/docs/RST-SYS-REQ-0020D_DOORs_Export.pdf},
}

@ARTICLE{Berlfein_2025,
       author = {{Berlfein}, Federico and {Mandelbaum}, Rachel and {Li}, Xiangchong and {Zhang}, Tianqing and {Dodelson}, Scott and {Markovic}, Katarina},
        title = "{Chromatic effects on the PSF and shear measurement for the Roman Space Telescope High-Latitude Wide Area Survey}",
      journal = {\mnras},
         year = 2025,
        month = sep,
       volume = {542},
       number = {2},
        pages = {608-628},
          doi = {10.1093/mnras/staf1255},
archivePrefix = {arXiv},
       eprint = {2505.00093},
 primaryClass = {astro-ph.CO},
       adsurl = {https://ui.adsabs.harvard.edu/abs/2025MNRAS.542..608B}
}

@INPROCEEDINGS{Cromey_2023,
       author = {{Cromey}, Benjamin and {Von Handorf}, Robert and {Pedroncelli}, John and {Delker}, Thomas and {Sabatke}, Derek},
        title = "{Science filter performance summary for the Roman Space Telescope Wide Field Instrument}",
    booktitle = {UV/Optical/IR Space Telescopes and Instruments: Innovative Technologies and Concepts XI},
         year = 2023,
       editor = {{Barto}, Allison A. and {Keller}, Fanny and {Stahl}, H. Philip},
       series = {Society of Photo-Optical Instrumentation Engineers (SPIE) Conference Series},
       volume = {12676},
        month = oct,
          eid = {126760R},
        pages = {126760R},
          doi = {10.1117/12.2676488},
       adsurl = {https://ui.adsabs.harvard.edu/abs/2023SPIE12676E..0RC}
}

@BOOK{Born_1980,
       author = {{Born}, Max and {Wolf}, Emil},
        title = "{Principles of Optics Electromagnetic Theory of Propagation, Interference and Diffraction of Light}",
         year = 1980,
         publisher = "Cambridge University Press",
       adsurl = {https://ui.adsabs.harvard.edu/abs/1980poet.book.....B}
}

@ARTICLE{Noll_1976,
       author = {{Noll}, R.~J.},
        title = "{Zernike polynomials and atmospheric turbulence.}",
      journal = {Journal of the Optical Society of America (1917-1983)},
         year = 1976,
        month = mar,
       volume = {66},
        pages = {207-211},
          doi = {10.1364/JOSA.66.000207},
       adsurl = {https://ui.adsabs.harvard.edu/abs/1976JOSA...66..207N}
}

@misc{Sahlmann_2019,
  author       = {Johannes Sahlmann and
                  Shannon Osborne and
                  Colin Cox and
                  charlesrp and
                  David Law and
                  Marshall Perrin and
                  marthaboyer and
                  Joseph Hunkeler},
  title        = {{spacetelescope/pysiaf: Introduce check against
                   online version}},
  month        = oct,
  year         = 2019,
  publisher    = {Zenodo},
  version      = {v0.5.1},
  doi          = {10.5281/zenodo.3516964},
  url          = {https://doi.org/10.5281/zenodo.3516964}
}

@misc{Eifler_2023,
  author       = {Eifler, Tim and Hirata, Christopher and others},
  title        = {Optimizing the Roman HLWAS},
  year         = {2023},
  howpublished = {Roman Core Community Survey White Paper},
  note         = {High Latitude Wide Area Survey (Imaging)},
  url          = {https://asd.gsfc.nasa.gov/roman/white_papers/2023/070_Eifler_HLWAS.pdf}
}

@ARTICLE{Eifler_2024,
       author = {{Eifler}, Tim and {Fang}, Xiao and {Krause}, Elisabeth and {Hirata}, Christopher M. and {Xu}, Jiachuan and {Benabed}, Karim and {Ferraro}, Simone and {Miranda}, Vivian and {Pranjal R.}, S. and {Ay{\c{c}}oberry}, Emma and {Dubois}, Yohan},
        title = "{Cosmology from weak lensing, galaxy clustering, CMB lensing and tSZ: II. Optimizing Roman survey design for CMB cross-correlation science}",
      journal = {arXiv e-prints},
         year = 2024,
        month = nov,
          eid = {arXiv:2411.04088},
        pages = {arXiv:2411.04088},
          doi = {10.48550/arXiv.2411.04088},
archivePrefix = {arXiv},
       eprint = {2411.04088},
 primaryClass = {astro-ph.CO},
       adsurl = {https://ui.adsabs.harvard.edu/abs/2024arXiv241104088E}
}

@misc{Linden_2023,
  author       = {von der Linden, Anja and Allen, Steven and Avestruz, Camille and Hilton, Lucie and Kelly, Patrick and Krause, Elisabeth and Mantz, Adam and McDonald, Michael and Miranda, Vivian and Ricci, Marina and Seitz, Stella and Shin, Tae-hyeon and Wright, Adam},
  title        = {A wide-area extension to the Roman HLIS for multiwavelength galaxy cluster science},
  year         = {2023},
  howpublished = {Roman Core Community Survey White Paper},
  url          = {https://asd.gsfc.nasa.gov/roman/white_papers/2023/084_Linden_HLWAS.pdf}
}

@ARTICLE{Astier_2013,
       author = {{Astier}, P. and {El Hage}, P. and {Guy}, J. and {Hardin}, D. and {Betoule}, M. and {Fabbro}, S. and {Fourmanoit}, N. and {Pain}, R. and {Regnault}, N.},
        title = "{Photometry of supernovae in an image series: methods and application to the SuperNova Legacy Survey (SNLS)}",
      journal = {\aap},
         year = 2013,
        month = sep,
       volume = {557},
          eid = {A55},
        pages = {A55},
          doi = {10.1051/0004-6361/201321668},
archivePrefix = {arXiv},
       eprint = {1306.5153},
 primaryClass = {astro-ph.IM},
       adsurl = {https://ui.adsabs.harvard.edu/abs/2013A&A...557A..55A}
}

@ARTICLE{Lee_2023,
       author = {{Lee}, J. and {Acevedo}, M. and {Sako}, M. and {Vincenzi}, M. and {Brout}, D. and {Sanchez}, B. and {Chen}, R. and {Davis}, T.~M. and {Jarvis}, M. and {Scolnic}, D. and {Qu}, H. and {Galbany}, L. and {Kessler}, R. and {Lasker}, J. and {Sullivan}, M. and {Wiseman}, P. and {Aguena}, M. and {Allam}, S. and {Alves}, O. and {Andrade-Oliveira}, F. and {Bertin}, E. and {Bocquet}, S. and {Brooks}, D. and {Burke}, D.~L. and {Carnero Rosell}, A. and {Carrasco Kind}, M. and {Carretero}, J. and {Costanzi}, M. and {da Costa}, L.~N. and {Pereira}, M.~E.~S. and {De Vicente}, J. and {Desai}, S. and {Diehl}, H.~T. and {Doel}, P. and {Everett}, S. and {Ferrero}, I. and {Friedel}, D. and {Frieman}, J. and {Garc{\'\i}a-Bellido}, J. and {Gerdes}, D.~W. and {Gruen}, D. and {Gruendl}, R.~A. and {Gutierrez}, G. and {Hinton}, S.~R. and {Hollowood}, D.~L. and {Honscheid}, K. and {James}, D.~J. and {Kent}, S. and {Kuehn}, K. and {Kuropatkin}, N. and {Mena-Fern{\'a}ndez}, J. and {Miquel}, R. and {Ogando}, R.~L.~C. and {Palmese}, A. and {Pieres}, A. and {Malag{\'o}n}, A.~A. Plazas and {Raveri}, M. and {Reil}, K. and {Rodriguez-Monroy}, M. and {Sanchez}, E. and {Scarpine}, V. and {Sevilla-Noarbe}, I. and {Smith}, M. and {Suchyta}, E. and {Tarle}, G. and {To}, C. and {Weaverdyck}, N. and {DES Collaboration}},
        title = "{The Dark Energy Survey Supernova Program: Corrections on Photometry Due to Wavelength-dependent Atmospheric Effects}",
      journal = {\aj},
         year = 2023,
        month = jun,
       volume = {165},
       number = {6},
          eid = {222},
        pages = {222},
          doi = {10.3847/1538-3881/acca15},
archivePrefix = {arXiv},
       eprint = {2304.01858},
 primaryClass = {astro-ph.CO},
       adsurl = {https://ui.adsabs.harvard.edu/abs/2023AJ....165..222L}
}

@ARTICLE{Jimenez_2015,
       author = {{Jim{\'e}nez-Teja}, Y. and {Ben{\'\i}tez}, N. and {Molino}, A. and {Fernandes}, C.~A.~C.},
        title = "{Accurate PSF-matched photometry and photometric redshifts for the extreme deep field with the Chebyshev-Fourier functions}",
      journal = {\mnras},
         year = 2015,
        month = oct,
       volume = {453},
       number = {1},
        pages = {1136-1146},
          doi = {10.1093/mnras/stv1612},
archivePrefix = {arXiv},
       eprint = {1507.04750},
 primaryClass = {astro-ph.GA},
       adsurl = {https://ui.adsabs.harvard.edu/abs/2015MNRAS.453.1136J}
}

@ARTICLE{Kitching_2010,
       author = {{Kitching}, Thomas and {Balan}, Sreekumar and {Bernstein}, Gary and {Bethge}, Matthias and {Bridle}, Sarah and {Courbin}, Frederic and {Gentile}, Marc and {Heavens}, Alan and {Hirsch}, Michael and {Hosseini}, Reshad and {Kiessling}, Alina and {Amara}, Adam and {Kirk}, Donnacha and {Kuijken}, Konrad and {Mandelbaum}, Rachel and {Moghaddam}, Baback and {Nurbaeva}, Guldariya and {Paulin-Henriksson}, Stephane and {Rassat}, Anais and {Rhodes}, Jason and {Sch{\"o}lkopf}, Bernhard and {Shawe-Taylor}, John and {Gill}, Mandeep and {Shmakova}, Marina and {Taylor}, Andy and {Velander}, Malin and {van Waerbeke}, Ludovic and {Witherick}, Dugan and {Wittman}, David and {Harmeling}, Stefan and {Heymans}, Catherine and {Massey}, Richard and {Rowe}, Barnaby and {Schrabback}, Tim and {Voigt}, Lisa},
        title = "{Gravitational Lensing Accuracy Testing 2010 (GREAT10) Challenge Handbook}",
      journal = {arXiv e-prints},
         year = 2010,
        month = sep,
          eid = {arXiv:1009.0779},
        pages = {arXiv:1009.0779},
          doi = {10.48550/arXiv.1009.0779},
archivePrefix = {arXiv},
       eprint = {1009.0779},
 primaryClass = {astro-ph.CO},
       adsurl = {https://ui.adsabs.harvard.edu/abs/2010arXiv1009.0779K}
}

@ARTICLE{Akeson_2019,
       author = {{Akeson}, Rachel and {Armus}, Lee and {Bachelet}, Etienne and {Bailey}, Vanessa and {Bartusek}, Lisa and {Bellini}, Andrea and {Benford}, Dominic and {Bennett}, David and {Bhattacharya}, Aparna and {Bohlin}, Ralph and {Boyer}, Martha and {Bozza}, Valerio and {Bryden}, Geoffrey and {Calchi Novati}, Sebastiano and {Carpenter}, Kenneth and {Casertano}, Stefano and {Choi}, Ami and {Content}, David and {Dayal}, Pratika and {Dressler}, Alan and {Dor{\'e}}, Olivier and {Fall}, S. Michael and {Fan}, Xiaohui and {Fang}, Xiao and {Filippenko}, Alexei and {Finkelstein}, Steven and {Foley}, Ryan and {Furlanetto}, Steven and {Kalirai}, Jason and {Gaudi}, B. Scott and {Gilbert}, Karoline and {Girard}, Julien and {Grady}, Kevin and {Greene}, Jenny and {Guhathakurta}, Puragra and {Heinrich}, Chen and {Hemmati}, Shoubaneh and {Hendel}, David and {Henderson}, Calen and {Henning}, Thomas and {Hirata}, Christopher and {Ho}, Shirley and {Huff}, Eric and {Hutter}, Anne and {Jansen}, Rolf and {Jha}, Saurabh and {Johnson}, Samson and {Jones}, David and {Kasdin}, Jeremy and {Kelly}, Patrick and {Kirshner}, Robert and {Koekemoer}, Anton and {Kruk}, Jeffrey and {Lewis}, Nikole and {Macintosh}, Bruce and {Madau}, Piero and {Malhotra}, Sangeeta and {Mandel}, Kaisey and {Massara}, Elena and {Masters}, Daniel and {McEnery}, Julie and {McQuinn}, Kristen and {Melchior}, Peter and {Melton}, Mark and {Mennesson}, Bertrand and {Peeples}, Molly and {Penny}, Matthew and {Perlmutter}, Saul and {Pisani}, Alice and {Plazas}, Andr{\'e}s and {Poleski}, Radek and {Postman}, Marc and {Ranc}, Cl{\'e}ment and {Rauscher}, Bernard and {Rest}, Armin and {Roberge}, Aki and {Robertson}, Brant and {Rodney}, Steven and {Rhoads}, James and {Rhodes}, Jason and {Ryan}, Jr., Russell and {Sahu}, Kailash and {Sand}, David and {Scolnic}, Dan and {Seth}, Anil and {Shvartzvald}, Yossi and {Siellez}, Karelle and {Smith}, Arfon and {Spergel}, David and {Stassun}, Keivan and {Street}, Rachel and {Strolger}, Louis-Gregory and {Szalay}, Alexander and {Trauger}, John and {Troxel}, M.~A. and {Turnbull}, Margaret and {van der Marel}, Roeland and {von der Linden}, Anja and {Wang}, Yun and {Weinberg}, David and {Williams}, Benjamin and {Windhorst}, Rogier and {Wollack}, Edward and {Wu}, Hao-Yi and {Yee}, Jennifer and {Zimmerman}, Neil},
        title = "{The Wide Field Infrared Survey Telescope: 100 Hubbles for the 2020s}",
      journal = {arXiv e-prints},
         year = 2019,
        month = feb,
          eid = {arXiv:1902.05569},
        pages = {arXiv:1902.05569},
          doi = {10.48550/arXiv.1902.05569},
archivePrefix = {arXiv},
       eprint = {1902.05569},
 primaryClass = {astro-ph.IM},
       adsurl = {https://ui.adsabs.harvard.edu/abs/2019arXiv190205569A}
}

@ARTICLE{Troxel_2021,
       author = {{Troxel}, M.~A. and {Long}, H. and {Hirata}, C.~M. and {Choi}, A. and {Jarvis}, M. and {Mandelbaum}, R. and {Wang}, K. and {Yamamoto}, M. and {Hemmati}, S. and {Capak}, P.},
        title = "{A synthetic Roman Space Telescope High-Latitude Imaging Survey: simulation suite and the impact of wavefront errors on weak gravitational lensing}",
      journal = {\mnras},
         year = 2021,
        month = feb,
       volume = {501},
       number = {2},
        pages = {2044-2070},
          doi = {10.1093/mnras/staa3658},
archivePrefix = {arXiv},
       eprint = {1912.09481},
 primaryClass = {astro-ph.IM},
       adsurl = {https://ui.adsabs.harvard.edu/abs/2021MNRAS.501.2044T}
}

@ARTICLE{McKechnie_1992,
       author = {{McKechnie}, T.~S.},
        title = "{Atmospheric turbulence and the resolution limits of large ground-based telescopes.}",
      journal = {Journal of the Optical Society of America A},
         year = 1992,
        month = nov,
       volume = {9},
       number = {11},
        pages = {1937-1954},
          doi = {10.1364/JOSAA.9.001937},
       adsurl = {https://ui.adsabs.harvard.edu/abs/1992JOSAA...9.1937M}
}

@ARTICLE{LSST_2009,
       author = {{LSST Science Collaboration} and {Abell}, Paul A. and {Allison}, Julius and {Anderson}, Scott F. and {Andrew}, John R. and {Angel}, J. Roger P. and {Armus}, Lee and {Arnett}, David and {Asztalos}, S.~J. and {Axelrod}, Tim S. and {Bailey}, Stephen and {Ballantyne}, D.~R. and {Bankert}, Justin R. and {Barkhouse}, Wayne A. and {Barr}, Jeffrey D. and {Barrientos}, L. Felipe and {Barth}, Aaron J. and {Bartlett}, James G. and {Becker}, Andrew C. and {Becla}, Jacek and {Beers}, Timothy C. and {Bernstein}, Joseph P. and {Biswas}, Rahul and {Blanton}, Michael R. and {Bloom}, Joshua S. and {Bochanski}, John J. and {Boeshaar}, Pat and {Borne}, Kirk D. and {Bradac}, Marusa and {Brandt}, W.~N. and {Bridge}, Carrie R. and {Brown}, Michael E. and {Brunner}, Robert J. and {Bullock}, James S. and {Burgasser}, Adam J. and {Burge}, James H. and {Burke}, David L. and {Cargile}, Phillip A. and {Chandrasekharan}, Srinivasan and {Chartas}, George and {Chesley}, Steven R. and {Chu}, You-Hua and {Cinabro}, David and {Claire}, Mark W. and {Claver}, Charles F. and {Clowe}, Douglas and {Connolly}, A.~J. and {Cook}, Kem H. and {Cooke}, Jeff and {Cooray}, Asantha and {Covey}, Kevin R. and {Culliton}, Christopher S. and {de Jong}, Roelof and {de Vries}, Willem H. and {Debattista}, Victor P. and {Delgado}, Francisco and {Dell'Antonio}, Ian P. and {Dhital}, Saurav and {Di Stefano}, Rosanne and {Dickinson}, Mark and {Dilday}, Benjamin and {Djorgovski}, S.~G. and {Dobler}, Gregory and {Donalek}, Ciro and {Dubois-Felsmann}, Gregory and {Durech}, Josef and {Eliasdottir}, Ardis and {Eracleous}, Michael and {Eyer}, Laurent and {Falco}, Emilio E. and {Fan}, Xiaohui and {Fassnacht}, Christopher D. and {Ferguson}, Harry C. and {Fernandez}, Yanga R. and {Fields}, Brian D. and {Finkbeiner}, Douglas and {Figueroa}, Eduardo E. and {Fox}, Derek B. and {Francke}, Harold and {Frank}, James S. and {Frieman}, Josh and {Fromenteau}, Sebastien and {Furqan}, Muhammad and {Galaz}, Gaspar and {Gal-Yam}, A. and {Garnavich}, Peter and {Gawiser}, Eric and {Geary}, John and {Gee}, Perry and {Gibson}, Robert R. and {Gilmore}, Kirk and {Grace}, Emily A. and {Green}, Richard F. and {Gressler}, William J. and {Grillmair}, Carl J. and {Habib}, Salman and {Haggerty}, J.~S. and {Hamuy}, Mario and {Harris}, Alan W. and {Hawley}, Suzanne L. and {Heavens}, Alan F. and {Hebb}, Leslie and {Henry}, Todd J. and {Hileman}, Edward and {Hilton}, Eric J. and {Hoadley}, Keri and {Holberg}, J.~B. and {Holman}, Matt J. and {Howell}, Steve B. and {Infante}, Leopoldo and {Ivezic}, Zeljko and {Jacoby}, Suzanne H. and {Jain}, Bhuvnesh and {R} and {Jedicke} and {Jee}, M. James and {Garrett Jernigan}, J. and {Jha}, Saurabh W. and {Johnston}, Kathryn V. and {Jones}, R. Lynne and {Juric}, Mario and {Kaasalainen}, Mikko and {Styliani} and {Kafka} and {Kahn}, Steven M. and {Kaib}, Nathan A. and {Kalirai}, Jason and {Kantor}, Jeff and {Kasliwal}, Mansi M. and {Keeton}, Charles R. and {Kessler}, Richard and {Knezevic}, Zoran and {Kowalski}, Adam and {Krabbendam}, Victor L. and {Krughoff}, K. Simon and {Kulkarni}, Shrinivas and {Kuhlman}, Stephen and {Lacy}, Mark and {Lepine}, Sebastien and {Liang}, Ming and {Lien}, Amy and {Lira}, Paulina and {Long}, Knox S. and {Lorenz}, Suzanne and {Lotz}, Jennifer M. and {Lupton}, R.~H. and {Lutz}, Julie and {Macri}, Lucas M. and {Mahabal}, Ashish A. and {Mandelbaum}, Rachel and {Marshall}, Phil and {May}, Morgan and {McGehee}, Peregrine M. and {Meadows}, Brian T. and {Meert}, Alan and {Milani}, Andrea and {Miller}, Christopher J. and {Miller}, Michelle and {Mills}, David and {Minniti}, Dante and {Monet}, David and {Mukadam}, Anjum S. and {Nakar}, Ehud and {Neill}, Douglas R. and {Newman}, Jeffrey A. and {Nikolaev}, Sergei and {Nordby}, Martin and {O'Connor}, Paul and {Oguri}, Masamune and {Oliver}, John and {Olivier}, Scot S. and {Olsen}, Julia K. and {Olsen}, Knut and {Olszewski}, Edward W. and {Oluseyi}, Hakeem and {Padilla}, Nelson D. and {Parker}, Alex and {Pepper}, Joshua and {Peterson}, John R. and {Petry}, Catherine and {Pinto}, Philip A. and {Pizagno}, James L. and {Popescu}, Bogdan and {Prsa}, Andrej and {Radcka}, Veljko and {Raddick}, M. Jordan and {Rasmussen}, Andrew and {Rau}, Arne and {Rho}, Jeonghee and {Rhoads}, James E. and {Richards}, Gordon T. and {Ridgway}, Stephen T. and {Robertson}, Brant E. and {Roskar}, Rok and {Saha}, Abhijit and {Sarajedini}, Ata and {Scannapieco}, Evan and {Schalk}, Terry and {Schindler}, Rafe and {Schmidt}, Samuel},
        title = "{LSST Science Book, Version 2.0}",
      journal = {arXiv e-prints},
         year = 2009,
        month = dec,
          eid = {arXiv:0912.0201},
        pages = {arXiv:0912.0201},
          doi = {10.48550/arXiv.0912.0201},
archivePrefix = {arXiv},
       eprint = {0912.0201},
 primaryClass = {astro-ph.IM},
       adsurl = {https://ui.adsabs.harvard.edu/abs/2009arXiv0912.0201L}
}

@ARTICLE{Berlfein_2026,
       author = {{Berlfein}, Federico and {Mandelbaum}, Rachel and {Xu}, Jiachuan and {Zhang}, Tianqing},
        title = "{Optimizing the Roman Space Telescope High-Latitude Wide Area Survey for mitigating chromatic PSF effects on shear measurement}",
      journal = {arXiv e-prints},
         year = 2026,
        month = mar,
          eid = {arXiv:2603.15763},
        pages = {arXiv:2603.15763},
          doi = {10.48550/arXiv.2603.15763},
archivePrefix = {arXiv},
       eprint = {2603.15763},
 primaryClass = {astro-ph.CO},
       adsurl = {https://ui.adsabs.harvard.edu/abs/2026arXiv260315763B}
}

@INPROCEEDINGS{Pasquale_2018,
       author = {{Pasquale}, Bert A. and {Casey}, Thomas and {Marx}, Catherine and {Gao}, Guangjun and {Armani}, Nerses and {Content}, David and {Hagopian}, John and {Jurling}, Alden and {Jackson}, Clifton and {Liu}, Alice and {Whipple}, Art and {Murray}, Jacob},
        title = "{Optical design and predicted performance of the WFIRST phase-b imaging optics assembly and wide field instrument}",
    booktitle = {Current Developments in Lens Design and Optical Engineering XIX},
         year = 2018,
       editor = {{Johnson}, R. Barry and {Mahajan}, Virendra N. and {Thibault}, Simon},
       series = {Society of Photo-Optical Instrumentation Engineers (SPIE) Conference Series},
       volume = {10745},
        month = sep,
          eid = {107450K},
        pages = {107450K},
          doi = {10.1117/12.2325859},
       adsurl = {https://ui.adsabs.harvard.edu/abs/2018SPIE10745E..0KP}
}

@ARTICLE{Switzer_2025,
       author = {{Switzer}, Eric R. and {Bray}, Evan and {Will}, Scott D. and {Cromey}, Benjamin and {Gao}, Guangjun and {Groff}, Tyler D. and {Jurling}, Alden S. and {Kruk}, Jeffrey and {Marx}, Catherine T. and {Morey}, Peter A. and {Patel}, Jessica and {Quijada}, Manuel A. and {Rizzo}, Maxime J. and {Schlieder}, Joshua E. and {Wollack}, Edward J.},
        title = "{Laboratory characterization of widefield filter transmission for the Nancy Grace Roman Space Telescope's Wide Field Instrument}",
      journal = {\ao},
         year = 2025,
        month = dec,
       volume = {64},
       number = {35},
        pages = {10525},
          doi = {10.1364/AO.569503},
       adsurl = {https://ui.adsabs.harvard.edu/abs/2025ApOpt..6410525S}
}

@ARTICLE{Baron_2023,
       author = {{Baron}, M. and {Sassolas}, B. and {Pinard}, L. and {Ealet}, A.},
        title = "{Numerical modelling for retrieval of the coating thickness variations from wavefront errors measurements}",
      journal = {Optics Express},
         year = 2023,
        month = sep,
       volume = {31},
       number = {20},
        pages = {32968},
          doi = {10.1364/OE.494683},
       adsurl = {https://ui.adsabs.harvard.edu/abs/2023OExpr..3132968B}
}

@ARTICLE{EuclidSchirmer_2022,
       author = {{Euclid Collaboration} and {Schirmer}, M. and {Jahnke}, K. and {Seidel}, G. and {Aussel}, H. and {Bodendorf}, C. and {Grupp}, F. and {Hormuth}, F. and {Wachter}, S. and {Appleton}, P.~N. and {Barbier}, R. and {Brinchmann}, J. and {Carrasco}, J.~M. and {Castander}, F.~J. and {Coupon}, J. and {De Paolis}, F. and {Franco}, A. and {Ganga}, K. and {Hudelot}, P. and {Jullo}, E. and {Lan{\c{c}}on}, A. and {Nucita}, A.~A. and {Paltani}, S. and {Smadja}, G. and {Strafella}, F. and {Venancio}, L.~M.~G. and {Weiler}, M. and {Amara}, A. and {Auphan}, T. and {Auricchio}, N. and {Balestra}, A. and {Bender}, R. and {Bonino}, D. and {Branchini}, E. and {Brescia}, M. and {Capobianco}, V. and {Carbone}, C. and {Carretero}, J. and {Casas}, R. and {Castellano}, M. and {Cavuoti}, S. and {Cimatti}, A. and {Cledassou}, R. and {Congedo}, G. and {Conselice}, C.~J. and {Conversi}, L. and {Copin}, Y. and {Corcione}, L. and {Costille}, A. and {Courbin}, F. and {Da Silva}, A. and {Degaudenzi}, H. and {Douspis}, M. and {Dubath}, F. and {Dupac}, X. and {Dusini}, S. and {Ealet}, A. and {Farrens}, S. and {Ferriol}, S. and {Fosalba}, P. and {Frailis}, M. and {Franceschi}, E. and {Franzetti}, P. and {Fumana}, M. and {Garilli}, B. and {Gillard}, W. and {Gillis}, B. and {Giocoli}, C. and {Grazian}, A. and {Guzzo}, L. and {Haugan}, S.~V.~H. and {Hoekstra}, H. and {Holmes}, W. and {Hornstrup}, A. and {K{\"u}mmel}, M. and {Kermiche}, S. and {Kiessling}, A. and {Kilbinger}, M. and {Kitching}, T. and {Kohley}, R. and {Kunz}, M. and {Kurki-Suonio}, H. and {Laureijs}, R. and {Ligori}, S. and {Lilje}, P.~B. and {Lloro}, I. and {Maciaszek}, T. and {Maiorano}, E. and {Mansutti}, O. and {Marggraf}, O. and {Markovic}, K. and {Marulli}, F. and {Massey}, R. and {Maurogordato}, S. and {Mellier}, Y. and {Meneghetti}, M. and {Merlin}, E. and {Meylan}, G. and {Moresco}, M. and {Moscardini}, L. and {Munari}, E. and {Nakajima}, R. and {Nichol}, R.~C. and {Niemi}, S.~M. and {Padilla}, C. and {Pasian}, F. and {Pedersen}, K. and {Percival}, W.~J. and {Pettorino}, V. and {Pires}, S. and {Poncet}, M. and {Popa}, L. and {Pozzetti}, L. and {Prieto}, E. and {Raison}, F. and {Rhodes}, J. and {Rix}, H.-W. and {Roncarelli}, M. and {Rossetti}, E. and {Saglia}, R. and {Sartoris}, B. and {Scaramella}, R. and {Schneider}, P. and {Secroun}, A. and {Serrano}, S. and {Sirignano}, C. and {Sirri}, G. and {Stanco}, L. and {Tallada-Cresp{\'\i}}, P. and {Taylor}, A.~N. and {Teplitz}, H.~I. and {Tereno}, I. and {Toledo-Moreo}, R. and {Torradeflot}, F. and {Trifoglio}, M. and {Valentijn}, E.~A. and {Valenziano}, L. and {Wang}, Y. and {Weller}, J. and {Zamorani}, G. and {Zoubian}, J. and {Andreon}, S. and {Bardelli}, S. and {Boucaud}, A. and {Camera}, S. and {Farinelli}, R. and {Graci{\'a}-Carpio}, J. and {Maino}, D. and {Medinaceli}, E. and {Mei}, S. and {Morisset}, N. and {Polenta}, G. and {Renzi}, A. and {Romelli}, E. and {Tenti}, M. and {Vassallo}, T. and {Zacchei}, A. and {Zucca}, E. and {Baccigalupi}, C. and {Balaguera-Antol{\'\i}nez}, A. and {Biviano}, A. and {Blanchard}, A. and {Borgani}, S. and {Bozzo}, E. and {Burigana}, C. and {Cabanac}, R. and {Cappi}, A. and {Carvalho}, C.~S. and {Casas}, S. and {Castignani}, G. and {Colodro-Conde}, C. and {Cooray}, A.~R. and {Courtois}, H.~M. and {Crocce}, M. and {Cuby}, J.-G. and {Davini}, S. and {de la Torre}, S. and {Di Ferdinando}, D. and {Escartin}, J.~A. and {Farina}, M. and {Ferreira}, P.~G. and {Finelli}, F. and {Fotopoulou}, S. and {Galeotta}, S. and {Garcia-Bellido}, J. and {Gaztanaga}, E. and {George}, K. and {Gozaliasl}, G. and {Hook}, I.~M. and {Ili{\'c}}, S. and {Kansal}, V. and {Kashlinsky}, A. and {Keihanen}, E. and {Kirkpatrick}, C.~C. and {Lindholm}, V. and {Mainetti}, G. and {Maoli}, R. and {Martinelli}, M. and {Martinet}, N. and {Maturi}, M.},
        title = "{Euclid preparation. XVIII. The NISP photometric system}",
      journal = {\aap},
         year = 2022,
        month = jun,
       volume = {662},
          eid = {A92},
        pages = {A92},
          doi = {10.1051/0004-6361/202142897},
archivePrefix = {arXiv},
       eprint = {2203.01650},
 primaryClass = {astro-ph.IM},
       adsurl = {https://ui.adsabs.harvard.edu/abs/2022A&A...662A..92E}
}



\appendix

\section{Dependence of defocus on pre-existing aberrations}\label{app:defocus}

\begin{figure*}
    \centering
    \includegraphics[width=0.99\linewidth]{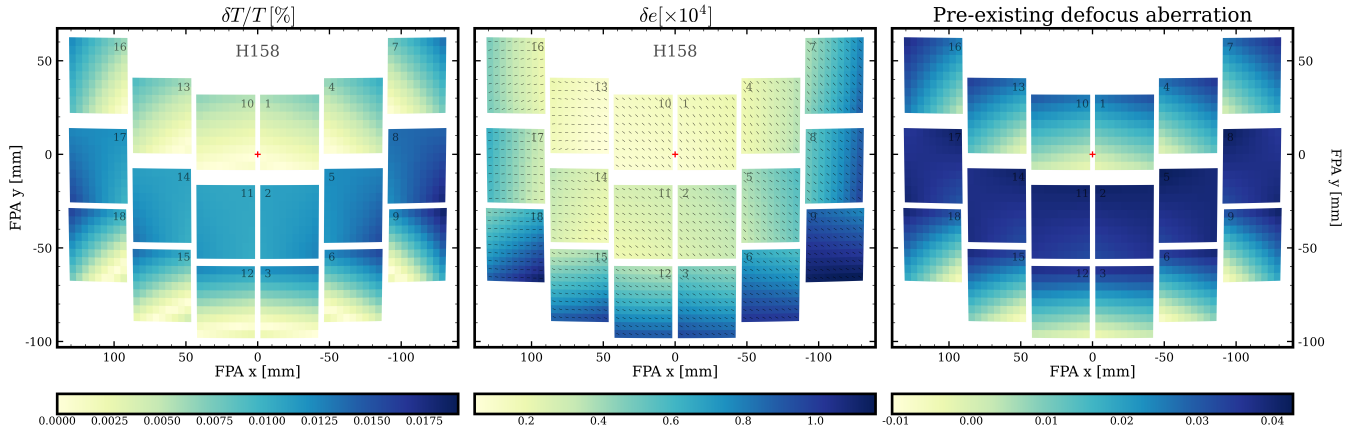}
    \caption{Similar to Fig.~\ref{fig:fov_dependence} but for defocus, with an additional panel on the right showing the pre-existing defocus aberration across the focal plane. More specifically, the right panel is showing the negative of the fourth zernike coefficient $a_4$. We clearly see incoherence and detector dependence of the PSF size residuals across the field of view. This is due to the effect of the added defocus on the PSF size being dependent on the pre-existing defocus aberrations, which can vary between detectors. This dependence can be seen by comparing the left and right panels. The same cannot be said about the ellipticity residuals, which do not seem to depend on pre-existing aberrations like the PSF size does. We also note that the PSF size residuals can be negative, which happens when the effective longitudinal shift due to refraction causes the out-of-focus point to move closer to the detector plane.}
    \label{fig:defocus_fov}

\end{figure*}

In Sec.~\ref{subsubsec:spatial_dep} 
we noted that the spatial dependence of the PSF size residuals induced by the added defocus term is much less smooth than that produced by decentering. In particular, the amplitude of the defocus-induced size residual can vary abruptly from detector to detector. Fig.~\ref{fig:defocus_fov} shows this dependence of the PSF size residuals on the specific detector. This behavior is not driven primarily by the spatial dependence of the longitudinal chromatic shift itself. Rather, it arises because the impact of an additional defocus perturbation depends on the pre-existing aberrations of the underlying Roman PSF. In practice, we find that the spatial pattern of the size residuals induced by the added defocus closely tracks the spatial pattern of the pre-existing size aberrations in the native Roman PSF model, which can be seen directly by comparing the left and right panels of Fig.~\ref{fig:defocus_fov}. The flux residuals, although not shown explicitly, track the size residuals and are considerably smaller than the $1\%$ requirement. 

The origin of this behavior can be understood from a simple Fourier-optics argument. Let the pupil-plane complex field be
\begin{equation}
U(\boldsymbol{\rho}) = P(\boldsymbol{\rho})\,e^{i\phi(\boldsymbol{\rho})},
\end{equation}
where \(P\) is the pupil transmission, $\rho$ is the normalized pupil radius, and \(\phi\) is the pupil phase. 
The focal-plane PSF is \(I(\boldsymbol{\theta}) = |\Psi(\boldsymbol{\theta})|^2\), where \(\Psi\) is the Fraunhofer transform of \(U\) and \(\boldsymbol{\theta}=(\theta_1,\theta_2)\) denotes the angular coordinate in the focal plane, measured relative to the PSF centroid so that \(\bar{\boldsymbol{\theta}}=0\). With this choice of origin, we define the normalized trace of the second-moment matrix, \(R^2 = Q_{11}+Q_{22}\), written in angular coordinates, as 
\begin{equation}
R^2 \equiv
\frac{\int |\boldsymbol{\theta}|^2 I(\boldsymbol{\theta})\, \mathrm{d}^2\theta}
{\int I(\boldsymbol{\theta})\, \mathrm{d}^2\theta}.
\end{equation}
Parseval's theorem gives
\begin{equation}
R^2
=
\left(\frac{\lambda f}{2\pi}\right)^2
\frac{\int |\nabla_{\rho} U(\boldsymbol{\rho})|^2\, d^2\rho}
{\int |U(\boldsymbol{\rho})|^2\, d^2\rho}.
\label{eq:R2_parseval}
\end{equation}
Using
\begin{equation}
\nabla U
=
\nabla\!\left(Pe^{i\phi}\right)
=
e^{i\phi}\left(\nabla P + iP\nabla\phi\right),
\end{equation}
we obtain
\begin{equation}
|\nabla U|^2 = |\nabla P|^2 + P^2|\nabla\phi|^2,
\end{equation}
since the cross term cancels in the dot product. Because \(|U|^2=P^2\), Equation~(\ref{eq:R2_parseval}) becomes
\begin{equation}
R^2
=
\left(\frac{\lambda f}{2\pi}\right)^2
\left[
\frac{\int |\nabla P|^2\,d^2\rho}{\int P^2\,d^2\rho}
+
\frac{\int P^2|\nabla\phi|^2\,d^2\rho}{\int P^2\,d^2\rho}
\right].
\end{equation}
For a hard-edged uniform pupil, the \(|\nabla P|^2\) term is a pupil-edge constant and does not change when defocus is added. The change in \(R^2\) therefore depends only on the phase term. Writing the pupil phase as
\begin{equation}
\phi(\boldsymbol{\rho}) = 2\pi W(\boldsymbol{\rho}),
\end{equation}
the aberration-dependent part of \(R^2\) is proportional to the pupil average of \(|\nabla W|^2\). The change in \(R^2\) produced by modifying the wavefront from \(W\) to \(W'\) is then
\begin{equation}
\Delta R^2
=
(\lambda f)^2
\left[
\left\langle |\nabla W'|^2 \right\rangle_P
-
\left\langle |\nabla W|^2 \right\rangle_P
\right],
\label{eq:dR2_general}
\end{equation}
where \(\langle \cdot \rangle_P\) denotes the pupil-weighted average.

Now consider adding an incremental defocus term \(\Delta a_4\) to an existing wavefront,
\begin{equation}
W'(\boldsymbol{\rho}) = W(\boldsymbol{\rho}) + \Delta a_4\, Z_4(\boldsymbol{\rho}),
\end{equation}
where \(Z_4\) is the Noll defocus mode. Expanding Equation~(\ref{eq:dR2_general}) gives
\begin{equation}
\Delta R^2
=
(\lambda f)^2
\left[
2\Delta a_4 \left\langle \nabla W \cdot \nabla Z_4 \right\rangle_P
+
(\Delta a_4)^2 \left\langle |\nabla Z_4|^2 \right\rangle_P
\right].
\label{eq:dR2_expand}
\end{equation}
Thus, the change in PSF size squared is not determined only by the newly added defocus term \(\Delta a_4\); it also depends linearly on the coupling between the existing wavefront \(W\) and the added defocus through the cross term \(\langle \nabla W \cdot \nabla Z_4 \rangle_P\). If the pre-existing wavefront already contains defocus, so that
\begin{equation}
W(\boldsymbol{\rho}) = W_{\rm other}(\boldsymbol{\rho}) + a_4 Z_4(\boldsymbol{\rho}),
\end{equation}
and the remaining aberrations are approximately uncorrelated with \(Z_4\), then
\begin{equation}
\left\langle \nabla W \cdot \nabla Z_4 \right\rangle_P
\simeq
a_4 \left\langle |\nabla Z_4|^2 \right\rangle_P.
\end{equation}
Equation~(\ref{eq:dR2_expand}) then reduces to
\begin{equation}
\Delta R^2
\simeq
(\lambda f)^2
\left(2a_4\Delta a_4 + (\Delta a_4)^2\right)
\left\langle |\nabla Z_4|^2 \right\rangle_P.
\label{eq:dR2_final}
\end{equation}

Equation~(\ref{eq:dR2_final}) shows explicitly why the size residual induced by an added defocus term depends on the pre-existing aberration state of the PSF. In the regime relevant here, the cross term \(2a_4\Delta a_4\) generally dominates over the quadratic term \((\Delta a_4)^2\), so the induced size change is expected to scale approximately with the native defocus already present in the Roman PSF model. This explains the behavior seen in Fig.~\ref{fig:defocus_fov}, 
where the resulting PSF size residual is modulated by the detector-dependent pattern of the pre-existing aberrations, producing a more irregular spatial structure. Conversely, the shape residuals in Fig.~\ref{fig:defocus_fov} 
do not show the same strong dependence on the pre-existing aberration pattern. Instead, they follow the expected field-angle dependence more closely, with the residuals generally increasing toward larger distances from the center of the focal plane. 


\bsp	
\label{lastpage}
\end{document}